\documentclass{aa}  

\usepackage{graphicx}
\usepackage{txfonts}
\usepackage{xcolor}
\usepackage{soul}
\usepackage{hyperref}

\usepackage{multirow}
\usepackage{mathrsfs}
\usepackage{gensymb}
\usepackage{amsmath}

\begin{document} 

   \title{A clustering-based search for substructures in the Galactic plane and bulge using RR Lyrae stars as tracers}

   \author{N. Cristi-Cambiaso\inst{1, 2}
          \and
          C. Navarrete\inst{1,3}
          \and
          M. Catelan\inst{2, 4}
          \and
          M. Zoccali\inst{2, 4}
          \and
          C. Quezada\inst{2, 4}
          }

   \institute{Université Côte d’Azur, Observatoire de la Côte d’Azur, CNRS, Laboratoire Lagrange, Bd de l’Observatoire, CS 34229, 06304, Nice Cedex 4, France \\ \email{nicolas.cristi@oca.eu}
            \and
            Instituto de Astrofísica, Pontificia Universidad Católica de Chile, Av. Vicuña Mackenna 4860, 7820436 Macul, Santiago, Chile
            \and
            Centro de Investigación en Ciencias del Espacio y Física Teórica, Universidad Central de Chile, Avenida Francisco de Aguirre 0405, La Serena, Chile
            \and 
            Millennium Institute of Astrophysics, Nuncio Monseñor Sotero Sanz 100, Of. 104, Providencia, Santiago, Chile
             }

   \date{Received xx; accepted yy}

  \abstract
   {Although many globular clusters (GCs) have been identified in the Galaxy, their population is estimated to be incomplete, especially in regions with significant crowding and/or interstellar extinction, such as the Galactic bulge and plane. RR Lyrae stars, as bright standard candles and tracers of old populations, hold immense potential in the search for GCs in these regions. Furthermore, large catalogs of RR Lyrae stars in these areas have become available in recent years.}
   {We aim to build a sample of RR Lyrae stars with six-dimensional information (three-dimensional positions, proper motions, and metallicities) in the Galactic plane and bulge, and to exploit it with a hierarchical clustering algorithm to search for new Galactic substructures.}
   {We build a sample of fundamental-mode RR Lyrae (RRab) stars in the Galactic plane and bulge with positions, distances, proper motions, and photometric metallicity estimates, using data from the \textit{Gaia} and VVV surveys. Using a clustering algorithm calibrated to optimize the recovery of GCs, we form groups of RRab stars with similar positions in the six-dimensional space studied. Finally, to identify the most promising RRab groups among the many artifacts produced by the clustering algorithm, we compare their properties with those of known GCs.}
   {We find many RRab groups associated with known Galactic GCs. Additionally, we estimate the first RR Lyrae-based distances for the GCs BH~140 and NGC~5986, further constraining their positions in the Milky Way. We detect small groups of $2 - 3$ RRab stars, located at distances of up to $\sim 25$ kpc, that are not associated with any known GC, but exhibit GC-like distributions across all six parameters analyzed. Several of these groups \textemdash mostly pairs \textemdash are found toward the Galactic bulge, but have distinct proper motions or distances, indicating that they may not belong to the bulge population.}
   {By exploiting an RRab sample in the Galactic plane and bulge with a hierarchical clustering algorithm, we identify dozens of groups displaying GC-like properties, which are excellent candidates for further follow-up observations. Furthermore, future radial velocity measurements could evaluate if the RRab members of our groups are truly moving together.}

   \keywords{Galaxy: structure - (\textit{Galaxy:}) globular clusters: general - Stars: variables: RR Lyrae}

\titlerunning{Search for substructures in the Galactic plane and bulge using RR Lyrae stars} 

   \maketitle

\defcitealias{Dekany2022}{DG22}
%

\section{Introduction}
\label{sec:introduction}

Galaxies grow through cosmic time by merging with smaller satellite galaxies and accreting material over time \citep[see, e.g.,][]{White1991}. This mechanism is a key feature of the Lambda Cold Dark Matter paradigm, which proposes that the growth of galaxies is driven by the gravitational interactions between dark matter halos, a process that is commonly referred to as the hierarchical model for galaxy formation \citep{WhiteRees1978}.
Hierarchical clustering can significantly influence the properties of a galaxy, such as enhancing its star formation \citep[see, e.g.,][]{Mihos1996, Lambas2003}, changing its morphology \citep[e.g.,][]{Lotz2008, Hopkins2009}, or triggering an active galactic nucleus \citep[e.g.,][]{Hopkins2008}. Therefore, understanding this hierarchical growth is essential for explaining the large-scale structure of the Universe and the formation of complex galaxy systems.

Plenty of evidence suggests that the formation history of the Milky Way (MW) includes many interactions with less massive companions, although the exact number, masses, and timeline of these events are still under debate \citep[see][and references therein, for thorough reviews on the formation history of the MW]{Helmi2020, Deason2024}.
These smaller galaxies can host their own families of globular clusters (GCs), which can be destroyed by the tidal forces exerted on them during the galactic merging events, thus depositing their stars into the MW. Indeed, several works have found that disrupted GCs have contributed to the build-up of the MW halo and bulge field components, though the exact amount remains the subject of debate \citep[see, e.g.,][]{Altmann2005, Fernandez-Trincado2019, Koch2019, Hanke2020, Ferraro2021, Belokurov2023, Xu2024}.
Several GCs have been found in the process of being stripped, often exhibiting symmetric tidal tails emerging from the (still visible) cluster core \citep[see, e.g.,][]{Sollima2020}; examples include Palomar 5 \citep[e.g.,][]{Starkman2020}, $\omega$~Centauri \citep[e.g.,][]{Ibata2019}, NGC~288 \citep[e.g.,][]{Grillmair2025}, and NGC~7492 \citep[e.g.,][]{Navarrete2017}.
However, some GCs may endure the tidal forces during the merger and survive to contribute to the present-day MW GC population \citep{Brodie2006,Trujillo-Gomez2021}, as ex-situ GCs.
To tell these GCs apart from those formed in the MW (in-situ), ages, chemical abundances, and dynamical properties can be used in combination with numerical simulations \citep[see, e.g.,][]{Massari2019,Forbes2020,Belokurov2024}.
With this information, such clusters allow for a reconstruction of the MW's formation history \citep[e.g.,][]{Monty2024,Valenzuela2024}.

Many GCs have been discovered in the MW bulge in the past decade \citep[e.g.,][]{Gran2022, Bica2024}, thanks in large part to the near-infrared (NIR) photometry provided by the Vista Variables in the Vía Láctea ESO Public Survey \citep[VVV;][]{VVV2010, Saito2012, Minniti2020} and its eXtension, VVVx \citep{VVVx2018, Saito2024}. Additionally, although the photometry provided by the \textit{Gaia} survey \citep{Gaia2018, Gaia2023} in this region is severely affected by the dust along the line of sight \citep[see, e.g.,][]{GaiaMathias2023}, the \textit{Gaia} proper motions (PMs) have played a crucial role in the search for new GCs, as they allow for a separation of the cluster members from the field stars.
Recently, \citet{Gran2024} pointed out that the sample of MW GC sample is still incomplete, estimating that $\sim 20$ GCs likely remain undiscovered behind the Galactic bulge and disk.

To probe old stellar populations such as Galactic GCs, RR Lyrae stars are particularly useful \citep[see][and references therein]{Smith2004, Catelan2009, Catelan2015}.
RR Lyrae are low-mass ($\sim 0.6 - 0.8 \, M_{\odot}$) and old ($\gtrsim 10$ Gyr) stars in the core helium-burning stage of their evolution (i.e., the horizontal branch, HB), which radially pulsate with stable periods ($P$) in the range between $0.2$ and $1.0$ days while in the instability strip. These pulsators follow a well-defined period-luminosity-metallicity (PLZ) relation, particularly in the NIR \citep[see, e.g.,][]{Longmore1990, Bono2001, Catelan2004, Marconi2015, Narloch2024, Prudil2024}, which allows for the absolute magnitudes ($M$), and therefore, the distances, to be inferred from their pulsation period.
Conveniently, the metal content of RRab stars correlates with the shape of their light curve \citep[as first discovered by][]{Kovacs1995}, which allows for a photometric metallicity ([Fe/H]$_{\rm phot}$) to be inferred, as a first approximation, in the absence of spectroscopic observations \citep{Jurcsik1996}.

Even in systems where they are commonly found, such as metal-poor GCs, dwarf spheroidal (dSph) galaxies, and even the halo field, these core He-burning stars still represent only ``the tip of the iceberg'', as they are much less numerous than the much fainter (and less evolved) main-sequence and subgiant branch stars. In the general (metal-rich) bulge and disk fields, they are vastly outnumbered not only by main-sequence stars, but also by other evolved tracers, such as red clump and red giant branch (RGB) stars, because metal-rich populations, irrespective of their ages, are extremely inefficient in generating RR Lyrae stars \citep{Taam1976, Layden1995, Dekany2018, Savino2020}. Thus, even a pair of RR Lyrae stars with similar chemical and kinematic properties may serve as a strong indicator of an underlying substructure \citep[e.g.,][]{Baker2015, Mateu2018, Medina2024}, as also demonstrated by known GCs which have only two RR Lyrae members \citep[e.g., NGC~288 and NGC~7492][]{Clement2001, Prudil2024b, CruzReyes2024}. This, combined with their relatively high brightness \citep[$M_{V} \simeq 0.6$ mag;][and references therein]{Catelan2015}, photometric metallicities, and PLZ relation-based distances, makes RR Lyrae stars ideal for the search of (moderately metal-poor) GCs hidden in or behind the Galactic bulge and disk.

To search for groups of RR Lyrae stars with similar positions, kinematics, and metallicities, clustering algorithms such as friends-of-friends \citep{Starkenburg2009}, Density-Based Spatial Clustering of Applications with Noise \citep[DBSCAN,][]{DBSCAN}, or Hierarchical DBSCAN \citep[HDBSCAN,][]{Campello2013, McInnes2017} are commonly used \citep[see, e.g.,][]{Wang2022, Cabrera-Garcia2024, Sheng2024}. 
These clustering algorithms are designed to detect groups in datasets by computing distances between data points in the chosen parameter space, but they vary in their approaches to defining clusters and handling noise.

In this work, we employ a hierarchical clustering algorithm to study RR Lyrae stars with 3D positions, kinematic information (i.e., PMs), and photometric metallicities near the Galactic plane and bulge, aiming to discover and characterize new Galactic GCs.
In Sect.~\ref{sec:data}, we describe the construction of our RR Lyrae catalog, while Sect.~\ref{sec:methods} details the preparation of an RR Lyrae sample with kinematics and photometric metallicities, and the clustering procedure employed to identify substructure candidates. Section~\ref{sec:results} describes the main findings of this study, including the most promising groups of RR Lyrae stars found. Finally, Sect. \ref{sec:conclusions} presents the discussion and conclusions drawn from our research.

\section{Data}
\label{sec:data}

In this study, we mainly utilize the RR Lyrae stars included in the third data release (DR3) of the European Space Agency (ESA) \textit{Gaia} mission \citep{Gaia2016, Gaia2023}.
As listed in \citet{Gaia2023}, \textit{Gaia} DR3 provides positions, optical photometry, PMs, and parallaxes for $1.5$ billion sources. Additionally, photometric time series are provided for around $11$ million sources, with variability classifications assigned to approximately $10.5$ million of these. 

A sample of 271\,779 RR Lyrae stars characterized and validated with \textit{Gaia} DR3 data is provided in \citet{Clementini2023}.
This sample includes periods, Fourier parameters of the $G$-band light curves, [$G$, $G_{\rm RP}$, $G_{\rm BP}$]-band (intensity-averaged) magnitudes, photometric metallicities, foreground extinction values, among other parameters \citep[see the full list in Table 12 of][]{Clementini2023}. From this sample, 175\,350 stars are classified as fundamental-mode pulsators (ab-type), 94\,422 are pulsating in the first-overtone (c-type), and 2007 are double-mode RR Lyrae (d-type). In this study, we focus on RRab stars, which are less likely to include contaminants from other variable types. Additionally, since our goal is to study regions near the Galactic plane and bulge, we discard stars based on their Galactic latitude, only keeping RRab stars within $-15 \leq b(\degree) \leq 15$. This cut leaves 93\,961 RRab stars in this sample. To retrieve PM measurements for these stars, we perform a crossmatch between the RR Lyrae sample and the main \textit{Gaia} DR3 catalog \citep[\texttt{gaiadr3.gaia\_source},][]{Gaia2023}.

By comparing the validated RR Lyrae sample with the fourth phase of the Optical Gravitational Lensing Experiment \citep[OGLE-IV,][]{OGLE-IV}, \citet{Clementini2023} estimate the completeness of their catalog in the Galactic bulge area. This comparison is carried out in two circular regions with $3\degree$ radius centered at $\ell = 0.9 \degree$ and $b = 6.1 \degree$ (bulge-down) and at $\ell = 1.0 \degree$ and $b=-7.6 \degree$ (bulge-up). They find completeness values of $79 \%$ and $82 \%$ in the bulge-up and bulge-down regions, respectively.
Since \textit{Gaia}'s optical photometry is strongly affected by dust absorption, we expect their completeness to decrease significantly in regions within a few degrees of the Galactic plane, where interstellar extinction is most prominent \citep[see, e.g.,][]{GaiaMathias2023}; therefore, complementing the \textit{Gaia} survey with longer wavelength photometry, which is less affected by dust absorption, can be highly beneficial for our study.

With this in mind, we utilize the NIR photometry made available by the VVV ESO Public Survey. Specifically, we make use of the RRab samples classified and characterized by \citet{Molnar2022} and \citet{ZoccaliQuezada2024}, using data from the VVV survey and, in the case of the former, its extension, VVVx.

\begin{figure*}
    \begin{center}
    \includegraphics[width=1.6\columnwidth]{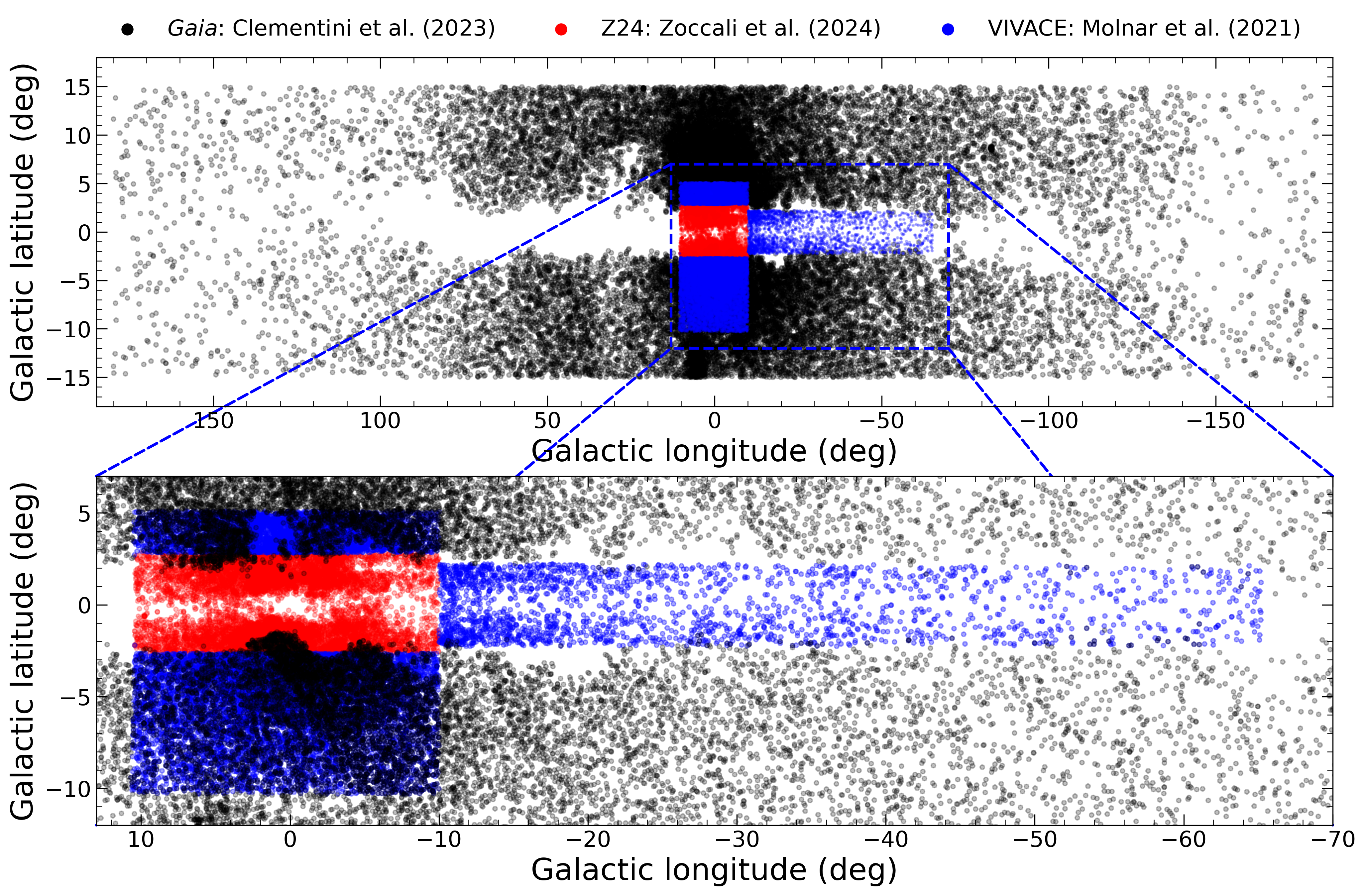}
    \end{center}
    \caption{Sky distribution in Galactic coordinates of our RRab samples after the quality cuts described in Sect.~\ref{sec:quality_cuts}. RRab stars from \textit{Gaia} DR3 are shown as black circles, while those from Z24 and VIVACE are shown as red and blue circles, respectively. The lower panel is a zoom-in of the upper one, as highlighted with blue dashed lines. For a better visualization, stars in the VVV samples are plotted on top in the upper panel, while stars in the \textit{Gaia} sample are plotted on top in the lower panel.}
    \label{fig:sample_sky}
\end{figure*}

The VIrac VAriable Classification Ensemble (VIVACE), presented in \citet{Molnar2022}, provides ~39\,000 RRab stars (with classification probability > $0.9$) distributed throughout the VVV footprint (obtained using both VVV and VVVx data), covering the Galactic bulge and the southern disk of the MW. The catalog provided by \citet[][hereafter, Z24]{ZoccaliQuezada2024}, on the other hand, includes 16\,488 RRab stars in the inner bulge region within $-10\degree \lesssim \ell \lesssim 10\degree$ and $-2.8\degree \lesssim b \lesssim 2.8\degree$.
As the Z24 catalog has been optimized to prioritize high purity over completeness, it includes fewer stars than the VIVACE RRab sample; however, the Z24 catalog ensures fewer contaminants from other variable types and reduced uncertainties in their analysis. Taking this into account, we adopt the Z24 catalog as our RRab sample in the inner bulge region of the VVV footprint, while using the VIVACE sample for the rest of the VVV footprint.
The spatial distributions of our adopted RRab samples (after quality cuts described in Sect.~\ref{sec:quality_cuts}) are illustrated in Fig.~\ref{fig:sample_sky}.

For the RRab stars in the VIVACE sample, we retrieved PM estimates from the second version of the VVV Infrared Astrometric Catalogue \citep[VIRAC2,][]{VIRAC2}, which is based on a point-spread function (PSF) fitting of nearly a decade of VVV and VVVx images.
On the other hand, although the public Z24 catalog does not include PMs, the authors have estimated them using the multi-epoch VVV PSF photometry \citep[see][for details on the PM determination]{ContrerasRamos2017}, and their derived PMs have been made available and incorporated into our analysis.

The PMs we use for the stars in both the VIVACE and Z24 catalogs were calibrated to the \textit{Gaia} DR3 astrometric system \citep[see][for more details about this calibration for the Z24 catalog]{ZoccaliRojas24}. Therefore, there is no systematic difference between the PMs in the VVV and \textit{Gaia} samples.
In order to adopt the most precise PM estimates available, which leads to more robust groups being found with our clustering algorithm (see Sect.~\ref{sec:clustering}), we crossmatch the VIVACE and Z24 samples with the \textit{Gaia} sample and adopt, for each star, the PM estimate (from either \textit{Gaia} or VVV data) with the smallest uncertainty.

\section{Methods}
\label{sec:methods}

From the 93\,961 RRab stars selected in the Galactic plane area from the \textit{Gaia} sample (i.e., those within $-15 \leq b(\degree) \leq 15$), we additionally discard those that do not have estimates for their phase difference $\varphi_{31}$ (required to estimate photometric metallicities, see Sect.~\ref{sec:FeH}), intensity-averaged $G_{\rm RP}$-band magnitude (used to estimate distances, see Sect.~\ref{sec:dist}), or PMs. These criteria leave us with 41\,410 RRab stars in our \textit{Gaia} sample.

\subsection{Calculating metallicities}
\label{sec:FeH}

Although the \textit{Gaia} sample includes [Fe/H]$_{\rm phot}$ values for tens of thousands of stars, these estimates have been found to be unreliable in both \citet{Jurcsik2023} and \citet{Muraveva2025}, where revised estimates for these metallicities are also provided.

We compute metallicities for the RRab stars in our \textit{Gaia} sample using the relation presented in \citet[][Eq. 2]{Jurcsik2023}, which was validated with a calibrating sample comprised of spectroscopic [Fe/H] measurements from the literature of $35$ Galactic GCs (see their Sect. 2.2 for more details about their calibrating sample). We determine uncertainties in these metallicities by propagating the errors in the coefficients of the [Fe/H]$_{\rm phot}$ relation, along with the errors in the periods (which are generally negligible) and in $\varphi_{31}$ \citep[provided by][]{Clementini2023}. Additionally, the root-mean-square (RMS = 0.152 dex) reported in \citet{Jurcsik2023} is added in quadrature to these uncertainties.

Z24 estimate photometric metallicities for their RRab sample by employing the methods developed by \citet[][hereafter DG22]{Dekany2022}.
This work utilizes deep learning to predict photometric metallicities of RRab stars, taking the light curve of the stars in either the optical \textit{Gaia} $G$-band or the NIR VISTA $K_{\rm s}$-band (Z24 use the latter) as input, with a validating sample of [Fe/H]$_{\rm phot}$ measurements obtained using $I$-band light curves \citep[see][]{Dekany2021}. Noticeably, the metallicity errors provided by this code display a distribution peaking at $\simeq 0.03$~dex, while those obtained for our \textit{Gaia} sample show a distribution peaking at $\simeq 0.25$~dex. This discrepancy arises because the uncertainties from \citetalias{Dekany2022} do not account for the errors in the metallicities of their validating sample or the spread of their predictions relative to that sample. To incorporate these factors, we consider the mean absolute error (MAE) between the $I$-band [Fe/H]$_{\rm phot}$ values of \citet{Dekany2021} and their spectroscopic calibrating sample ($0.16$~dex), and the MAE between the [Fe/H]$_{\rm phot}$ predictions of \citetalias{Dekany2022}, obtained using $K_{\rm s}$-band light curves, and their $I$-band [Fe/H]$_{\rm phot}$ calibrating sample ($0.11$~dex). These quantities are added in quadrature to the [Fe/H]$_{\rm phot}$ statistical errors provided by \citetalias{Dekany2022}.
With this, the distribution of the revised [Fe/H]$_{\rm phot}$ errors in the Z24 sample now has a steep rise (and peak) at $\simeq 0.2$~dex.

For the VIVACE RRab sample, we follow the steps of \citet{ZoccaliQuezada2024} and derive [Fe/H]$_{\rm phot}$ values using the algorithm presented in \citetalias{Dekany2022}, with $K_{\rm s}$-band light curves as input. Additionally, we apply the same correction for the errors described above.

\subsection{Estimating distances}
\label{sec:dist}

The RRab sample provided by \citet{Clementini2023} includes parallaxes, which can be used to derive distances. However, parallax angles become very small at large distances, which leads to very significant uncertainties.
Therefore, to derive reliable distances for our \textit{Gaia} RRab sample, we employ the $G_{\rm RP}$-band PLZ relation derived by \citet[][Eq. 20]{Prudil2024}.
The $G_{\rm RP}$ band relation was chosen because this is the \textit{Gaia} band covering the longest wavelengths, where the luminosity of RR Lyrae stars is less dependent on their metallicity, and more so on their period \citep[e.g.,][]{Catelan2004, Prudil2024}. Since the periods are obtained with much greater precision than the metallicities, this leads to more precise distance estimations.
Additionally, the $G_{\rm RP}$-band magnitudes are less affected by interstellar extinction, and its the relation has a smaller intrinsic scatter \citep[$\varepsilon = 0.107$ mag,][]{Prudil2024} than those they obtained using the $G$ ($\varepsilon = 0.118$ mag) and $G_{\rm BP}$ ($\varepsilon = 0.126$ mag) bands. We compute the errors in the absolute magnitudes by propagating the errors in the quantities used for their estimation (period and metallicity), the errors of the PLZ coefficients, and adding the PLZ spread of $0.107$ mag in quadrature.

To compute distances for our \textit{Gaia} sample, we adopt the two-dimensional interstellar reddening map produced by \protect\citet[][hereafter, SFD]{Schlegel1998}. This map provides $E(B-V)$ reddening values, which we convert to interstellar extinction values in the $V$ band ($\mathcal{A}_V$) by using an absolute-to-selective extinction ratio of $R_V = 2.742$ \citep[from][]{Schlafly2011}. This value of $R_V$ accounts for a recalibration of the SFD map using the extinction law from \citet{Fitzpatrick1999} and $R_V = 3.1$.
Finally, we convert these extinction values to the $G_{\rm RP}$ band ($\mathcal{A}_{G_{\rm RP}}$) using a relative extinction of $\mathcal{A}_{G_{\rm RP}} / \mathcal{A}_V = 0.589 \pm 0.004$ \citep{Wang2019}.
We assume a $15\%$ uncertainty for the extinction values, which we propagate into the distance estimation, in addition to the uncertainties from the absolute and apparent magnitudes.

For the VIVACE and Z24 samples, we compute predicted absolute magnitudes in the $J$ and $K_{\rm s}$ bands using the relations presented in \citet[][Eqs.~16~and~18]{Prudil2024}. Using these magnitudes, we compute the intrinsic colors of the stars, ($J - K_{\rm s}$)$_0$, which we then subtract from their observed colors, $J - K_{\rm s}$, to obtain individual reddening estimates, $E(J - K_{\rm s})$. Finally, using a selective-to-total extinction ratio of $A_{K_{\rm s}}/E(J-K_{\rm s})~=~0.465 \pm 0.022$ from \citet{Minniti2020}, we obtain the $K_{\rm s}$-band extinction values and compute the distances of the RRab stars using their $K_{\rm s}$-band absolute magnitudes.

\subsection{Quality cuts}
\label{sec:quality_cuts}

With our samples now including metallicities, distances ($d$), and PM values ($\mu_{\ell} \cos b$ and $\mu_{\ell}$), we remove the stars with large errors (which are represented by $\sigma$) or possibly unreliable values in these quantities. Specifically, we keep stars with metallicities in the range of $-2.5 \, {\rm dex} \leq {\rm [Fe/H]} \leq 0 \, {\rm dex}$, metallicity errors $\sigma_{\rm [Fe/H]} \leq 0.5 \, {\rm dex}$, relative distance errors $\sigma_d / d \leq 0.2$, and PM errors ($\sigma_{\mu_{\ell} \cos b}$, $\sigma_{\mu_{\ell}}$) $\leq 1 \, {\rm mas \, yr}^{-1}$.


Additionally, in the \textit{Gaia} sample, we remove stars with renormalised unit weight error (RUWE) values larger than $1.4$, as this indicates poor quality in the astrometric solution \citep[see][and references therein]{Gaia2021}. For such stars in the VIVACE and Z24 samples, the PMs from VVV photometry are adopted (if they satisfy the cuts mentioned above).

Using these criteria, our final \textit{Gaia}, Z24, and VIVACE samples consist of 34\,829, 11\,265, and 18\,389 RRab stars, respectively, with six-dimensional information available (including three-dimensional positions, two-dimensional kinematics, and metallicities).
As shown in Fig.~\ref{fig:sample_sky}, these samples slightly overlap with each other on the sky, so some stars are included in more than one of these. Specifically, our \textit{Gaia} and VIVACE samples share 7\,136 RRab stars (most of which are located in the outer regions of the bulge), the \textit{Gaia} and Z24 samples share 799 stars, and the Z24 and VIVACE samples share only 6 stars located in the edges of the inner bulge region covered by Z24, as we removed the inner-bulge stars from the VIVACE sample for this study.

\subsection{Hierarchical clustering}
\label{sec:clustering}

To identify and characterize potential new GCs or overdensities, we search for groups of RRab stars exhibiting similar positions, kinematics, and metal content using the HDBSCAN algorithm.
HDBSCAN is an unsupervised clustering algorithm that transforms the input parameter space, builds the minimum spanning tree and a cluster hierarchy for the input sample, then condenses the cluster tree and extracts the clusters from this tree (for a detailed description of the algorithm, we refer the reader to the HDBSCAN documentation\footnote{\url{https://hdbscan.readthedocs.io/en/latest/index.html}}).

HDBSCAN has been used in various fields of astrophysics, such as studying clustering properties among a sample of young stellar objects \citep[e.g.,][]{Vioque2023}, variable star classification \citep[e.g.,][]{Pantoja2022}, and finding substructures, through their kinematic and/or chemical properties, in the halo of the MW \citep{Koppelman2019b, Lovdal2022, Shank2022a, Shank2022b, Shank2023, Ou2023, Kim2025}.
Furthermore, several low-mass bulge GCs were first discovered by \citet{Gran2022} using DBSCAN.
Additionally, HDBSCAN has been shown to have a high fidelity and purity (compared to other unsupervised clustering algorithms) when applied to an open cluster sample \citep{Hunt2021} and a sample of accreted ultra-faint dwarf galaxies \citep{Brauer2022}.

In this study, we apply HDBSCAN to the six-dimensional space defined by ($\ell$, $b$), ($\mu_{\ell} \cos b$, $\mu_b$), $d$, and [Fe/H]. Radial velocities (RVs) are not included since only a small fraction of our RRab stars ($842$, $2 \%$ of our \textit{Gaia} sample) have \textit{Gaia} DR3 RV measurements, and restricting our analysis to stars with available RV information would greatly reduce our sample size. Due to systematic differences between the \textit{Gaia}, Z24, and VIVACE samples \textemdash such as different sky coverages \textemdash we analyze them separately.
Before applying HDBSCAN, we subtract the median of each parameter and scale the data according to their interquartile range, using the \texttt{RobustScaler} function from the \texttt{scikit-learn} Python package \citep{sklearn}. This scaling ensures that differences in the order of magnitude and level of spread among the six dimensions do not influence their importance in the clustering process.

Parameter selection in HDBSCAN is critical, as it greatly influences the properties and, most importantly, the fidelity of the resulting groups. HDBSCAN offers several adjustable parameters, with the most important being \texttt{min\_cluster\_size}, which determines the smallest group of objects that can be considered as a cluster. In our study, we set this value to $2$, as even a pair of RRab stars can indicate a significant substructure \citep[see, e.g.,][]{Baker2015}, and there are several known GCs which only have $2$ RR Lyrae stars, such as NGC 288 and NGC 7492 \citep[see, e.g.,][]{Clement2001, Prudil2024b, CruzReyes2024}.
Additionally, in order to obtain groups with a small number of RRab star members, we use the cluster selection method \texttt{leaf} (instead of the default option, \texttt{eom}), which selects leaf nodes from the cluster hierarchy tree.
Other key parameters, such as \texttt{cluster\_selection\_epsilon}, which sets the distance threshold below which clusters are not split up further, and \texttt{min\_samples}, which determines the minimum number of points that must be present within the neighbourhood of a point to consider it a ``core'' point (building block for clusters), further refine the clustering criteria.

Additionally, after the interquartile scaling, we apply scaling factors (denoted by $S$) to each variable. 
This helps account for how compact (or spread) GCs can be in the six quantities used, when compared to the range of values these quantities take in our RRab samples.
We calibrate the HDBSCAN parameters (\texttt{min\_samples} and \texttt{cluster\_selection\_epsilon}) and the scaling factors to optimize the recovery of confirmed RRab members of known GCs. The GCs used for these calibrations and the literature references adopted for the identification of their confirmed RRab members are presented in Appendix~\ref{app:GC_members}.
To carry out this calibration, we define two key metrics: purity ($P$) and cohesion ($C$). Purity is defined as
\begin{equation}
    P_{{\rm GC}_{i}} = \frac{N_{{\rm GC}_{i}}^{\text{max}}}{N_T},
\end{equation}
\noindent where $N_{{\rm GC}_{i}}^{\text{max}}$ is the number of member stars from GC$_{i}$ in the HDBSCAN group with the most stars from GC$_{i}$, and $N_T$ is the total number of stars in that group. Thus, a higher purity indicates that GC members are recovered with fewer outliers (i.e., stars not classified as members in the literature).
Cohesion, on the other hand, measures the extent to which the stars of a GC are grouped together in a single HDBSCAN group, and it is defined as

\begin{equation}
    C_{{\rm GC}_{i}} = \frac{N_{{\rm GC}_{i}}^{\text{max}}}{N_{{\rm GC}_{i}}},
\end{equation}

\noindent where $N_{{\rm GC}_{i}}$ is the total number of stars in GC$_{i}$ that are in our sample. Thus, a higher cohesion indicates a more unified recovery of the members of a GC.

To provide a balanced assessment of both large and small GCs, we compute a weighted average purity and a weighted average cohesion, where we weigh each GC by $W_{{\rm GC}_{i}} = \log(N_{{\rm GC}_{i}})$. The weighted average purity and cohesion are then given by

\begin{align}
    \bar{P} &= \frac{\sum_{{\rm GC}_{i}} W_{{\rm GC}_{i}} P_{{\rm GC}_{i}}}{\sum_{{\rm GC}_{i}} W_{{\rm GC}_{i}}} \text{,} \nonumber\\
    \bar{C} &= \frac{\sum_{{\rm GC}_{i}} W_{{\rm GC}_{i}} C_{{\rm GC}_{i}}}{\sum_{{\rm GC}_{i}} W_{{\rm GC}_{i}}}.
\end{align}



Finally, to find the HDBSCAN parameters and scaling factors that maximize $\bar{P} + \bar{C}$, we perform a Bayesian optimization of these metrics using the \texttt{gp\_minimize} function provided in the \texttt{scikit-optimize}\footnote{\url{https://scikit-optimize.github.io/stable/}.} Python library.
Scaling factors equal to zero were also considered (i.e., where a parameter is not used in the clustering). The final parameters found to maximize the $\bar{P} + \bar{C}$ metric in each of our samples are listed in Table~\ref{tab:HDBSCAN_parameters}, with any HDBSCAN parameters not listed being set to their default value. 

\begin{table}
    \caption{HDBSCAN parameters, scaling factors, and minimum stabilities used for the clustering in the \textit{Gaia}, Z24, and VIVACE samples.}
    \resizebox{\columnwidth}{!}{%
    \centering
    \begin{tabular}{c|c|c|c}
        \hline
        \hline
        Parameter & \textit{Gaia} & Z24 & VIVACE \\
        \hline
        \texttt{min\_cluster\_size} & $2$ & $2$& $2$ \\
        \texttt{min\_samples} & $1$ & $1$ & $1$ \\
        \texttt{cluster\_selection\_epsilon} & $0.37$ & $1.50$ & $0.43$ \\
        \texttt{cluster\_selection\_method} & leaf & leaf & leaf \\
        \hline
        $S_{\ell}$ & $300.0$ & $46.4$ & $169.5$ \\
        $S_b$ & $51.2$ & $44.0$ & $30.0$ \\
        $S_{\mu_{\ell} \cos b}$ & $0.57$ & $3.00$ & $1.35$ \\
        $S_{\mu_b}$ & $0.57$ & $3.00$ & $1.35$ \\
        $S_{d}$ & $1.08$ & $1.27$ & $0.53$ \\
        $S_{\rm [Fe/H]}$ & $0.00$ & $1.85$ & $0.13$ \\
        Minimum stability & $0.650$ & $0.625$ & $0.325$ \\
        \hline
        $\bar{P}$ & $0.909$ & $0.909$ & $0.817$ \\
        $\bar{C}$ & $0.832$ & $0.957$ & $0.881$ \\
        $\bar{P} + \bar{C}$ & $1.741$ & $1.866$ & $1.699$ \\
        \hline
    \end{tabular}}
    \label{tab:HDBSCAN_parameters}
\end{table}

Additionally, to ensure that our groups are robust to the observational errors, we incorporate a Monte Carlo (MC) sampling technique in conjunction with the HDBSCAN clustering procedure. Namely, we resample the parameters of all the RRab stars using Gaussian distributions and run the HDBSCAN algorithm (with the calibrated HDBSCAN parameters and scaling factors shown in Table~\ref{tab:HDBSCAN_parameters}) on these resampled parameters. This process is repeated $1\,000$ times for each of our samples.

Then, for each star, we calculate the percentage of MC realizations in which it is grouped with each of the other stars in the sample. Finally, we form groups of stars that are clustered together in at least a minimum fraction of the realizations.
This minimum fraction, representing how consistently stars need to be clustered together to be identified as a group by our algorithm, is referred to as the minimum ``stability'', and has a range of possible values of [$0 - 1$].
The optimal value of the minimum stability for each sample is obtained by grouping the stars according to different minimum stabilities, and finding the value for which $\bar{P} + \bar{C}$ is at its maximum.

\section{Results}
\label{sec:results}

As shown in Table~\ref{tab:HDBSCAN_parameters}, our calibrated HDBSCAN-MC clustering algorithm leads to high purity and cohesion metrics (in the recovery of GC members) across the three samples of RRab stars analyzed, as $\bar{P}$ and $\bar{C}$ values of roughly $0.9$ are consistently obtained. This indicates that, on average, we retrieve for each cluster around $90 \%$ of its members in the same group, with only one outlier for every ten bona fide members. Thus, our algorithm is capable of efficiently grouping GC members while mostly separating them from the field RRab stars. Additionally, although GCs with more RR Lyrae members are given larger weights for the calibration of the clustering parameters, we recover smaller GCs with good purity and cohesion, e.g., GCs that have $2 - 3$ RRab members being recovered with purities of $\simeq 0.75$ and cohesions of $1.0$.
This excellent GC recovery would not be possible without using the \texttt{leaf} cluster selection method, which selects the finest splits in the density tree.
However, our algorithm also detects thousands of other RRab groups, with $76.4 \%$ of the \textit{Gaia} sample clustered into $9813$ groups, $26.8 \%$ of the Z24 sample clustered into $1408$ groups, and $73.8 \%$ of the VIVACE sample clustered into $4750$ groups.
In Sect.~\ref{sec:compact_groups_3_stars}, we introduce metrics to efficiently identify the GC-like groups among the other possible false positives or subdivisions of large Galactic substructures (e.g., the Galactic bulge).
This post-processing allows us to exploit the fine sensitivity of \texttt{leaf} while dealing with the excess of false positives it produces.

When calculating our purity metric, we assume that all stars not included as members in the literature are outliers, which is not necessarily true, as new members could be found with our methods. In Appendix~\ref{app:new_GC_members}, we explore the groups that contain confirmed GC members, searching for RR Lyrae stars within them that are not included in the literature, but exhibit similar properties to the confirmed members.

\subsection{Distance and [Fe/H]$_{\rm phot}$ estimates}
\label{sec:GC_distanceFeH}

The distance and metallicity of most of the GCs in our calibrating sample (see Table~\ref{tab:calibrating_GCs}) are well established \citep[see, e.g.,][]{Harris2010, Baumgardt2021}. Therefore, to determine how reliable the distances and metallicities of our RRab groups can be, we estimate mean values for the GCs in our samples, and compare these with values from the literature.
For this comparison, we only consider GCs with at least two confirmed members that satisfy our selection criteria described in Appendix~\ref{app:GC_members} (see Table~\ref{tab:calibrating_GCs}).
However, we note that NGC~6441 (which is used for the calibration of the clustering parameters in the \textit{Gaia} and VIVACE samples) was not included in this comparison, as it is a metal-rich GC \citep[$-0.46$ dex,][]{Harris2010} with long-period RR Lyrae variables \citep{Layden1999, Pritzl2000, Pritzl2003, Corwin2006}, whose metallicities are not reliably reproduced by photometric relations \citep[see, e.g.,][]{Clementini2005, Jurcsik2023, Prudil2024b, Kunder2024}.

\begin{figure*}
    \begin{center}
    \includegraphics[width=1.6\columnwidth]{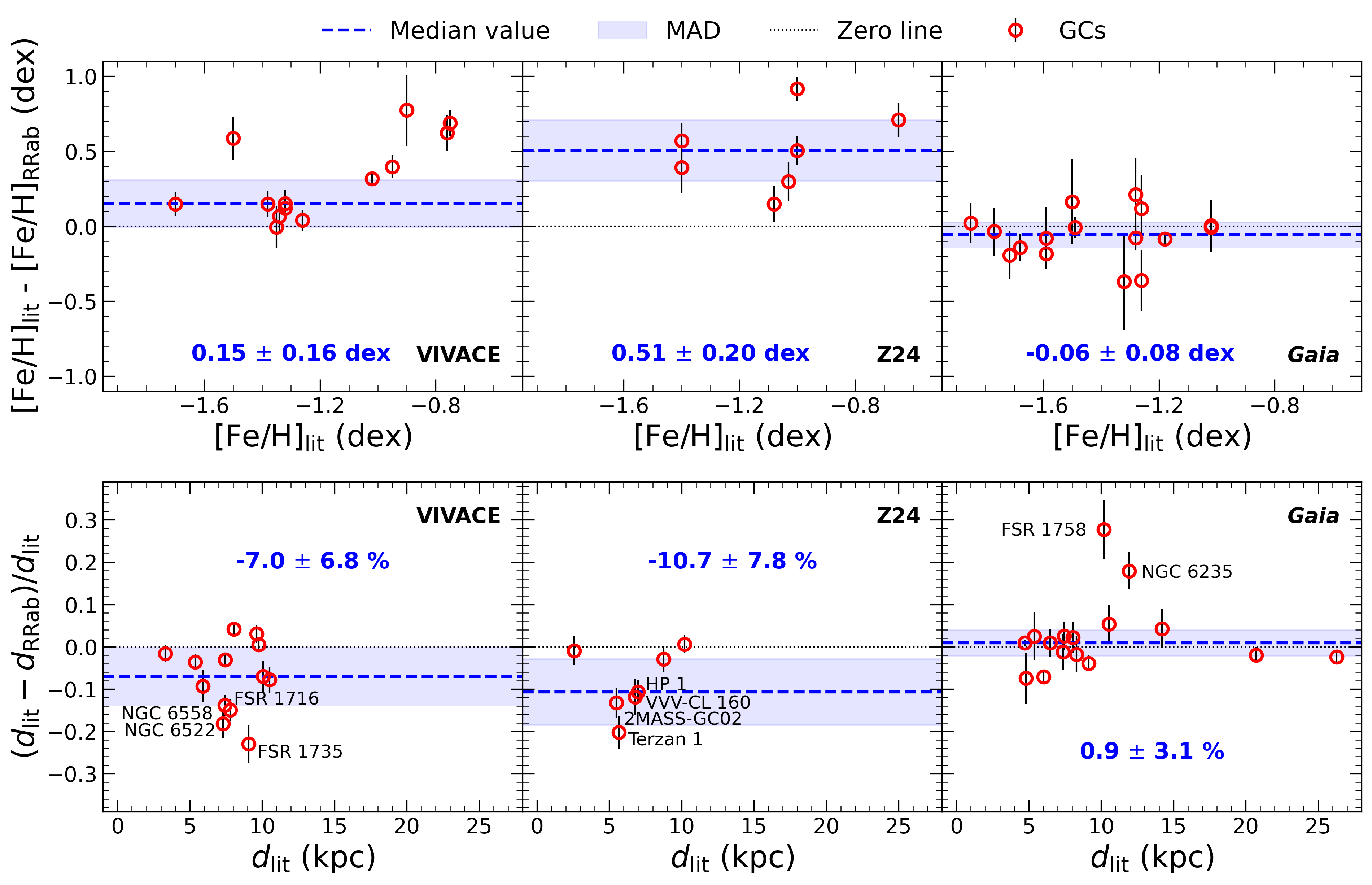}
    \end{center}
    \caption{Comparison of the clusters' [Fe/H] (upper panels) and distance (lower panels) values derived from their RRab stars and their values from the literature. Clusters in the VIVACE, Z24, and \textit{Gaia} samples are shown in the left, center, and right panels, respectively. The black dotted lines show the zero functions, while the median and MAD of the y-axis values in each panel are highlighted in blue text and as a blue line and shaded region, respectively. The y-axis error bars shown in the upper panels correspond to the errors in our [Fe/H] estimates, while the y-axis error bars in the lower panels correspond to the errors in the RRab-based distances divided by the nominal values from the literature. Clusters with relative distance differences greater than $10 \%$ have their names highlighted in the lower panels.}
    \label{fig:ClusterFeHsDistances}
\end{figure*}

We estimate mean distance and metallicity values for these GCs using the photometric metallicities (Sect.~\ref{sec:FeH}) and distances (Sect.~\ref{sec:dist}) we computed for these confirmed members. Errors in these estimates are obtained by propagating those in their RRab members.
In Fig.~\ref{fig:ClusterFeHsDistances}, we compare our distance and [Fe/H]$_{\rm phot}$ estimates with those available in the literature.
The literature [Fe/H] values of the clusters are extracted from the \citet{Harris2010} catalog, except for VVV-CL160 \citep[$-1.4 \pm 0.2$ dex, from][]{Minniti2021VVVCL160}, BH~140 ($-1.716 \pm 0.023$ dex, from \citeauthor{Prudil2024b} \citeyear{Prudil2024b}), FSR~1735 \citep[$-0.9 \pm 0.2$ dex, from][]{Carballo-Bello2016}, FSR~1716 \citep[$-1.38 \pm 0.2$ dex, from][]{Koch2017}, and FSR~1758 \citep[$-1.5 \pm 0.3$ dex, from][]{Barbá2019}.
The literature distances of the clusters and their errors are extracted from the GGCD. All of the distances and metallicities from the literature adopted for these GCs are included in Table~\ref{tab:calibrating_GCs}. 

As shown in Fig.~\ref{fig:ClusterFeHsDistances}, the [Fe/H]$_{\rm phot}$ estimates obtained with \textit{Gaia} data for the GCs show practically no systematic offset when compared with the [Fe/H] estimates from the literature, with a median difference of $-0.06$ dex and a median absolute deviation (from the median of the differences, hereafter MAD) of $0.08$ dex. Therefore, even though these estimates are not used in the clustering ($S_{\rm [Fe/H]}^{Gaia} = 0$, see Table~\ref{tab:HDBSCAN_parameters}), this shows that they can be used to characterize the metal content of our RRab groups.

In contrast to the \textit{Gaia} sample, the [Fe/H]$_{\rm phot}$ estimates obtained for the GCs in the VIVACE and Z24 samples show more noticeable systematic offsets when compared to the literature values, with median differences (and MAD values) of $0.15 \pm 0.16$ dex and $0.51 \pm 0.20$ dex, respectively. Particularly, known metal-rich GCs ([Fe/H] $\sim -1.0$ dex) in these samples have their metallicity consistently underestimated. This systematic offset \citepalias[which originates from the photometric metallicity relation provided by][]{Dekany2022} has been previously exposed in the literature \citep{Jurcsik2023, Kunder2024} through comparisons with spectroscopic [Fe/H] estimates.
However, since the calibrated scaling factor $S_{\rm [Fe/H]}$ is found not to be zero in both the VIVACE or Z24 samples (see Table~\ref{tab:HDBSCAN_parameters}), the use of these estimates in the clustering procedure still improves the recovery of known GCs in the samples. This is because the [Fe/H] estimates of members of the same GC (although biased from their literature values in the Z24 and VIVACE samples) remain similar to each other. For the \textit{Gaia} sample (excluding NGC 6715, since it has a large spread in metallicity), a GC has on average $0.8$ RRab stars which do not agree (within their uncertainty) with the mean [Fe/H] of its members. For the Z24 and VIVACE samples, this number is $0.1$ and $1.1$ RR Lyrae stars per cluster, respectively.

At first glance, the distances obtained for the GCs display a similar behavior. \textit{Gaia} distances (which have a median relative distance difference of $0.9 \%$ and a MAD of $3.1 \%$) are very precise, with only FSR~1758 and NGC~6235 showing a distance difference larger than $10 \%$.
The large difference for FSR~1758 is caused by our use of a larger reddening value, $E(B-V) = 0.95$ mag \citep[from the dust map by][]{Schlafly2011}, than previous works ($0.37$ mag and $0.8$ mag in \citeauthor{Barbá2019} \citeyear{Barbá2019} and \citeauthor{Baumgardt2021} \citeyear{Baumgardt2021}, respectively). When accounting for this difference, our distance estimate is within the error bar of both previous estimates.
In the case of NGC~6235, we also use a slightly larger $E(B-V)$ value ($0.36$ mag) than previous works \citep[$0.31$ mag,][]{Baumgardt2021}, but this difference does not fully explain the distance difference we find.
On the other hand, the distances for the GCs in the VIVACE and Z24 samples \textemdash with median relative distance differences (and MAD values) of $-7.0 \pm 6.8 \, \%$ and $-10.7 \pm 7.8 \, \%$, respectively \textemdash seem to be slightly overestimated when compared to the values provided by \citet{Baumgardt2021}.
The origin of this systematic offset is explored further in Appendix~\ref{app:GC_dists_VVV}, where we also argue that these distances can be used to characterize the RRab groups we find.

Among the GCs we use for calibrating our clustering parameters, there is no RR Lyrae-based distance estimate for two of them: BH~140 and NGC~5986. We estimated distances for both GCs using their confirmed RRab members in our \textit{Gaia} sample, obtaining values of $5.17 \pm 0.29$ kpc and $9.97 \pm 0.47$ kpc for BH~140 (two stars) and NGC~5986 (three stars), respectively. These estimates are in good agreement, to within the errors, with the distances provided by \citet{Baumgardt2021}, $4.81 \pm 0.25$ kpc for BH~140 and $10.54 \pm 0.13$ kpc for NGC~5986, and help further constrain their positions in the MW.

\subsection{Most compact groups}
\label{sec:GC_candidates}

Although we recover known GC members with good purity and cohesion (see Table~\ref{tab:HDBSCAN_parameters}), we find thousands of RRab groups within our samples that are most likely either chance alignments, stochastic subdivisions of large Galactic substructures (e.g., the Galactic bulge), or simply artifacts produced by our algorithm. In Appendix~\ref{app:min_3}, we explore the impact of choosing a larger value for \texttt{min\_cluster\_size} on the number and properties of the groups recovered.
To distinguish the most compact, GC-like groups, among the thousands, we define several metrics:

\begin{itemize}

    \item Sky compactness ($\delta_{\rm sky}$): Measures how compactly the members of a group are distributed in the sky. It is computed as $\delta_{\rm sky} (\degree)= \sqrt{\Sigma_{\ell}^2 + \Sigma_b^2}$, where $\Sigma_{\ell}$ and $\Sigma_b$ correspond to the standard deviations in the Galactic coordinates (in degrees) of the group members.

    \item PM compactness ($\delta_{\rm PM}$): Measures how compactly the members of a group are distributed in the PM space. It is computed as $\delta_{\rm sky}$, but using the standard deviations in the PMs (in mas yr$^{-1}$) of the group members.

    \item Distance compactness ($\delta_{\rm dist} / {\rm dist}$): Measures how compactly the distances of a group's members are distributed. It is computed as the standard deviation of the group's member distances divided by their mean value.

    \item Metallicity compactness ($\delta_{\rm [Fe/H]}$): Measures how compactly the metallicities of a group's members are distributed. It is computed as the standard deviation of the group's member metallicities.

\end{itemize}

\begin{figure*}
    \begin{center}
    \includegraphics[width=1.8\columnwidth]{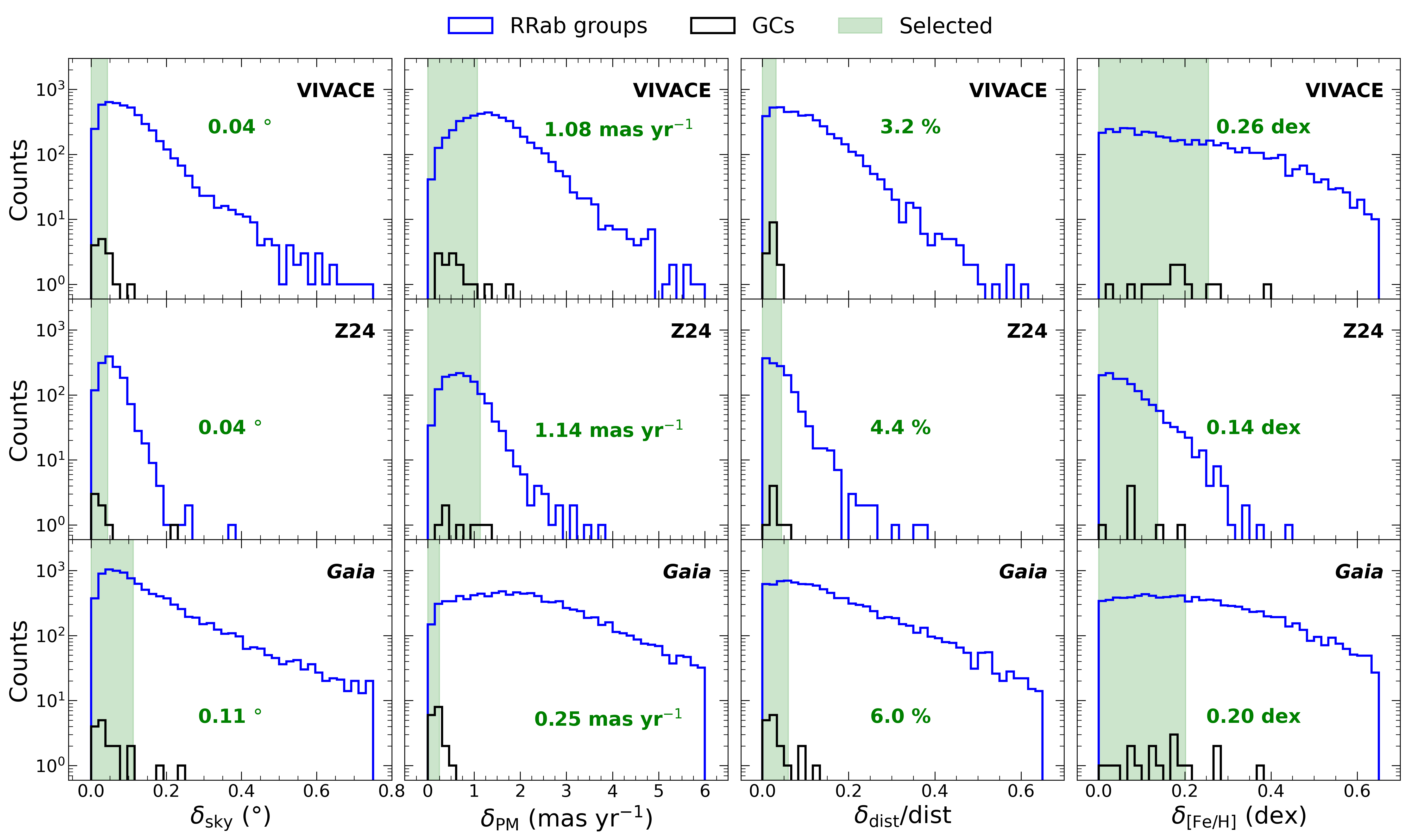}
    \end{center}
    \caption{Metric distributions of all the RRab groups (blue histograms) obtained, compared with the metrics of the cleaned GCs in our calibrating sample (black histograms) in the VIVACE, Z24, and \textit{Gaia} samples (upper, middle, and lower panels, respectively). Metric regions that satisfy our cuts are displayed with light-green shaded areas, with the cut threshold of each panel highlighted in green text.}
    \label{fig:metric_distributions}
\end{figure*}

We compare the distributions in these metrics of the cleaned GCs (see Appendix~\ref{app:GC_members}) with those of the RRab groups, as shown in Fig.~\ref{fig:metric_distributions}.
Here, we make cuts to remove groups that are not as compact as the GCs (recovered in the respective sample) in the sky, PM space, metallicities, and distances. Particularly, we define the cut thresholds as the $80$th percentile of the GCs' aforementioned metric distributions in each sample. By doing this, we make sure that the cuts employed account for how compact the GCs are in each parameter space, within each sample.
Note that this implies, by construction, that some bona fide GCs (even massive and RR Lyrae-rich ones, particularly those that occupy a relatively large area in the sky) would not be recovered. This includes eight out of $17$ GCs in the {\em Gaia} sample (most notably the very massive GCs NGC~3201 and NGC~6441), six out of $14$ in the VIVACE sample (including NGC~6656 and, once again, NGC 6441), and three out of seven in the Z24 sample (including Terzan~1 and NGC~6544).
Additionally, our strict quality cuts and selection criteria leave out the low-mass (candidate) GC Patchick 99 \citep[][and references therein]{Garro2021, Butler2024}, even though two of its three proposed RR Lyrae members are included in the {\em Gaia} sample (only one passes the quality cuts) and two of these RR Lyrae are included in the VIVACE sample. In the latter case, both RR Lyrae pass the quality cuts, but they are not included in our final list of cluster candidates due to their insufficient compactness in the sky. Other relatively low-mass GCs that are similarly not recovered include 2MASS-GC02 \citep{Hurt2000} and FSR~1735 \citep{Froebrich2007}. These cuts, although highly restrictive, allow us to focus on the RR Lyrae groups with the least dispersion in the 6D space analysed.

Applying these cuts and removing groups associated with the GCs used for the parameter calibration (i.e., those that contain known GC members), we are left with $79$, $278$, and $99$ RRab groups in our \textit{Gaia}, Z24, and VIVACE samples, respectively. 
The large majority of these groups are comprised of two RRab stars (which are discussed in Sect.~\ref{sec:compact_groups_2_stars}), but we focus first on groups comprised of at least three RRab stars, which can be more statistically significant.

\subsubsection{Groups with at least three stars}
\label{sec:compact_groups_3_stars}

We find $12$, $10$, and five groups with at least three RRab stars in the \textit{Gaia}, Z24, and VIVACE samples, respectively.
These groups are assigned identifiers that begin with G (\textit{Gaia}), Z (Z24), and V (VIVACE), followed by a number indicating their stability rank (e.g., G1 corresponds to the most stable group in the \textit{Gaia} sample). A summary of these groups' properties is presented in Table~\ref{tab:compact_groups}.

\begin{figure*}
    \begin{center}
    \includegraphics[width=1.6\columnwidth]{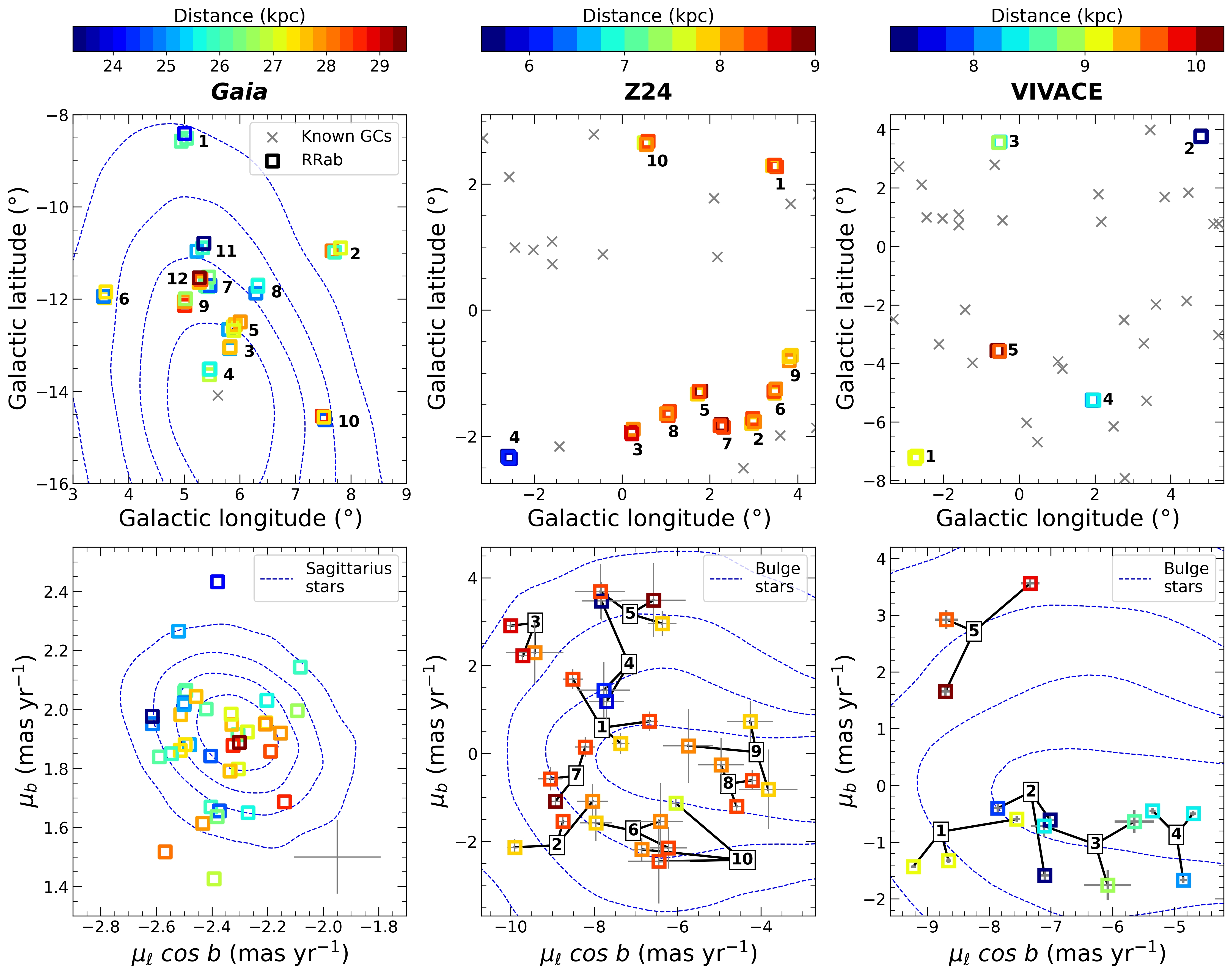}
    \end{center}
    \caption{Distribution of the compact groups with at least three RRab stars in the sky (upper panels) and PM-space (lower panels). Groups found in the \textit{Gaia}, Z24, and VIVACE samples are shown in the left, middle, and right panels, respectively. Stars in the compact groups are colored according to their heliocentric distances. For easier identification of the groups, their IDs are highlighted in the upper panels, and stars are connected by black lines to a box with their group's ID in the lower panels (the latter is not done for the \textit{Gaia} groups since all of their stars have very similar PMs). Blue-dashed lines represent the density contours of Sgr dSph stars (in the left panels) and bulge stars (in the middle and right lower panels), while gray crosses represent known GCs. The mean PM error in \textit{Gaia} are shown in the lower-right corner.}
    \label{fig:compact_groups}
\end{figure*}

In Fig.~\ref{fig:compact_groups} we present the distribution of these groups in the sky and PM-space, where we compare their positions with the known Galactic GCs \protect\citep[retrieved from][]{Baumgardt2021, Bica2024}. This comparison shows that none of the compact groups obtained is associated to a known GC.
However, in Fig.~\ref{fig:compact_groups} we additionally compare the PMs and positions of the \textit{Gaia} groups with density contours\footnote{Contours shown in Figs.~\ref{fig:compact_groups},~and~\ref{fig:compact_pairs_Z24_VIVACE} were computed using the \texttt{kdeplot} method of the \texttt{Seaborn} Python library \citep{Seaborn}, with \texttt{levels} of $0.15, 0.5, 0.85$, and \texttt{cumulative} set to False.} of stars from the core of the Sgr dSph galaxy (see Appendix~\ref{app:contours} for a description of how these stars were retrieved), which clearly shows that all of these compact \textit{Gaia} groups belong to the Sgr dSph galaxy.

Since the groups we find in the Z24 and VIVACE samples are in the direction of the Galactic center \citep[which is at a distance of $\sim8$ kpc from the Sun; see, e.g., Sect.~3.2 of][for a thorough review]{Bland-Hawthorn2016}, we also compare their PMs in Fig.~\ref{fig:compact_groups} with density contours of bulge stars (see Appendix~\ref{app:contours}).
This comparison shows that most of the compact groups found in these samples have distances and PMs consistent with those of bulge stars, and thus are very likely artificial subdivisions of the bulge. However, there are four groups that stand out from the rest: Z3, Z4, Z5, and V5.

Although both Z3 and Z5 have mean distances ($\langle d \rangle$) consistent with that of the Galactic bulge ($\langle d \rangle \simeq 8.4$ kpc, see Table~\ref{tab:compact_groups}), they are comprised of three RRab stars that are close to each other in the sky (both groups have $\delta_{\rm sky} \simeq 2$ arcmin), with similar PMs that are at the edge of the bulge's PM contours, which means these groups are less likely to be spurious bulge subdivision.
Z4, on the other hand, is comprised of three stars very near to each other in the sky, of which two share very similar PMs. Although the PMs of these stars are similar to those of the bulge, they are at a heliocentric distance of $\sim 6$ kpc, which puts them closer to the Sun than most bulge RR Lyrae stars \citep[see, e.g.,][]{Prudil2025a}, making it less likely for these groups to be associated with the bulge population. The V5 group shows the least similar PMs to the bulge among the five groups, while also being the one furthest away from the Galactic center, with a mean heliocentric distance of $9.9$~kpc.

Additionally, we searched for RV estimates in the literature for the RR Lyrae stars within our groups and pairs, as an independent validation. Even though scarce, we found measurements for the three stars in V5 \citep[from ][]{Kunder2016}. These RVs are $71$, $84$, and $49$ km s$^{-1}$ (with uncertainties of $\sim 5 - 10$ km s$^{-1}$), demonstrating that at least two of them have similar RVs, in addition to their similar PMs, distances, photometric metallicities, and spatial distribution. This RV consistency places this group as an excellent candidate for follow-up observations. It provides an encouraging first validation of our clustering algorithm.

\subsubsection{Groups with two stars}
\label{sec:compact_groups_2_stars}

As mentioned previously, even a pair of RRab stars that satisfies the strict cuts displayed in Fig.~\ref{fig:metric_distributions} could indicate a significant substructure. However, a pair of RRab stars is a much less robust detection than a group of three stars, especially when positioned in the direction of a large structure such as the Galactic bulge or the Sgr dSph galaxy. Therefore, we must identify and discard pairs of stars that are likely subdivisions of these structures.

\begin{figure}
    \begin{center}
    \includegraphics[width=0.9\columnwidth]{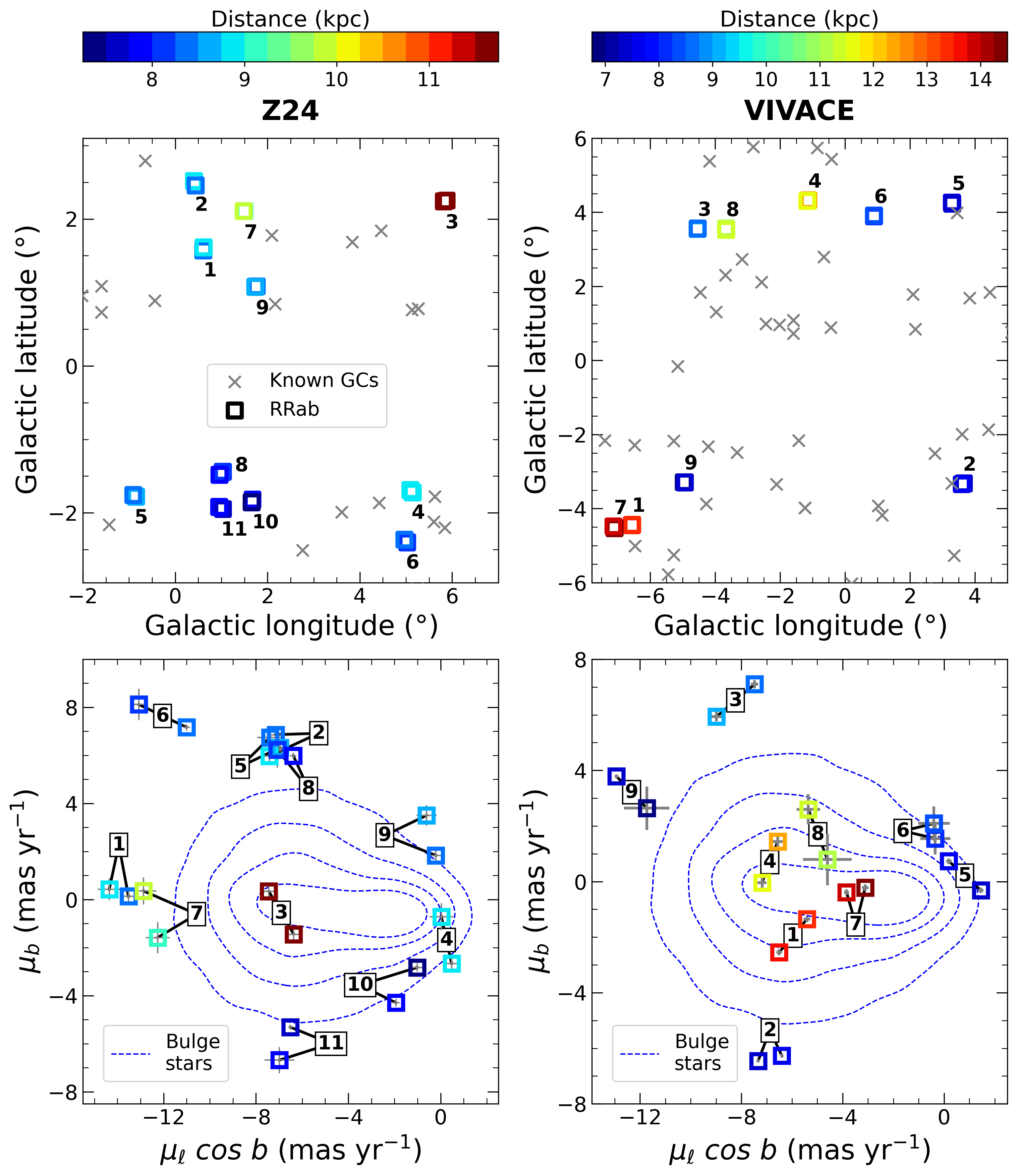}
    \end{center}
    \caption{Same as Fig.~\ref{fig:compact_groups}, but for the compact pairs of RRab stars in the Z24 and VIVACE samples that were not identified as potentially associated to the Galactic bulge or the Sgr dSph galaxy.}
    \label{fig:compact_pairs_Z24_VIVACE}
\end{figure}

\begin{figure}
    \begin{center}
    \includegraphics[width=0.9\columnwidth]{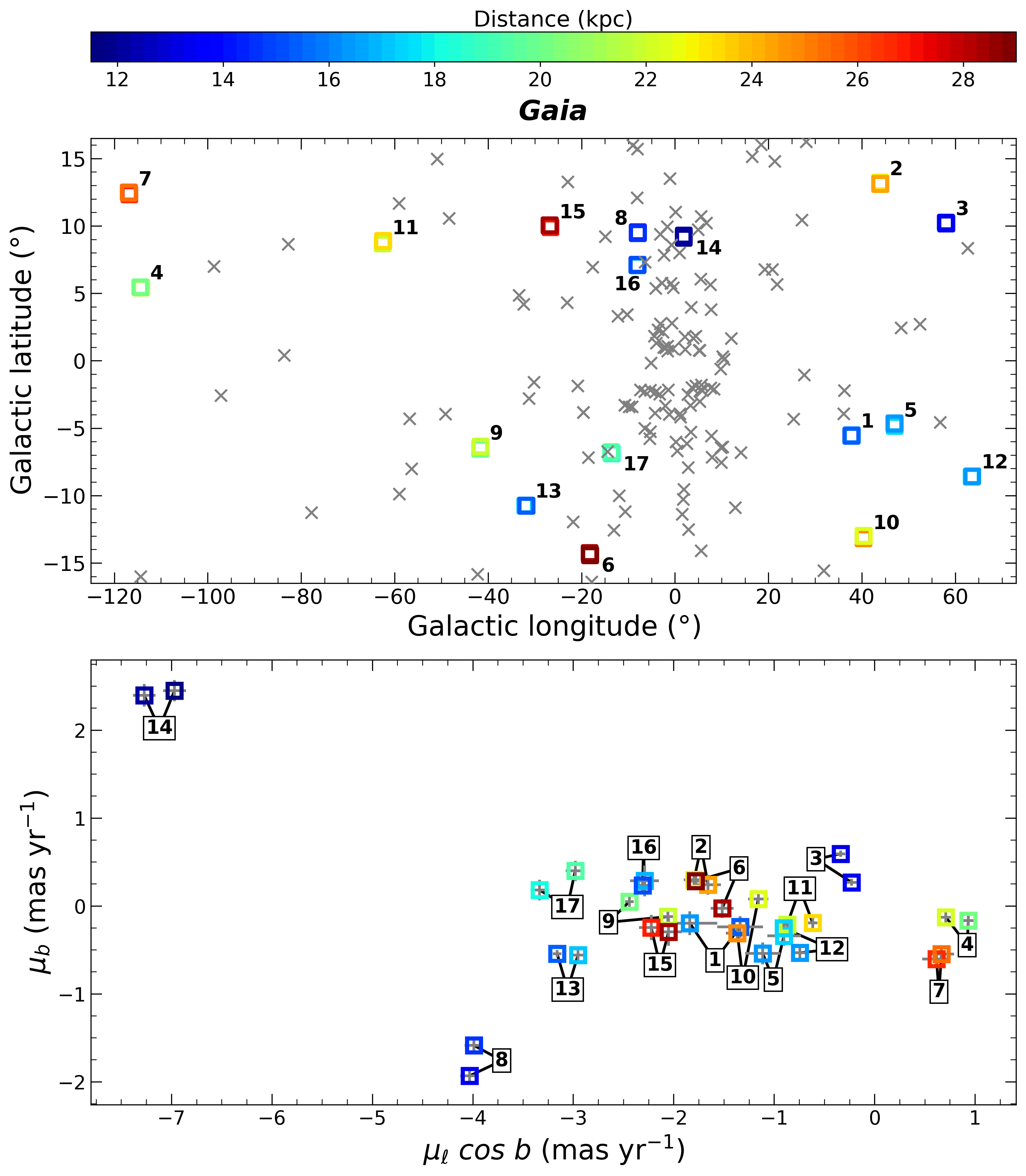}
    \end{center}
    \caption{Same as Fig.~\ref{fig:compact_groups}, but for the compact pairs of RRab stars in the \textit{Gaia} sample that were not identified as potentially associated to the Galactic bulge or the Sgr dSph galaxy.}
    \label{fig:compact_pairs_Gaia}
\end{figure}

We conservatively identify Galactic bulge pairs as those that have $-12 < \ell \, (\degree) < 12$, $-10 < b \, (\degree) < 10$, $5 < d \, ({\rm kpc})< 11$, and whose mean PM vector lies within $6$ mas yr$^{-1}$ of ($\mu_{\ell}$ cos $b$, $\mu_b$) = ($-6.5$, $0$) mas yr$^{-1}$ \citep[approximately the mean PMs of the Galactic bulge; see, e.g.,][]{Clarke2019, Olivares2024}.
Pairs potentially associated with Sgr, on the other hand, are conservatively identified as those with an angular distance smaller than $8 \degree$ from $\ell = 5.607 \degree$ and $b = -14.087 \degree$ (the mean coordinates of NGC~6715, the core of Sgr, from the GGCD), a PM vector within $2$ mas yr$^{-1}$ of ($\mu_{\ell}$ cos $b$, $\mu_b$) = ($-2.3$, $1.9$) mas yr$^{-1}$ (where most Sgr dSph stars are located in PM-space; see Fig.~\ref{fig:compact_groups}), and a mean distance in the range of $[20 - 30]$ kpc].
These pairs are discarded, which leaves $17$, $11$, and $9$ RRab pairs in our \textit{Gaia}, Z24, and VIVACE samples, respectively.
These pairs have identifiers G, Z, or V, followed by a number indicating their stability rank and a ``p'' label (indicating they are pairs). We present a summary of their properties in Table~\ref{tab:compact_pairs} and we showcase the positions and PMs of these pairs in Figs.~\ref{fig:compact_pairs_Z24_VIVACE}~and~\ref{fig:compact_pairs_Gaia}.
As shown in Fig.~\ref{fig:compact_pairs_Z24_VIVACE}, we find multiple compact pairs of RRab stars in our Z24 and VIVACE samples that are in the direction of the bulge, but have distinct PMs or are located in front of or behind it.
On the other hand, the compact pairs found in the \textit{Gaia} sample (shown in Fig.~\ref{fig:compact_pairs_Gaia}) are scattered across a large region of the Galactic plane, with heliocentric distances reaching as far as $30$ kpc.
As ensured by our cuts shown in Fig.~\ref{fig:metric_distributions}, all of these pairs display GC-like compactness in their positions, PMs, and distances.
However, RV measurements of these RRab stars (and of those comprising the groups discussed in Sect.~\ref{sec:compact_groups_3_stars}) will be crucial to evaluate whether the members of the groups and pairs we find truly move together in space, and are thus physically connected.

In addition to comparing our RRab groups and pairs with the known Galactic GCs, the Galactic bulge, and the Sgr dSph galaxy (as shown in Figs.~\ref{fig:compact_groups},~\ref{fig:compact_pairs_Z24_VIVACE},~and~\ref{fig:compact_pairs_Gaia}), we also compare their positions with the stellar stream tracks included in the \textit{galstreams} compilation by \cite{Mateu2023} and with the RR Lyrae substructure candidates proposed in \citet{Torrealba2015}. We find no matches between these and our RRab groups or pairs.

\section{Summary and conclusions}
\label{sec:conclusions}

GCs provide crucial insights into the formation history of the MW, and although many have been identified in the Galaxy, their population is still expected to be incomplete. Finding these yet undiscovered Galactic GCs would add crucial pieces to the puzzle that is the formation of the MW.
RR Lyrae stars, as standard candles and tracers of old populations, hold immense potential in this search for undiscovered GCs.

In this work, we build samples of RRab stars with spatial coordinates, PMs, distances, and photometric metallicities derived from \textit{Gaia} and VVV (plus VVVx) data. We exploit these samples to search for groups of RRab stars in the Galactic plane and bulge displaying GC-like properties. This is done with the hierarchical clustering algorithm HDBSCAN, which we calibrate to optimize the recovery of known GCs.

Employing our clustering procedure, we recover known GCs (that have confirmed members in our samples) with excellent purity and cohesion. We use these GCs to judge the quality of our metallicity and distance estimates, and to select the most promising GC-like candidates among the thousands of groups our clustering procedure initially finds. All the GC-like groups we find with a significant number of RR Lyrae stars are associated with already known GCs. Furthermore, due to their low masses \citep[see, e.g.,][and references therein]{Bica2024}, the most recently discovered GCs tend to have few (if any) RR Lyrae stars \citep[e.g.,][]{Gran2022, Pace2023}. This suggests that the lack of new clusters in our results may be due to the remaining population of undiscovered GCs being primarily low-mass, which lowers the likelihood of them hosting RR Lyrae stars. These undiscovered GCs could also be metal-rich (i.e., with ${\rm [Fe/H]} > -1$~dex), which would further reduce this likelihood, given their typically very red HB morphologies \citep[][and references therein]{Catelan2009}. A well-known example of this is 47 Tuc (NGC~104), a massive ($8.5 \times 10^5 \, M_{\odot}$, from the GGCD), moderately metal-rich \citep[${\rm [Fe/H]} \approx -0.76$~dex; e.g.,][]{Carretta2009, Cordero2014, Marino2016, Kolomiecas2024} Galactic GC with only one RR Lyrae member \citep[][]{Carney1993,Storm1994,Clement2001}\footnote{Here we used the online version of the \citet{Clement2001} catalog. Its 47~Tuc entry was last updated in 2024, and is available online at \url{https://www.astro.utoronto.ca/~cclement/cat/C0021m723}.}.

By comparing the groups of two and three RR Lyrae displaying GC-like properties with Galactic substructures, we find that $\simeq 40$ of them (which are scattered throughout the Galactic plane and bulge, and cover a large range of distances) are not associated to any known GC, to the Galactic bulge field, or with the Sgr dSph galaxy.
If these are indeed real groups, there should be significant stellar populations associated with them. These populations should be searched for and studied, both dynamically and chemically. We searched for other tracers, such as blue horizontal branch stars, since they are metal-poor stars for which relatively accurate distance estimates can be obtained through color-magnitude relations. Nonetheless, there is no catalogue in the literature covering the same distance range as our sample and, more importantly, the available catalogues specifically avoid the Galactic plane and focus only on the halo \citep[see e.g.,][]{Ju24, Culpan24, Amarante2024, Wu2025}. Future spectroscopic surveys of blue horizontal branch stars in the bulge and plane could help recover the substructures through kinematics, even if the distance estimates are not as accurate as those of RR Lyrae stars.

One of the limitations of our work is that only $34\,829$ stars (those that have estimates for the parameters required for our analysis and satisfy our quality cuts described in Sect.~\ref{sec:quality_cuts}) out of $\sim 94\, 000$ RRab stars in our initial \textit{Gaia} sample that are in the Galactic plane ($-15 \leq b(\degree) \leq 15$) were analysed with our clustering algorithm. This leads to large regions of the Galactic plane where our final \textit{Gaia} sample does not reach or has poor completeness (see Fig.~\ref{fig:sample_sky}), and, although the inclusion of the Z24 and VIVACE samples helps cover the bulge and southern disk of the MW, our analysis is still significantly limited by the removal of this large fraction of the \textit{Gaia} sample, as possible RRab members of new GC candidates could have been excluded from our clustering.

This work provides several promising candidates for future follow-up studies searching for undiscovered Galactic GCs. Additionally, it shows the potential of using a clustering algorithm (with careful calibration and false positive detection) to search for compact GC-like objects in the most challenging regions of the MW. This potential was already demonstrated in works such as \citet{Gran2022}, where new low-mass bulge GCs were discovered using the DBSCAN algorithm (without using RV estimates or metallicities), and later confirmed spectroscopically.

None of the stars in the groups we find have an RV estimate in \textit{Gaia} DR3, as these groups are at large distances (see Tables~\ref{tab:compact_groups}~and~\ref{tab:compact_pairs}) and in regions of high interstellar extinction. These stars have $G$-band magnitudes fainter than $\sim16$~mag (with RV estimates being available for stars brighter than $G\simeq 15$ mag).
The acquisition of additional spectroscopic data is a necessary component to validate the remaining groups and pairs.
Therefore, spectroscopic RV follow-ups of these groups will be essential to validate the groups and pairs detected in this work.
\textit{Gaia} DR4 is expected to provide RV estimates for a significantly larger number of RR Lyrae stars, which could allow us to include this additional dimension in our algorithm without drastically reducing our sample size, as well as to use them to further investigate the robustness of our candidate groups and help distinguish whether they are indeed co-moving within the Galaxy.
Additionally, future \textit{Gaia} data releases, with longer baselines and more observations, should have reduced PM errors and a larger number of stars we can analyse with our algorithm, thus extending our sky coverage and improving our chances of finding new GC candidates.

\begin{acknowledgements}
We gratefully acknowledge the constructive comments
and suggestions provided by the referee.
We warmly thank P. W. Lucas for his help accessing VIRAC2 data prior to publication and for his helpful comments. 
Support for this project is provided by ANID's FONDECYT Regular grants \#1171273 and 1231637; ANID's Millennium Science Initiative through grants ICN12\textunderscore 009 and AIM23-0001, awarded to the Millennium Institute of Astrophysics (MAS); and ANID's Basal project FB210003. NCC acknowledges support from SOCHIAS grant ``Beca Adelina'' and funding from the AERIAL labcom and the ANR as part of project ANR-24-LCV1-0002.
The research for this work made use of Astropy \citep{Astropy2013, Astropy2018}, Matplotlib \citep{Matplotlib}, NumPy \citep{Numpy}, SciPy \citep{Scipy}, Jupyter \citep{Jupyter}, Seaborn \citep{Seaborn}, and pandas \citep{Pandas}.
\end{acknowledgements}

\bibliographystyle{aa}
\bibliography{bibliography}

@ARTICLE{Culpan24,
       author = {{Culpan}, R. and {Dorsch}, M. and {Geier}, S. and {Pelisoli}, I. and {Heber}, U. and {Kub{\'a}tov{\'a}}, B. and {Cabezas}, M.},
        title = "{Probing the inner Galactic halo with blue horizontal-branch stars. Gaia DR3-based catalogue with atmospheric and stellar parameters}",
      journal = {\aap},
         year = 2024,
        month = may,
       volume = {685},
          eid = {A134},
        pages = {A134},
          doi = {10.1051/0004-6361/202348323},
archivePrefix = {arXiv},
       eprint = {2402.09779},
 primaryClass = {astro-ph.SR},
       adsurl = {https://ui.adsabs.harvard.edu/abs/2024A&A...685A.134C}
}

@ARTICLE{Amarante2024,
       author = {{Amarante}, Jo{\~a}o A.~S. and {Koposov}, Sergey E. and {Laporte}, Chervin F.~P.},
        title = "{Mapping the anisotropic Galactic stellar halo with blue horizontal branch stars}",
      journal = {\aap},
         year = 2024,
        month = oct,
       volume = {690},
          eid = {A166},
        pages = {A166},
          doi = {10.1051/0004-6361/202450351},
archivePrefix = {arXiv},
       eprint = {2404.09825},
 primaryClass = {astro-ph.GA},
       adsurl = {https://ui.adsabs.harvard.edu/abs/2024A&A...690A.166A}
}

@ARTICLE{Ju24,
       author = {{Ju}, Jie and {Cui}, Wenyuan and {Huo}, Zhenyan and {Liu}, Chao and {Xue}, Xiangxiang and {Liu}, Jiaming and {Feng}, Shuai and {Sun}, Mingxu and {Li}, Linlin},
        title = "{Identification of Blue Horizontal-branch Stars from LAMOST DR5}",
      journal = {\apjs},
         year = 2024,
        month = jan,
       volume = {270},
       number = {1},
          eid = {11},
        pages = {11},
          doi = {10.3847/1538-4365/ad0df9},
archivePrefix = {arXiv},
       eprint = {2311.16430},
 primaryClass = {astro-ph.GA},
       adsurl = {https://ui.adsabs.harvard.edu/abs/2024ApJS..270...11J}
}

@ARTICLE{Mateu2023,
       author = {{Mateu}, Cecilia},
        title = "{galstreams: A library of Milky Way stellar stream footprints and tracks}",
      journal = {\mnras},
         year = 2023,
        month = apr,
       volume = {520},
       number = {4},
        pages = {5225-5258},
          doi = {10.1093/mnras/stad321},
archivePrefix = {arXiv},
       eprint = {2204.10326},
 primaryClass = {astro-ph.GA},
       adsurl = {https://ui.adsabs.harvard.edu/abs/2023MNRAS.520.5225M}
}

@ARTICLE{Clementini2023,
       author = {{Clementini}, G. and {Ripepi}, V. and {Garofalo}, A. and {Molinaro}, R. and {Muraveva}, T. and {Leccia}, S. and {Rimoldini}, L. and {Holl}, B. and {Jevardat de Fombelle}, G. and {Sartoretti}, P. and {Marchal}, O. and {Audard}, M. and {Nienartowicz}, K. and {Andrae}, R. and {Marconi}, M. and {Szabados}, L. and {Evans}, D.~W. and {Lecoeur-Taibi}, I. and {Mowlavi}, N. and {Musella}, I. and {Eyer}, L.},
        title = "{Gaia Data Release 3. Specific processing and validation of all-sky RR Lyrae and Cepheid stars: The RR Lyrae sample}",
      journal = {\aap},
         year = 2023,
        month = jun,
       volume = {674},
          eid = {A18},
        pages = {A18},
          doi = {10.1051/0004-6361/202243964},
archivePrefix = {arXiv},
       eprint = {2206.06278},
 primaryClass = {astro-ph.SR},
       adsurl = {https://ui.adsabs.harvard.edu/abs/2023A&A...674A..18C}
}

@article{Prudil2024,
	author = {{Prudil}, Z. and {Kunder, A.} and {Dékány, I.} and {Koch-Hansen, A. J.}},
	title = {The Galactic bulge exploration - I. The period–absolute magnitude–metallicity relations for RR Lyrae stars for GBP, V, G, GRP, I, J, H, and Ks passbands using Gaia DR3 parallaxes},
	DOI= "10.1051/0004-6361/202347338",
	url= "https://doi.org/10.1051/0004-6361/202347338",
	journal = {\aap},
	year = 2024,
	volume = 684,
	pages = "A176",
}

@ARTICLE{Kunder2024,
       author = {{Kunder}, Andrea and {Prudil}, Zdenek and {Skaggs}, Claire and {Reggiani}, Henrique and {Nataf}, David M. and {Hughes}, Joanne and {Covey}, Kevin R. and {Devine}, Kathryn},
        title = "{The Galactic Bulge Exploration. III. Calcium Triplet Metallicities for RR Lyrae Stars}",
      journal = {\aj},
         year = 2024,
        month = sep,
       volume = {168},
       number = {3},
          eid = {139},
        pages = {139},
          doi = {10.3847/1538-3881/ad6262},
archivePrefix = {arXiv},
       eprint = {2407.01515},
 primaryClass = {astro-ph.GA},
       adsurl = {https://ui.adsabs.harvard.edu/abs/2024AJ....168..139K}
}

@ARTICLE{Prudil2024b,
       author = {{Prudil}, Z. and {Arellano Ferro}, A.},
        title = "{On the membership of variable stars in Galactic globular clusters: the Oosterhoff gap}",
      journal = {\mnras},
         year = 2024,
        month = nov,
       volume = {534},
       number = {4},
        pages = {3654-3664},
          doi = {10.1093/mnras/stae2335},
archivePrefix = {arXiv},
       eprint = {2410.06435},
 primaryClass = {astro-ph.GA},
       adsurl = {https://ui.adsabs.harvard.edu/abs/2024MNRAS.534.3654P}
}

@ARTICLE{Vioque2023,
       author = {{Vioque}, Miguel and {Cavieres}, Manuel and {Pantaleoni Gonz{\'a}lez}, Michelangelo and {Ribas}, {\'A}lvaro and {Oudmaijer}, Ren{\'e} D. and {Mendigut{\'\i}a}, Ignacio and {Kilian}, Lena and {C{\'a}novas}, H{\'e}ctor and {Kuhn}, Michael A.},
        title = "{Clustering Properties of Intermediate and High-mass Young Stellar Objects}",
      journal = {\aj},
         year = 2023,
        month = nov,
       volume = {166},
       number = {5},
          eid = {183},
        pages = {183},
          doi = {10.3847/1538-3881/acf75f},
archivePrefix = {arXiv},
       eprint = {2309.00678},
 primaryClass = {astro-ph.SR},
       adsurl = {https://ui.adsabs.harvard.edu/abs/2023AJ....166..183V}
}

@ARTICLE{Catelan2004,
       author = {{Catelan}, M. and {Pritzl}, Barton J. and {Smith}, Horace A.},
        title = "{The RR Lyrae Period-Luminosity Relation. I. Theoretical Calibration}",
      journal = {\apjs},
         year = 2004,
        month = oct,
       volume = {154},
       number = {2},
        pages = {633-649},
          doi = {10.1086/422916},
archivePrefix = {arXiv},
       eprint = {astro-ph/0406067},
 primaryClass = {astro-ph},
       adsurl = {https://ui.adsabs.harvard.edu/abs/2004ApJS..154..633C}
}

@ARTICLE{McInnes2017,
       author = {{McInnes}, Leland and {Healy}, John and {Astels}, Steve},
        title = "{hdbscan: Hierarchical density based clustering}",
      journal = {The Journal of Open Source Software},
         year = 2017,
        month = mar,
       volume = {2},
       number = {11},
          eid = {205},
        pages = {205},
          doi = {10.21105/joss.00205},
       adsurl = {https://ui.adsabs.harvard.edu/abs/2017JOSS....2..205M}
}

@ARTICLE{Dekany2021,
       author = {{D{\'e}k{\'a}ny}, Istv{\'a}n and {Grebel}, Eva K. and {Pojma{\'n}ski}, Grzegorz},
        title = "{Metallicity Estimation of RR Lyrae Stars From Their I-Band Light Curves}",
      journal = {\apj},
         year = 2021,
        month = oct,
       volume = {920},
       number = {1},
          eid = {33},
        pages = {33},
          doi = {10.3847/1538-4357/ac106f},
archivePrefix = {arXiv},
       eprint = {2107.05983},
 primaryClass = {astro-ph.SR},
       adsurl = {https://ui.adsabs.harvard.edu/abs/2021ApJ...920...33D}
}

@InProceedings{Campello2013,
    author="Campello, Ricardo J. G. B.
    and Moulavi, Davoud
    and Sander, Joerg",
    editor="Pei, Jian
    and Tseng, Vincent S.
    and Cao, Longbing
    and Motoda, Hiroshi
    and Xu, Guandong",
    title="Density-Based Clustering Based on Hierarchical Density Estimates",
    booktitle="Advances in Knowledge Discovery and Data Mining",
    year="2013",
    publisher="Springer Berlin Heidelberg",
    address="Berlin, Heidelberg",
    pages="160--172",
    isbn="978-3-642-37456-2"
}

@ARTICLE{Helmi2020,
       author = {{Helmi}, Amina},
        title = "{Streams, Substructures, and the Early History of the Milky Way}",
      journal = {\araa},
         year = 2020,
        month = aug,
       volume = {58},
        pages = {205-256},
          doi = {10.1146/annurev-astro-032620-021917},
archivePrefix = {arXiv},
       eprint = {2002.04340},
 primaryClass = {astro-ph.GA},
       adsurl = {https://ui.adsabs.harvard.edu/abs/2020ARA&A..58..205H}
}

@ARTICLE{Jurcsik2023,
       author = {{Jurcsik}, Johanna and {Hajdu}, Gergely},
        title = "{Photometric metallicities of fundamental-mode RR Lyr stars from Gaia G band photometry of globular-cluster variables}",
      journal = {\mnras},
         year = 2023,
        month = nov,
       volume = {525},
       number = {3},
        pages = {3486-3498},
          doi = {10.1093/mnras/stad2510},
archivePrefix = {arXiv},
       eprint = {2308.08929},
 primaryClass = {astro-ph.SR},
       adsurl = {https://ui.adsabs.harvard.edu/abs/2023MNRAS.525.3486J}
}

@ARTICLE{Catelan2009,
       author = {{Catelan}, M.},
        title = "{Horizontal branch stars: the interplay between observations and theory, and insights into the formation of the Galaxy}",
      journal = {\apss},
         year = 2009,
        month = apr,
       volume = {320},
       number = {4},
        pages = {261-309},
          doi = {10.1007/s10509-009-9987-8},
archivePrefix = {arXiv},
       eprint = {astro-ph/0507464},
 primaryClass = {astro-ph},
       adsurl = {https://ui.adsabs.harvard.edu/abs/2009Ap&SS.320..261C}
}

@ARTICLE{Gaia2018,
       author = {{Gaia Collaboration} and {Brown}, A.~G.~A. and {Vallenari}, A. and {Prusti}, T. and {de Bruijne}, J.~H.~J. and {Babusiaux}, C. and {Bailer-Jones}, C.~A.~L. and {Biermann}, M. and {Evans}, D.~W. and {Eyer}, L. and {Jansen}, F. and {Jordi}, C. and {Klioner}, S.~A. and {Lammers}, U. and {Lindegren}, L. and {Luri}, X. and {Mignard}, F. and {Panem}, C. and {Pourbaix}, D. and {Randich}, S. and {Sartoretti}, P. and {Siddiqui}, H.~I. and {Soubiran}, C. and {van Leeuwen}, F. and {Walton}, N.~A. and {Arenou}, F. and {Bastian}, U. and {Cropper}, M. and {Drimmel}, R. and {Katz}, D. and {Lattanzi}, M.~G. and {Bakker}, J. and {Cacciari}, C. and {Casta{\~n}eda}, J. and {Chaoul}, L. and {Cheek}, N. and {De Angeli}, F. and {Fabricius}, C. and {Guerra}, R. and {Holl}, B. and {Masana}, E. and {Messineo}, R. and {Mowlavi}, N. and {Nienartowicz}, K. and {Panuzzo}, P. and {Portell}, J. and {Riello}, M. and {Seabroke}, G.~M. and {Tanga}, P. and {Th{\'e}venin}, F. and {Gracia-Abril}, G. and {Comoretto}, G. and {Garcia-Reinaldos}, M. and {Teyssier}, D. and {Altmann}, M. and {Andrae}, R. and {Audard}, M. and {Bellas-Velidis}, I. and {Benson}, K. and {Berthier}, J. and {Blomme}, R. and {Burgess}, P. and {Busso}, G. and {Carry}, B. and {Cellino}, A. and {Clementini}, G. and {Clotet}, M. and {Creevey}, O. and {Davidson}, M. and {De Ridder}, J. and {Delchambre}, L. and {Dell'Oro}, A. and {Ducourant}, C. and {Fern{\'a}ndez-Hern{\'a}ndez}, J. and {Fouesneau}, M. and {Fr{\'e}mat}, Y. and {Galluccio}, L. and {Garc{\'\i}a-Torres}, M. and {Gonz{\'a}lez-N{\'u}{\~n}ez}, J. and {Gonz{\'a}lez-Vidal}, J.~J. and {Gosset}, E. and {Guy}, L.~P. and {Halbwachs}, J. -L. and {Hambly}, N.~C. and {Harrison}, D.~L. and {Hern{\'a}ndez}, J. and {Hestroffer}, D. and {Hodgkin}, S.~T. and {Hutton}, A. and {Jasniewicz}, G. and {Jean-Antoine-Piccolo}, A. and {Jordan}, S. and {Korn}, A.~J. and {Krone-Martins}, A. and {Lanzafame}, A.~C. and {Lebzelter}, T. and {L{\"o}ffler}, W. and {Manteiga}, M. and {Marrese}, P.~M. and {Mart{\'\i}n-Fleitas}, J.~M. and {Moitinho}, A. and {Mora}, A. and {Muinonen}, K. and {Osinde}, J. and {Pancino}, E. and {Pauwels}, T. and {Petit}, J. -M. and {Recio-Blanco}, A. and {Richards}, P.~J. and {Rimoldini}, L. and {Robin}, A.~C. and {Sarro}, L.~M. and {Siopis}, C. and {Smith}, M. and {Sozzetti}, A. and {S{\"u}veges}, M. and {Torra}, J. and {van Reeven}, W. and {Abbas}, U. and {Abreu Aramburu}, A. and {Accart}, S. and {Aerts}, C. and {Altavilla}, G. and {{\'A}lvarez}, M.~A. and {Alvarez}, R. and {Alves}, J. and {Anderson}, R.~I. and {Andrei}, A.~H. and {Anglada Varela}, E. and {Antiche}, E. and {Antoja}, T. and {Arcay}, B. and {Astraatmadja}, T.~L. and {Bach}, N. and {Baker}, S.~G. and {Balaguer-N{\'u}{\~n}ez}, L. and {Balm}, P. and {Barache}, C. and {Barata}, C. and {Barbato}, D. and {Barblan}, F. and {Barklem}, P.~S. and {Barrado}, D. and {Barros}, M. and {Barstow}, M.~A. and {Bartholom{\'e} Mu{\~n}oz}, S. and {Bassilana}, J. -L. and {Becciani}, U. and {Bellazzini}, M. and {Berihuete}, A. and {Bertone}, S. and {Bianchi}, L. and {Bienaym{\'e}}, O. and {Blanco-Cuaresma}, S. and {Boch}, T. and {Boeche}, C. and {Bombrun}, A. and {Borrachero}, R. and {Bossini}, D. and {Bouquillon}, S. and {Bourda}, G. and {Bragaglia}, A. and {Bramante}, L. and {Breddels}, M.~A. and {Bressan}, A. and {Brouillet}, N. and {Br{\"u}semeister}, T. and {Brugaletta}, E. and {Bucciarelli}, B. and {Burlacu}, A. and {Busonero}, D. and {Butkevich}, A.~G. and {Buzzi}, R. and {Caffau}, E. and {Cancelliere}, R. and {Cannizzaro}, G. and {Cantat-Gaudin}, T. and {Carballo}, R. and {Carlucci}, T. and {Carrasco}, J.~M. and {Casamiquela}, L. and {Castellani}, M. and {Castro-Ginard}, A. and {Charlot}, P. and {Chemin}, L. and {Chiavassa}, A. and {Cocozza}, G. and {Costigan}, G. and {Cowell}, S. and {Crifo}, F. and {Crosta}, M. and {Crowley}, C. and {Cuypers}, J. and {Dafonte}, C. and {Damerdji}, Y. and {Dapergolas}, A. and {David}, P. and {David}, M. and {de Laverny}, P. and {De Luise}, F. and {De March}, R. and {de Martino}, D. and {de Souza}, R. and {de Torres}, A. and {Debosscher}, J. and {del Pozo}, E. and {Delbo}, M. and {Delgado}, A. and {Delgado}, H.~E. and {Di Matteo}, P. and {Diakite}, S. and {Diener}, C. and {Distefano}, E. and {Dolding}, C. and {Drazinos}, P. and {Dur{\'a}n}, J. and {Edvardsson}, B. and {Enke}, H. and {Eriksson}, K. and {Esquej}, P. and {Eynard Bontemps}, G. and {Fabre}, C. and {Fabrizio}, M. and {Faigler}, S. and {Falc{\~a}o}, A.~J. and {Farr{\`a}s Casas}, M. and {Federici}, L. and {Fedorets}, G. and {Fernique}, P. and {Figueras}, F. and {Filippi}, F. and {Findeisen}, K. and {Fonti}, A. and {Fraile}, E. and {Fraser}, M. and {Fr{\'e}zouls}, B. and {Gai}, M. and {Galleti}, S. and {Garabato}, D. and {Garc{\'\i}a-Sedano}, F. and {Garofalo}, A. and {Garralda}, N. and {Gavel}, A. and {Gavras}, P. and {Gerssen}, J. and {Geyer}, R. and {Giacobbe}, P. and {Gilmore}, G. and {Girona}, S. and {Giuffrida}, G. and {Glass}, F. and {Gomes}, M. and {Granvik}, M. and {Gueguen}, A. and {Guerrier}, A. and {Guiraud}, J. and {Guti{\'e}rrez-S{\'a}nchez}, R. and {Haigron}, R. and {Hatzidimitriou}, D. and {Hauser}, M. and {Haywood}, M. and {Heiter}, U. and {Helmi}, A. and {Heu}, J. and {Hilger}, T. and {Hobbs}, D. and {Hofmann}, W. and {Holland}, G. and {Huckle}, H.~E. and {Hypki}, A. and {Icardi}, V. and {Jan{\ss}en}, K. and {Jevardat de Fombelle}, G. and {Jonker}, P.~G. and {Juh{\'a}sz}, {\'A}. L. and {Julbe}, F. and {Karampelas}, A. and {Kewley}, A. and {Klar}, J. and {Kochoska}, A. and {Kohley}, R. and {Kolenberg}, K. and {Kontizas}, M. and {Kontizas}, E. and {Koposov}, S.~E. and {Kordopatis}, G. and {Kostrzewa-Rutkowska}, Z. and {Koubsky}, P. and {Lambert}, S. and {Lanza}, A.~F. and {Lasne}, Y. and {Lavigne}, J. -B. and {Le Fustec}, Y. and {Le Poncin-Lafitte}, C. and {Lebreton}, Y. and {Leccia}, S. and {Leclerc}, N. and {Lecoeur-Taibi}, I. and {Lenhardt}, H. and {Leroux}, F. and {Liao}, S. and {Licata}, E. and {Lindstr{\o}m}, H.~E.~P. and {Lister}, T.~A. and {Livanou}, E. and {Lobel}, A. and {L{\'o}pez}, M. and {Managau}, S. and {Mann}, R.~G. and {Mantelet}, G. and {Marchal}, O. and {Marchant}, J.~M. and {Marconi}, M. and {Marinoni}, S. and {Marschalk{\'o}}, G. and {Marshall}, D.~J. and {Martino}, M. and {Marton}, G. and {Mary}, N. and {Massari}, D. and {Matijevi{\v{c}}}, G. and {Mazeh}, T. and {McMillan}, P.~J. and {Messina}, S. and {Michalik}, D. and {Millar}, N.~R. and {Molina}, D. and {Molinaro}, R. and {Moln{\'a}r}, L. and {Montegriffo}, P. and {Mor}, R. and {Morbidelli}, R. and {Morel}, T. and {Morris}, D. and {Mulone}, A.~F. and {Muraveva}, T. and {Musella}, I. and {Nelemans}, G. and {Nicastro}, L. and {Noval}, L. and {O'Mullane}, W. and {Ord{\'e}novic}, C. and {Ord{\'o}{\~n}ez-Blanco}, D. and {Osborne}, P. and {Pagani}, C. and {Pagano}, I. and {Pailler}, F. and {Palacin}, H. and {Palaversa}, L. and {Panahi}, A. and {Pawlak}, M. and {Piersimoni}, A.~M. and {Pineau}, F. -X. and {Plachy}, E. and {Plum}, G. and {Poggio}, E. and {Poujoulet}, E. and {Pr{\v{s}}a}, A. and {Pulone}, L. and {Racero}, E. and {Ragaini}, S. and {Rambaux}, N. and {Ramos-Lerate}, M. and {Regibo}, S. and {Reyl{\'e}}, C. and {Riclet}, F. and {Ripepi}, V. and {Riva}, A. and {Rivard}, A. and {Rixon}, G. and {Roegiers}, T. and {Roelens}, M. and {Romero-G{\'o}mez}, M. and {Rowell}, N. and {Royer}, F. and {Ruiz-Dern}, L. and {Sadowski}, G. and {Sagrist{\`a} Sell{\'e}s}, T. and {Sahlmann}, J. and {Salgado}, J. and {Salguero}, E. and {Sanna}, N. and {Santana-Ros}, T. and {Sarasso}, M. and {Savietto}, H. and {Schultheis}, M. and {Sciacca}, E. and {Segol}, M. and {Segovia}, J.~C. and {S{\'e}gransan}, D. and {Shih}, I. -C. and {Siltala}, L. and {Silva}, A.~F. and {Smart}, R.~L. and {Smith}, K.~W. and {Solano}, E. and {Solitro}, F. and {Sordo}, R. and {Soria Nieto}, S. and {Souchay}, J. and {Spagna}, A. and {Spoto}, F. and {Stampa}, U. and {Steele}, I.~A. and {Steidelm{\"u}ller}, H. and {Stephenson}, C.~A. and {Stoev}, H. and {Suess}, F.~F. and {Surdej}, J. and {Szabados}, L. and {Szegedi-Elek}, E. and {Tapiador}, D. and {Taris}, F. and {Tauran}, G. and {Taylor}, M.~B. and {Teixeira}, R. and {Terrett}, D. and {Teyssandier}, P. and {Thuillot}, W. and {Titarenko}, A. and {Torra Clotet}, F. and {Turon}, C. and {Ulla}, A. and {Utrilla}, E. and {Uzzi}, S. and {Vaillant}, M. and {Valentini}, G. and {Valette}, V. and {van Elteren}, A. and {Van Hemelryck}, E. and {van Leeuwen}, M. and {Vaschetto}, M. and {Vecchiato}, A. and {Veljanoski}, J. and {Viala}, Y. and {Vicente}, D. and {Vogt}, S. and {von Essen}, C. and {Voss}, H. and {Votruba}, V. and {Voutsinas}, S. and {Walmsley}, G. and {Weiler}, M. and {Wertz}, O. and {Wevers}, T. and {Wyrzykowski}, {\L}. and {Yoldas}, A. and {{\v{Z}}erjal}, M. and {Ziaeepour}, H. and {Zorec}, J. and {Zschocke}, S. and {Zucker}, S. and {Zurbach}, C. and {Zwitter}, T.},
        title = "{Gaia Data Release 2. Summary of the contents and survey properties}",
      journal = {\aap},
         year = 2018,
        month = aug,
       volume = {616},
          eid = {A1},
        pages = {A1},
          doi = {10.1051/0004-6361/201833051},
archivePrefix = {arXiv},
       eprint = {1804.09365},
 primaryClass = {astro-ph.GA},
       adsurl = {https://ui.adsabs.harvard.edu/abs/2018A&A...616A...1G}
}

@ARTICLE{Gaia2016,
       author = {{Gaia Collaboration} and {Prusti}, T. and {de Bruijne}, J.~H.~J. and {Brown}, A.~G.~A. and {Vallenari}, A. and {Babusiaux}, C. and {Bailer-Jones}, C.~A.~L. and {Bastian}, U. and {Biermann}, M. and {Evans}, D.~W. and {Eyer}, L. and {Jansen}, F. and {Jordi}, C. and {Klioner}, S.~A. and {Lammers}, U. and {Lindegren}, L. and {Luri}, X. and {Mignard}, F. and {Milligan}, D.~J. and {Panem}, C. and {Poinsignon}, V. and {Pourbaix}, D. and {Randich}, S. and {Sarri}, G. and {Sartoretti}, P. and {Siddiqui}, H.~I. and {Soubiran}, C. and {Valette}, V. and {van Leeuwen}, F. and {Walton}, N.~A. and {Aerts}, C. and {Arenou}, F. and {Cropper}, M. and {Drimmel}, R. and {H{\o}g}, E. and {Katz}, D. and {Lattanzi}, M.~G. and {O'Mullane}, W. and {Grebel}, E.~K. and {Holland}, A.~D. and {Huc}, C. and {Passot}, X. and {Bramante}, L. and {Cacciari}, C. and {Casta{\~n}eda}, J. and {Chaoul}, L. and {Cheek}, N. and {De Angeli}, F. and {Fabricius}, C. and {Guerra}, R. and {Hern{\'a}ndez}, J. and {Jean-Antoine-Piccolo}, A. and {Masana}, E. and {Messineo}, R. and {Mowlavi}, N. and {Nienartowicz}, K. and {Ord{\'o}{\~n}ez-Blanco}, D. and {Panuzzo}, P. and {Portell}, J. and {Richards}, P.~J. and {Riello}, M. and {Seabroke}, G.~M. and {Tanga}, P. and {Th{\'e}venin}, F. and {Torra}, J. and {Els}, S.~G. and {Gracia-Abril}, G. and {Comoretto}, G. and {Garcia-Reinaldos}, M. and {Lock}, T. and {Mercier}, E. and {Altmann}, M. and {Andrae}, R. and {Astraatmadja}, T.~L. and {Bellas-Velidis}, I. and {Benson}, K. and {Berthier}, J. and {Blomme}, R. and {Busso}, G. and {Carry}, B. and {Cellino}, A. and {Clementini}, G. and {Cowell}, S. and {Creevey}, O. and {Cuypers}, J. and {Davidson}, M. and {De Ridder}, J. and {de Torres}, A. and {Delchambre}, L. and {Dell'Oro}, A. and {Ducourant}, C. and {Fr{\'e}mat}, Y. and {Garc{\'\i}a-Torres}, M. and {Gosset}, E. and {Halbwachs}, J. -L. and {Hambly}, N.~C. and {Harrison}, D.~L. and {Hauser}, M. and {Hestroffer}, D. and {Hodgkin}, S.~T. and {Huckle}, H.~E. and {Hutton}, A. and {Jasniewicz}, G. and {Jordan}, S. and {Kontizas}, M. and {Korn}, A.~J. and {Lanzafame}, A.~C. and {Manteiga}, M. and {Moitinho}, A. and {Muinonen}, K. and {Osinde}, J. and {Pancino}, E. and {Pauwels}, T. and {Petit}, J. -M. and {Recio-Blanco}, A. and {Robin}, A.~C. and {Sarro}, L.~M. and {Siopis}, C. and {Smith}, M. and {Smith}, K.~W. and {Sozzetti}, A. and {Thuillot}, W. and {van Reeven}, W. and {Viala}, Y. and {Abbas}, U. and {Abreu Aramburu}, A. and {Accart}, S. and {Aguado}, J.~J. and {Allan}, P.~M. and {Allasia}, W. and {Altavilla}, G. and {{\'A}lvarez}, M.~A. and {Alves}, J. and {Anderson}, R.~I. and {Andrei}, A.~H. and {Anglada Varela}, E. and {Antiche}, E. and {Antoja}, T. and {Ant{\'o}n}, S. and {Arcay}, B. and {Atzei}, A. and {Ayache}, L. and {Bach}, N. and {Baker}, S.~G. and {Balaguer-N{\'u}{\~n}ez}, L. and {Barache}, C. and {Barata}, C. and {Barbier}, A. and {Barblan}, F. and {Baroni}, M. and {Barrado y Navascu{\'e}s}, D. and {Barros}, M. and {Barstow}, M.~A. and {Becciani}, U. and {Bellazzini}, M. and {Bellei}, G. and {Bello Garc{\'\i}a}, A. and {Belokurov}, V. and {Bendjoya}, P. and {Berihuete}, A. and {Bianchi}, L. and {Bienaym{\'e}}, O. and {Billebaud}, F. and {Blagorodnova}, N. and {Blanco-Cuaresma}, S. and {Boch}, T. and {Bombrun}, A. and {Borrachero}, R. and {Bouquillon}, S. and {Bourda}, G. and {Bouy}, H. and {Bragaglia}, A. and {Breddels}, M.~A. and {Brouillet}, N. and {Br{\"u}semeister}, T. and {Bucciarelli}, B. and {Budnik}, F. and {Burgess}, P. and {Burgon}, R. and {Burlacu}, A. and {Busonero}, D. and {Buzzi}, R. and {Caffau}, E. and {Cambras}, J. and {Campbell}, H. and {Cancelliere}, R. and {Cantat-Gaudin}, T. and {Carlucci}, T. and {Carrasco}, J.~M. and {Castellani}, M. and {Charlot}, P. and {Charnas}, J. and {Charvet}, P. and {Chassat}, F. and {Chiavassa}, A. and {Clotet}, M. and {Cocozza}, G. and {Collins}, R.~S. and {Collins}, P. and {Costigan}, G. and {Crifo}, F. and {Cross}, N.~J.~G. and {Crosta}, M. and {Crowley}, C. and {Dafonte}, C. and {Damerdji}, Y. and {Dapergolas}, A. and {David}, P. and {David}, M. and {De Cat}, P. and {de Felice}, F. and {de Laverny}, P. and {De Luise}, F. and {De March}, R. and {de Martino}, D. and {de Souza}, R. and {Debosscher}, J. and {del Pozo}, E. and {Delbo}, M. and {Delgado}, A. and {Delgado}, H.~E. and {di Marco}, F. and {Di Matteo}, P. and {Diakite}, S. and {Distefano}, E. and {Dolding}, C. and {Dos Anjos}, S. and {Drazinos}, P. and {Dur{\'a}n}, J. and {Dzigan}, Y. and {Ecale}, E. and {Edvardsson}, B. and {Enke}, H. and {Erdmann}, M. and {Escolar}, D. and {Espina}, M. and {Evans}, N.~W. and {Eynard Bontemps}, G. and {Fabre}, C. and {Fabrizio}, M. and {Faigler}, S. and {Falc{\~a}o}, A.~J. and {Farr{\`a}s Casas}, M. and {Faye}, F. and {Federici}, L. and {Fedorets}, G. and {Fern{\'a}ndez-Hern{\'a}ndez}, J. and {Fernique}, P. and {Fienga}, A. and {Figueras}, F. and {Filippi}, F. and {Findeisen}, K. and {Fonti}, A. and {Fouesneau}, M. and {Fraile}, E. and {Fraser}, M. and {Fuchs}, J. and {Furnell}, R. and {Gai}, M. and {Galleti}, S. and {Galluccio}, L. and {Garabato}, D. and {Garc{\'\i}a-Sedano}, F. and {Gar{\'e}}, P. and {Garofalo}, A. and {Garralda}, N. and {Gavras}, P. and {Gerssen}, J. and {Geyer}, R. and {Gilmore}, G. and {Girona}, S. and {Giuffrida}, G. and {Gomes}, M. and {Gonz{\'a}lez-Marcos}, A. and {Gonz{\'a}lez-N{\'u}{\~n}ez}, J. and {Gonz{\'a}lez-Vidal}, J.~J. and {Granvik}, M. and {Guerrier}, A. and {Guillout}, P. and {Guiraud}, J. and {G{\'u}rpide}, A. and {Guti{\'e}rrez-S{\'a}nchez}, R. and {Guy}, L.~P. and {Haigron}, R. and {Hatzidimitriou}, D. and {Haywood}, M. and {Heiter}, U. and {Helmi}, A. and {Hobbs}, D. and {Hofmann}, W. and {Holl}, B. and {Holland}, G. and {Hunt}, J.~A.~S. and {Hypki}, A. and {Icardi}, V. and {Irwin}, M. and {Jevardat de Fombelle}, G. and {Jofr{\'e}}, P. and {Jonker}, P.~G. and {Jorissen}, A. and {Julbe}, F. and {Karampelas}, A. and {Kochoska}, A. and {Kohley}, R. and {Kolenberg}, K. and {Kontizas}, E. and {Koposov}, S.~E. and {Kordopatis}, G. and {Koubsky}, P. and {Kowalczyk}, A. and {Krone-Martins}, A. and {Kudryashova}, M. and {Kull}, I. and {Bachchan}, R.~K. and {Lacoste-Seris}, F. and {Lanza}, A.~F. and {Lavigne}, J. -B. and {Le Poncin-Lafitte}, C. and {Lebreton}, Y. and {Lebzelter}, T. and {Leccia}, S. and {Leclerc}, N. and {Lecoeur-Taibi}, I. and {Lemaitre}, V. and {Lenhardt}, H. and {Leroux}, F. and {Liao}, S. and {Licata}, E. and {Lindstr{\o}m}, H.~E.~P. and {Lister}, T.~A. and {Livanou}, E. and {Lobel}, A. and {L{\"o}ffler}, W. and {L{\'o}pez}, M. and {Lopez-Lozano}, A. and {Lorenz}, D. and {Loureiro}, T. and {MacDonald}, I. and {Magalh{\~a}es Fernandes}, T. and {Managau}, S. and {Mann}, R.~G. and {Mantelet}, G. and {Marchal}, O. and {Marchant}, J.~M. and {Marconi}, M. and {Marie}, J. and {Marinoni}, S. and {Marrese}, P.~M. and {Marschalk{\'o}}, G. and {Marshall}, D.~J. and {Mart{\'\i}n-Fleitas}, J.~M. and {Martino}, M. and {Mary}, N. and {Matijevi{\v{c}}}, G. and {Mazeh}, T. and {McMillan}, P.~J. and {Messina}, S. and {Mestre}, A. and {Michalik}, D. and {Millar}, N.~R. and {Miranda}, B.~M.~H. and {Molina}, D. and {Molinaro}, R. and {Molinaro}, M. and {Moln{\'a}r}, L. and {Moniez}, M. and {Montegriffo}, P. and {Monteiro}, D. and {Mor}, R. and {Mora}, A. and {Morbidelli}, R. and {Morel}, T. and {Morgenthaler}, S. and {Morley}, T. and {Morris}, D. and {Mulone}, A.~F. and {Muraveva}, T. and {Musella}, I. and {Narbonne}, J. and {Nelemans}, G. and {Nicastro}, L. and {Noval}, L. and {Ord{\'e}novic}, C. and {Ordieres-Mer{\'e}}, J. and {Osborne}, P. and {Pagani}, C. and {Pagano}, I. and {Pailler}, F. and {Palacin}, H. and {Palaversa}, L. and {Parsons}, P. and {Paulsen}, T. and {Pecoraro}, M. and {Pedrosa}, R. and {Pentik{\"a}inen}, H. and {Pereira}, J. and {Pichon}, B. and {Piersimoni}, A.~M. and {Pineau}, F. -X. and {Plachy}, E. and {Plum}, G. and {Poujoulet}, E. and {Pr{\v{s}}a}, A. and {Pulone}, L. and {Ragaini}, S. and {Rago}, S. and {Rambaux}, N. and {Ramos-Lerate}, M. and {Ranalli}, P. and {Rauw}, G. and {Read}, A. and {Regibo}, S. and {Renk}, F. and {Reyl{\'e}}, C. and {Ribeiro}, R.~A. and {Rimoldini}, L. and {Ripepi}, V. and {Riva}, A. and {Rixon}, G. and {Roelens}, M. and {Romero-G{\'o}mez}, M. and {Rowell}, N. and {Royer}, F. and {Rudolph}, A. and {Ruiz-Dern}, L. and {Sadowski}, G. and {Sagrist{\`a} Sell{\'e}s}, T. and {Sahlmann}, J. and {Salgado}, J. and {Salguero}, E. and {Sarasso}, M. and {Savietto}, H. and {Schnorhk}, A. and {Schultheis}, M. and {Sciacca}, E. and {Segol}, M. and {Segovia}, J.~C. and {Segransan}, D. and {Serpell}, E. and {Shih}, I. -C. and {Smareglia}, R. and {Smart}, R.~L. and {Smith}, C. and {Solano}, E. and {Solitro}, F. and {Sordo}, R. and {Soria Nieto}, S. and {Souchay}, J. and {Spagna}, A. and {Spoto}, F. and {Stampa}, U. and {Steele}, I.~A. and {Steidelm{\"u}ller}, H. and {Stephenson}, C.~A. and {Stoev}, H. and {Suess}, F.~F. and {S{\"u}veges}, M. and {Surdej}, J. and {Szabados}, L. and {Szegedi-Elek}, E. and {Tapiador}, D. and {Taris}, F. and {Tauran}, G. and {Taylor}, M.~B. and {Teixeira}, R. and {Terrett}, D. and {Tingley}, B. and {Trager}, S.~C. and {Turon}, C. and {Ulla}, A. and {Utrilla}, E. and {Valentini}, G. and {van Elteren}, A. and {Van Hemelryck}, E. and {van Leeuwen}, M. and {Varadi}, M. and {Vecchiato}, A. and {Veljanoski}, J. and {Via}, T. and {Vicente}, D. and {Vogt}, S. and {Voss}, H. and {Votruba}, V. and {Voutsinas}, S. and {Walmsley}, G. and {Weiler}, M. and {Weingrill}, K. and {Werner}, D. and {Wevers}, T. and {Whitehead}, G. and {Wyrzykowski}, {\L}. and {Yoldas}, A. and {{\v{Z}}erjal}, M. and {Zucker}, S. and {Zurbach}, C. and {Zwitter}, T. and {Alecu}, A. and {Allen}, M. and {Allende Prieto}, C. and {Amorim}, A. and {Anglada-Escud{\'e}}, G. and {Arsenijevic}, V. and {Azaz}, S. and {Balm}, P. and {Beck}, M. and {Bernstein}, H. -H. and {Bigot}, L. and {Bijaoui}, A. and {Blasco}, C. and {Bonfigli}, M. and {Bono}, G. and {Boudreault}, S. and {Bressan}, A. and {Brown}, S. and {Brunet}, P. -M. and {Bunclark}, P. and {Buonanno}, R. and {Butkevich}, A.~G. and {Carret}, C. and {Carrion}, C. and {Chemin}, L. and {Ch{\'e}reau}, F. and {Corcione}, L. and {Darmigny}, E. and {de Boer}, K.~S. and {de Teodoro}, P. and {de Zeeuw}, P.~T. and {Delle Luche}, C. and {Domingues}, C.~D. and {Dubath}, P. and {Fodor}, F. and {Fr{\'e}zouls}, B. and {Fries}, A. and {Fustes}, D. and {Fyfe}, D. and {Gallardo}, E. and {Gallegos}, J. and {Gardiol}, D. and {Gebran}, M. and {Gomboc}, A. and {G{\'o}mez}, A. and {Grux}, E. and {Gueguen}, A. and {Heyrovsky}, A. and {Hoar}, J. and {Iannicola}, G. and {Isasi Parache}, Y. and {Janotto}, A. -M. and {Joliet}, E. and {Jonckheere}, A. and {Keil}, R. and {Kim}, D. -W. and {Klagyivik}, P. and {Klar}, J. and {Knude}, J. and {Kochukhov}, O. and {Kolka}, I. and {Kos}, J. and {Kutka}, A. and {Lainey}, V. and {LeBouquin}, D. and {Liu}, C. and {Loreggia}, D. and {Makarov}, V.~V. and {Marseille}, M.~G. and {Martayan}, C. and {Martinez-Rubi}, O. and {Massart}, B. and {Meynadier}, F. and {Mignot}, S. and {Munari}, U. and {Nguyen}, A. -T. and {Nordlander}, T. and {Ocvirk}, P. and {O'Flaherty}, K.~S. and {Olias Sanz}, A. and {Ortiz}, P. and {Osorio}, J. and {Oszkiewicz}, D. and {Ouzounis}, A. and {Palmer}, M. and {Park}, P. and {Pasquato}, E. and {Peltzer}, C. and {Peralta}, J. and {P{\'e}turaud}, F. and {Pieniluoma}, T. and {Pigozzi}, E. and {Poels}, J. and {Prat}, G. and {Prod'homme}, T. and {Raison}, F. and {Rebordao}, J.~M. and {Risquez}, D. and {Rocca-Volmerange}, B. and {Rosen}, S. and {Ruiz-Fuertes}, M.~I. and {Russo}, F. and {Sembay}, S. and {Serraller Vizcaino}, I. and {Short}, A. and {Siebert}, A. and {Silva}, H. and {Sinachopoulos}, D. and {Slezak}, E. and {Soffel}, M. and {Sosnowska}, D. and {Strai{\v{z}}ys}, V. and {ter Linden}, M. and {Terrell}, D. and {Theil}, S. and {Tiede}, C. and {Troisi}, L. and {Tsalmantza}, P. and {Tur}, D. and {Vaccari}, M. and {Vachier}, F. and {Valles}, P. and {Van Hamme}, W. and {Veltz}, L. and {Virtanen}, J. and {Wallut}, J. -M. and {Wichmann}, R. and {Wilkinson}, M.~I. and {Ziaeepour}, H. and {Zschocke}, S.},
        title = "{The Gaia mission}",
      journal = {\aap},
         year = 2016,
        month = nov,
       volume = {595},
          eid = {A1},
        pages = {A1},
          doi = {10.1051/0004-6361/201629272},
archivePrefix = {arXiv},
       eprint = {1609.04153},
 primaryClass = {astro-ph.IM},
       adsurl = {https://ui.adsabs.harvard.edu/abs/2016A&A...595A...1G}
}

@ARTICLE{VVV2010,
       author = {{Minniti}, D. and {Lucas}, P.~W. and {Emerson}, J.~P. and {Saito}, R.~K. and {Hempel}, M. and {Pietrukowicz}, P. and {Ahumada}, A.~V. and {Alonso}, M.~V. and {Alonso-Garcia}, J. and {Arias}, J.~I. and {Bandyopadhyay}, R.~M. and {Barb{\'a}}, R.~H. and {Barbuy}, B. and {Bedin}, L.~R. and {Bica}, E. and {Borissova}, J. and {Bronfman}, L. and {Carraro}, G. and {Catelan}, M. and {Clari{\'a}}, J.~J. and {Cross}, N. and {de Grijs}, R. and {D{\'e}k{\'a}ny}, I. and {Drew}, J.~E. and {Fari{\~n}a}, C. and {Feinstein}, C. and {Fern{\'a}ndez Laj{\'u}s}, E. and {Gamen}, R.~C. and {Geisler}, D. and {Gieren}, W. and {Goldman}, B. and {Gonzalez}, O.~A. and {Gunthardt}, G. and {Gurovich}, S. and {Hambly}, N.~C. and {Irwin}, M.~J. and {Ivanov}, V.~D. and {Jord{\'a}n}, A. and {Kerins}, E. and {Kinemuchi}, K. and {Kurtev}, R. and {L{\'o}pez-Corredoira}, M. and {Maccarone}, T. and {Masetti}, N. and {Merlo}, D. and {Messineo}, M. and {Mirabel}, I.~F. and {Monaco}, L. and {Morelli}, L. and {Padilla}, N. and {Palma}, T. and {Parisi}, M.~C. and {Pignata}, G. and {Rejkuba}, M. and {Roman-Lopes}, A. and {Sale}, S.~E. and {Schreiber}, M.~R. and {Schr{\"o}der}, A.~C. and {Smith}, M. and {Sodr{\'e}}, Jr., L. and {Soto}, M. and {Tamura}, M. and {Tappert}, C. and {Thompson}, M.~A. and {Toledo}, I. and {Zoccali}, M. and {Pietrzynski}, G.},
        title = "{VISTA Variables in the Via Lactea (VVV): The public ESO near-IR variability survey of the Milky Way}",
      journal = {\na},
         year = 2010,
        month = jul,
       volume = {15},
       number = {5},
        pages = {433-443},
          doi = {10.1016/j.newast.2009.12.002},
archivePrefix = {arXiv},
       eprint = {0912.1056},
 primaryClass = {astro-ph.GA},
       adsurl = {https://ui.adsabs.harvard.edu/abs/2010NewA...15..433M}
}

@ARTICLE{Gaia2023,
       author = {{Gaia Collaboration} and {Vallenari}, A. and {Brown}, A.~G.~A. and {Prusti}, T. and {de Bruijne}, J.~H.~J. and {Arenou}, F. and {Babusiaux}, C. and {Biermann}, M. and {Creevey}, O.~L. and {Ducourant}, C. and {Evans}, D.~W. and {Eyer}, L. and {Guerra}, R. and {Hutton}, A. and {Jordi}, C. and {Klioner}, S.~A. and {Lammers}, U.~L. and {Lindegren}, L. and {Luri}, X. and {Mignard}, F. and {Panem}, C. and {Pourbaix}, D. and {Randich}, S. and {Sartoretti}, P. and {Soubiran}, C. and {Tanga}, P. and {Walton}, N.~A. and {Bailer-Jones}, C.~A.~L. and {Bastian}, U. and {Drimmel}, R. and {Jansen}, F. and {Katz}, D. and {Lattanzi}, M.~G. and {van Leeuwen}, F. and {Bakker}, J. and {Cacciari}, C. and {Casta{\~n}eda}, J. and {De Angeli}, F. and {Fabricius}, C. and {Fouesneau}, M. and {Fr{\'e}mat}, Y. and {Galluccio}, L. and {Guerrier}, A. and {Heiter}, U. and {Masana}, E. and {Messineo}, R. and {Mowlavi}, N. and {Nicolas}, C. and {Nienartowicz}, K. and {Pailler}, F. and {Panuzzo}, P. and {Riclet}, F. and {Roux}, W. and {Seabroke}, G.~M. and {Sordo}, R. and {Th{\'e}venin}, F. and {Gracia-Abril}, G. and {Portell}, J. and {Teyssier}, D. and {Altmann}, M. and {Andrae}, R. and {Audard}, M. and {Bellas-Velidis}, I. and {Benson}, K. and {Berthier}, J. and {Blomme}, R. and {Burgess}, P.~W. and {Busonero}, D. and {Busso}, G. and {C{\'a}novas}, H. and {Carry}, B. and {Cellino}, A. and {Cheek}, N. and {Clementini}, G. and {Damerdji}, Y. and {Davidson}, M. and {de Teodoro}, P. and {Nu{\~n}ez Campos}, M. and {Delchambre}, L. and {Dell'Oro}, A. and {Esquej}, P. and {Fern{\'a}ndez-Hern{\'a}ndez}, J. and {Fraile}, E. and {Garabato}, D. and {Garc{\'\i}a-Lario}, P. and {Gosset}, E. and {Haigron}, R. and {Halbwachs}, J. -L. and {Hambly}, N.~C. and {Harrison}, D.~L. and {Hern{\'a}ndez}, J. and {Hestroffer}, D. and {Hodgkin}, S.~T. and {Holl}, B. and {Jan{\ss}en}, K. and {Jevardat de Fombelle}, G. and {Jordan}, S. and {Krone-Martins}, A. and {Lanzafame}, A.~C. and {L{\"o}ffler}, W. and {Marchal}, O. and {Marrese}, P.~M. and {Moitinho}, A. and {Muinonen}, K. and {Osborne}, P. and {Pancino}, E. and {Pauwels}, T. and {Recio-Blanco}, A. and {Reyl{\'e}}, C. and {Riello}, M. and {Rimoldini}, L. and {Roegiers}, T. and {Rybizki}, J. and {Sarro}, L.~M. and {Siopis}, C. and {Smith}, M. and {Sozzetti}, A. and {Utrilla}, E. and {van Leeuwen}, M. and {Abbas}, U. and {{\'A}brah{\'a}m}, P. and {Abreu Aramburu}, A. and {Aerts}, C. and {Aguado}, J.~J. and {Ajaj}, M. and {Aldea-Montero}, F. and {Altavilla}, G. and {{\'A}lvarez}, M.~A. and {Alves}, J. and {Anders}, F. and {Anderson}, R.~I. and {Anglada Varela}, E. and {Antoja}, T. and {Baines}, D. and {Baker}, S.~G. and {Balaguer-N{\'u}{\~n}ez}, L. and {Balbinot}, E. and {Balog}, Z. and {Barache}, C. and {Barbato}, D. and {Barros}, M. and {Barstow}, M.~A. and {Bartolom{\'e}}, S. and {Bassilana}, J. -L. and {Bauchet}, N. and {Becciani}, U. and {Bellazzini}, M. and {Berihuete}, A. and {Bernet}, M. and {Bertone}, S. and {Bianchi}, L. and {Binnenfeld}, A. and {Blanco-Cuaresma}, S. and {Blazere}, A. and {Boch}, T. and {Bombrun}, A. and {Bossini}, D. and {Bouquillon}, S. and {Bragaglia}, A. and {Bramante}, L. and {Breedt}, E. and {Bressan}, A. and {Brouillet}, N. and {Brugaletta}, E. and {Bucciarelli}, B. and {Burlacu}, A. and {Butkevich}, A.~G. and {Buzzi}, R. and {Caffau}, E. and {Cancelliere}, R. and {Cantat-Gaudin}, T. and {Carballo}, R. and {Carlucci}, T. and {Carnerero}, M.~I. and {Carrasco}, J.~M. and {Casamiquela}, L. and {Castellani}, M. and {Castro-Ginard}, A. and {Chaoul}, L. and {Charlot}, P. and {Chemin}, L. and {Chiaramida}, V. and {Chiavassa}, A. and {Chornay}, N. and {Comoretto}, G. and {Contursi}, G. and {Cooper}, W.~J. and {Cornez}, T. and {Cowell}, S. and {Crifo}, F. and {Cropper}, M. and {Crosta}, M. and {Crowley}, C. and {Dafonte}, C. and {Dapergolas}, A. and {David}, M. and {David}, P. and {de Laverny}, P. and {De Luise}, F. and {De March}, R. and {De Ridder}, J. and {de Souza}, R. and {de Torres}, A. and {del Peloso}, E.~F. and {del Pozo}, E. and {Delbo}, M. and {Delgado}, A. and {Delisle}, J. -B. and {Demouchy}, C. and {Dharmawardena}, T.~E. and {Di Matteo}, P. and {Diakite}, S. and {Diener}, C. and {Distefano}, E. and {Dolding}, C. and {Edvardsson}, B. and {Enke}, H. and {Fabre}, C. and {Fabrizio}, M. and {Faigler}, S. and {Fedorets}, G. and {Fernique}, P. and {Fienga}, A. and {Figueras}, F. and {Fournier}, Y. and {Fouron}, C. and {Fragkoudi}, F. and {Gai}, M. and {Garcia-Gutierrez}, A. and {Garcia-Reinaldos}, M. and {Garc{\'\i}a-Torres}, M. and {Garofalo}, A. and {Gavel}, A. and {Gavras}, P. and {Gerlach}, E. and {Geyer}, R. and {Giacobbe}, P. and {Gilmore}, G. and {Girona}, S. and {Giuffrida}, G. and {Gomel}, R. and {Gomez}, A. and {Gonz{\'a}lez-N{\'u}{\~n}ez}, J. and {Gonz{\'a}lez-Santamar{\'\i}a}, I. and {Gonz{\'a}lez-Vidal}, J.~J. and {Granvik}, M. and {Guillout}, P. and {Guiraud}, J. and {Guti{\'e}rrez-S{\'a}nchez}, R. and {Guy}, L.~P. and {Hatzidimitriou}, D. and {Hauser}, M. and {Haywood}, M. and {Helmer}, A. and {Helmi}, A. and {Sarmiento}, M.~H. and {Hidalgo}, S.~L. and {Hilger}, T. and {H{\l}adczuk}, N. and {Hobbs}, D. and {Holland}, G. and {Huckle}, H.~E. and {Jardine}, K. and {Jasniewicz}, G. and {Jean-Antoine Piccolo}, A. and {Jim{\'e}nez-Arranz}, {\'O}. and {Jorissen}, A. and {Juaristi Campillo}, J. and {Julbe}, F. and {Karbevska}, L. and {Kervella}, P. and {Khanna}, S. and {Kontizas}, M. and {Kordopatis}, G. and {Korn}, A.~J. and {K{\'o}sp{\'a}l}, {\'A}. and {Kostrzewa-Rutkowska}, Z. and {Kruszy{\'n}ska}, K. and {Kun}, M. and {Laizeau}, P. and {Lambert}, S. and {Lanza}, A.~F. and {Lasne}, Y. and {Le Campion}, J. -F. and {Lebreton}, Y. and {Lebzelter}, T. and {Leccia}, S. and {Leclerc}, N. and {Lecoeur-Taibi}, I. and {Liao}, S. and {Licata}, E.~L. and {Lindstr{\o}m}, H.~E.~P. and {Lister}, T.~A. and {Livanou}, E. and {Lobel}, A. and {Lorca}, A. and {Loup}, C. and {Madrero Pardo}, P. and {Magdaleno Romeo}, A. and {Managau}, S. and {Mann}, R.~G. and {Manteiga}, M. and {Marchant}, J.~M. and {Marconi}, M. and {Marcos}, J. and {Marcos Santos}, M.~M.~S. and {Mar{\'\i}n Pina}, D. and {Marinoni}, S. and {Marocco}, F. and {Marshall}, D.~J. and {Martin Polo}, L. and {Mart{\'\i}n-Fleitas}, J.~M. and {Marton}, G. and {Mary}, N. and {Masip}, A. and {Massari}, D. and {Mastrobuono-Battisti}, A. and {Mazeh}, T. and {McMillan}, P.~J. and {Messina}, S. and {Michalik}, D. and {Millar}, N.~R. and {Mints}, A. and {Molina}, D. and {Molinaro}, R. and {Moln{\'a}r}, L. and {Monari}, G. and {Mongui{\'o}}, M. and {Montegriffo}, P. and {Montero}, A. and {Mor}, R. and {Mora}, A. and {Morbidelli}, R. and {Morel}, T. and {Morris}, D. and {Muraveva}, T. and {Murphy}, C.~P. and {Musella}, I. and {Nagy}, Z. and {Noval}, L. and {Oca{\~n}a}, F. and {Ogden}, A. and {Ordenovic}, C. and {Osinde}, J.~O. and {Pagani}, C. and {Pagano}, I. and {Palaversa}, L. and {Palicio}, P.~A. and {Pallas-Quintela}, L. and {Panahi}, A. and {Payne-Wardenaar}, S. and {Pe{\~n}alosa Esteller}, X. and {Penttil{\"a}}, A. and {Pichon}, B. and {Piersimoni}, A.~M. and {Pineau}, F. -X. and {Plachy}, E. and {Plum}, G. and {Poggio}, E. and {Pr{\v{s}}a}, A. and {Pulone}, L. and {Racero}, E. and {Ragaini}, S. and {Rainer}, M. and {Raiteri}, C.~M. and {Rambaux}, N. and {Ramos}, P. and {Ramos-Lerate}, M. and {Re Fiorentin}, P. and {Regibo}, S. and {Richards}, P.~J. and {Rios Diaz}, C. and {Ripepi}, V. and {Riva}, A. and {Rix}, H. -W. and {Rixon}, G. and {Robichon}, N. and {Robin}, A.~C. and {Robin}, C. and {Roelens}, M. and {Rogues}, H.~R.~O. and {Rohrbasser}, L. and {Romero-G{\'o}mez}, M. and {Rowell}, N. and {Royer}, F. and {Ruz Mieres}, D. and {Rybicki}, K.~A. and {Sadowski}, G. and {S{\'a}ez N{\'u}{\~n}ez}, A. and {Sagrist{\`a} Sell{\'e}s}, A. and {Sahlmann}, J. and {Salguero}, E. and {Samaras}, N. and {Sanchez Gimenez}, V. and {Sanna}, N. and {Santove{\~n}a}, R. and {Sarasso}, M. and {Schultheis}, M. and {Sciacca}, E. and {Segol}, M. and {Segovia}, J.~C. and {S{\'e}gransan}, D. and {Semeux}, D. and {Shahaf}, S. and {Siddiqui}, H.~I. and {Siebert}, A. and {Siltala}, L. and {Silvelo}, A. and {Slezak}, E. and {Slezak}, I. and {Smart}, R.~L. and {Snaith}, O.~N. and {Solano}, E. and {Solitro}, F. and {Souami}, D. and {Souchay}, J. and {Spagna}, A. and {Spina}, L. and {Spoto}, F. and {Steele}, I.~A. and {Steidelm{\"u}ller}, H. and {Stephenson}, C.~A. and {S{\"u}veges}, M. and {Surdej}, J. and {Szabados}, L. and {Szegedi-Elek}, E. and {Taris}, F. and {Taylor}, M.~B. and {Teixeira}, R. and {Tolomei}, L. and {Tonello}, N. and {Torra}, F. and {Torra}, J. and {Torralba Elipe}, G. and {Trabucchi}, M. and {Tsounis}, A.~T. and {Turon}, C. and {Ulla}, A. and {Unger}, N. and {Vaillant}, M.~V. and {van Dillen}, E. and {van Reeven}, W. and {Vanel}, O. and {Vecchiato}, A. and {Viala}, Y. and {Vicente}, D. and {Voutsinas}, S. and {Weiler}, M. and {Wevers}, T. and {Wyrzykowski}, {\L}. and {Yoldas}, A. and {Yvard}, P. and {Zhao}, H. and {Zorec}, J. and {Zucker}, S. and {Zwitter}, T.},
        title = "{Gaia Data Release 3. Summary of the content and survey properties}",
      journal = {\aap},
         year = 2023,
        month = jun,
       volume = {674},
          eid = {A1},
        pages = {A1},
          doi = {10.1051/0004-6361/202243940},
archivePrefix = {arXiv},
       eprint = {2208.00211},
 primaryClass = {astro-ph.GA},
       adsurl = {https://ui.adsabs.harvard.edu/abs/2023A&A...674A...1G}
}

@article{ZoccaliQuezada2024,
	author = {{Zoccali}, M. and {Quezada, C.} and {Contreras Ramos, R.} and {Valenti, E.} and {Valenzuela-Navarro, A.} and {Olivares Carvajal, J.} and {Rojas Arriagada, A.} and {Minniti, J. H.} and {Gran, F.} and {De Leo, M.}},
	title = {VVV catalog of ab-type RR Lyrae in the inner Galactic bulge★},
	DOI= "10.1051/0004-6361/202450126",
	url= "https://doi.org/10.1051/0004-6361/202450126",
	journal = {\aap},
	year = 2024,
	volume = 689,
	pages = "A240",
}

@ARTICLE{ZoccaliRojas24,
       author = {{Zoccali}, M. and {Rojas-Arriagada}, A. and {Valenti}, E. and {Contreras Ramos}, R. and {Valenzuela-Navarro}, A. and {Salvo-Guajardo}, C.},
        title = "{Observed kinematics of the Milky Way nuclear stellar disk region}",
      journal = {\aap},
         year = 2024,
        month = apr,
       volume = {684},
          eid = {A214},
        pages = {A214},
          doi = {10.1051/0004-6361/202347923},
archivePrefix = {arXiv},
       eprint = {2402.16800},
 primaryClass = {astro-ph.GA},
       adsurl = {https://ui.adsabs.harvard.edu/abs/2024A&A...684A.214Z}
}

@ARTICLE{VVVx2018,
       author = {{Minniti}, Dante and {Saito}, Roberto K. and {Gonzalez}, Oscar A. and {Alonso-Garc{\'\i}a}, Javier and {Rejkuba}, Marina and {Barb{\'a}}, Rodolfo and {Irwin}, Mike and {Kammers}, Roberto and {Lucas}, Phillip W. and {Majaess}, Daniel and {Valenti}, Elena},
        title = "{A new near-IR window of low extinction in the Galactic plane}",
      journal = {\aap},
         year = 2018,
        month = aug,
       volume = {616},
          eid = {A26},
        pages = {A26},
          doi = {10.1051/0004-6361/201732099},
archivePrefix = {arXiv},
       eprint = {1804.07785},
 primaryClass = {astro-ph.GA},
       adsurl = {https://ui.adsabs.harvard.edu/abs/2018A&A...616A..26M}
}

@ARTICLE{Dekany2022,
       author = {{D{\'e}k{\'a}ny}, Istv{\'a}n and {Grebel}, Eva K.},
        title = "{Photometric Metallicity Prediction of Fundamental-mode RR Lyrae Stars in the Gaia Optical and K $_{ s }$ Infrared Wave Bands by Deep Learning}",
      journal = {\apjs},
         year = 2022,
        month = aug,
       volume = {261},
       number = {2},
          eid = {33},
        pages = {33},
          doi = {10.3847/1538-4365/ac74ba},
       adsurl = {https://ui.adsabs.harvard.edu/abs/2022ApJS..261...33D}
}

@article{Minniti2020,
	author = {{Minniti}, J. H. and {Sbordone, L.} and {Rojas-Arriagada, A.} and {Zoccali, M.} and {Contreras Ramos, R.} and {Minniti, D.} and {Marconi, M.} and {Braga, V. F.} and {Catelan, M.} and {Duffau, S.} and {Gieren, W.} and {Valcarce, A. A. R.}},
	title = {Using classical Cepheids to study the far side of the Milky Way disk⋆ - I. Spectroscopic classification and the metallicity gradient},
	DOI= "10.1051/0004-6361/202037575",
	url= "https://doi.org/10.1051/0004-6361/202037575",
	journal = {\aap},
	year = 2020,
	volume = 640,
	pages = "A92",
}

@article{ContrerasRamos2017,
	author = {{Contreras Ramos}, R. and {Zoccali, M.} and {Rojas, F.} and {Rojas-Arriagada, A.} and {Gárate, M.} and {Huijse, P.} and {Gran, F.} and {Soto, M.} and {Valcarce, A. A. R.} and {Estévez, P. A.} and {Minniti, D.}},
	title = {Proper motions in the VVV Survey: Results for more   than 15 million stars across NGC 6544⋆},
	DOI= "10.1051/0004-6361/201731462",
	url= "https://doi.org/10.1051/0004-6361/201731462",
	journal = {\aap},
	year = 2017,
	volume = 608,
	pages = "A140",
}

@ARTICLE{Schlafly2011,
       author = {{Schlafly}, Edward F. and {Finkbeiner}, Douglas P.},
        title = "{Measuring Reddening with Sloan Digital Sky Survey Stellar Spectra and Recalibrating SFD}",
      journal = {\apj},
         year = 2011,
        month = aug,
       volume = {737},
       number = {2},
          eid = {103},
        pages = {103},
          doi = {10.1088/0004-637X/737/2/103},
archivePrefix = {arXiv},
       eprint = {1012.4804},
 primaryClass = {astro-ph.GA},
       adsurl = {https://ui.adsabs.harvard.edu/abs/2011ApJ...737..103S}
}

@ARTICLE{Baumgardt2021,
       author = {{Baumgardt}, H. and {Vasiliev}, E.},
        title = "{Accurate distances to Galactic globular clusters through a combination of Gaia EDR3, HST, and literature data}",
      journal = {\mnras},
         year = 2021,
        month = aug,
       volume = {505},
       number = {4},
        pages = {5957-5977},
          doi = {10.1093/mnras/stab1474},
archivePrefix = {arXiv},
       eprint = {2105.09526},
 primaryClass = {astro-ph.GA},
       adsurl = {https://ui.adsabs.harvard.edu/abs/2021MNRAS.505.5957B}
}

@ARTICLE{Wang2019,
       author = {{Wang}, Shu and {Chen}, Xiaodian},
        title = "{The Optical to Mid-infrared Extinction Law Based on the APOGEE, Gaia DR2, Pan-STARRS1, SDSS, APASS, 2MASS, and WISE Surveys}",
      journal = {\apj},
         year = 2019,
        month = jun,
       volume = {877},
       number = {2},
          eid = {116},
        pages = {116},
          doi = {10.3847/1538-4357/ab1c61},
archivePrefix = {arXiv},
       eprint = {1904.04575},
 primaryClass = {astro-ph.GA},
       adsurl = {https://ui.adsabs.harvard.edu/abs/2019ApJ...877..116W}
}

@article{sklearn,
author = {Pedregosa, Fabian and Varoquaux, Gaël and Gramfort, Alexandre and Michel, Vincent and Thirion, Bertrand and Grisel, Olivier and Blondel, Mathieu and Prettenhofer, Peter and Weiss, Ron and Dubourg, Vincent and Vanderplas, Jake and Passos, Alexandre and Cournapeau, David and Brucher, Matthieu and Perrot, Matthieu and Duchesnay, Édouard},
journal = {Journal of Machine Learning Research},
pages = {2825--2830},
title = {{Scikit-learn: Machine Learning in Python}},
url = {https://jmlr.csail.mit.edu/papers/v12/pedregosa11a.html},
volume = {12},
year = {2011}
}

@ARTICLE{Clement2001,
       author = {{Clement}, Christine M. and {Muzzin}, Adam and {Dufton}, Quentin and {Ponnampalam}, Thivya and {Wang}, John and {Burford}, Jay and {Richardson}, Alan and {Rosebery}, Tara and {Rowe}, Jason and {Hogg}, Helen Sawyer},
        title = "{Variable Stars in Galactic Globular Clusters}",
      journal = {\aj},
         year = 2001,
        month = nov,
       volume = {122},
       number = {5},
        pages = {2587-2599},
          doi = {10.1086/323719},
archivePrefix = {arXiv},
       eprint = {astro-ph/0108024},
 primaryClass = {astro-ph},
       adsurl = {https://ui.adsabs.harvard.edu/abs/2001AJ....122.2587C}
}

@ARTICLE{WhiteRees1978,
       author = {{White}, S.~D.~M. and {Rees}, M.~J.},
        title = "{Core condensation in heavy halos: a two-stage theory for galaxy formation and clustering.}",
      journal = {\mnras},
         year = 1978,
        month = may,
       volume = {183},
        pages = {341-358},
          doi = {10.1093/mnras/183.3.341},
       adsurl = {https://ui.adsabs.harvard.edu/abs/1978MNRAS.183..341W}
}

@ARTICLE{Monty2024,
       author = {{Monty}, Stephanie and {Belokurov}, Vasily and {Sanders}, Jason L. and {Hansen}, Terese T. and {Sakari}, Charli M. and {McKenzie}, Madeleine and {Myeong}, GyuChul and {Davies}, Elliot Y. and {Ardern-Arentsen}, Anke and {Massari}, Davide},
        title = "{The ratio of [Eu/{\ensuremath{\alpha}}] differentiates accreted/in situ Milky Way stars across metallicities, as indicated by both field stars and globular clusters}",
      journal = {\mnras},
         year = 2024,
        month = sep,
       volume = {533},
       number = {2},
        pages = {2420-2440},
          doi = {10.1093/mnras/stae1895},
archivePrefix = {arXiv},
       eprint = {2405.08963},
 primaryClass = {astro-ph.GA},
       adsurl = {https://ui.adsabs.harvard.edu/abs/2024MNRAS.533.2420M}
}

@BOOK{Catelan2015,
     author    = {{Catelan}, M. and {Smith}, H.~A.},
     title     = "{Pulsating Stars}",
     publisher = "Wiley-VCH, Weinheim",
     year      = 2015,
     doi       = "10.1002/9783527655182", 
     isbn      = "9783527655182",
     adsurl    = {https://ui.adsabs.harvard.edu/abs/2015pust.book.....C}
}

@ARTICLE{White1991,
       author = {{White}, Simon D.~M. and {Frenk}, Carlos S.},
        title = "{Galaxy Formation through Hierarchical Clustering}",
      journal = {\apj},
         year = 1991,
        month = sep,
       volume = {379},
        pages = {52},
          doi = {10.1086/170483},
       adsurl = {https://ui.adsabs.harvard.edu/abs/1991ApJ...379...52W}
}

@ARTICLE{Brodie2006,
       author = {{Brodie}, Jean P. and {Strader}, Jay},
        title = "{Extragalactic Globular Clusters and Galaxy Formation}",
      journal = {\araa},
         year = 2006,
        month = sep,
       volume = {44},
       number = {1},
        pages = {193-267},
          doi = {10.1146/annurev.astro.44.051905.092441},
archivePrefix = {arXiv},
       eprint = {astro-ph/0602601},
 primaryClass = {astro-ph},
       adsurl = {https://ui.adsabs.harvard.edu/abs/2006ARA&A..44..193B}
}

@ARTICLE{Trujillo-Gomez2021,
       author = {{Trujillo-Gomez}, Sebastian and {Kruijssen}, J.~M. Diederik and {Reina-Campos}, Marta and {Pfeffer}, Joel L. and {Keller}, Benjamin W. and {Crain}, Robert A. and {Bastian}, Nate and {Hughes}, Meghan E.},
        title = "{The kinematics of globular cluster populations in the E-MOSAICS simulations and their implications for the assembly history of the Milky Way}",
      journal = {\mnras},
         year = 2021,
        month = may,
       volume = {503},
       number = {1},
        pages = {31-58},
          doi = {10.1093/mnras/stab341},
archivePrefix = {arXiv},
       eprint = {2005.02401},
 primaryClass = {astro-ph.GA},
       adsurl = {https://ui.adsabs.harvard.edu/abs/2021MNRAS.503...31T}
}

@article{Koppelman2019b,
	author = {{Koppelman}, Helmer H. and {Helmi, Amina} and {Massari, Davide} and {Price-Whelan, Adrian M.} and {Starkenburg, Tjitske K.}},
	title = {Multiple retrograde substructures in the Galactic halo: A shattered view of Galactic history},
	DOI= "10.1051/0004-6361/201936738",
	url= "https://doi.org/10.1051/0004-6361/201936738",
	journal = {\aap},
	year = 2019,
	volume = 631,
	pages = "L9",
}

@article{Massari2019,
	author = {{Massari}, D. and {Koppelman, H. H.} and {Helmi, A.}},
	title = {Origin of the system of globular clusters in the Milky Way},
	DOI= "10.1051/0004-6361/201936135",
	url= "https://doi.org/10.1051/0004-6361/201936135",
	journal = {\aap},
	year = 2019,
	volume = 630,
	pages = "L4",
}

@ARTICLE{Forbes2020,
       author = {{Forbes}, Duncan A.},
        title = "{Reverse engineering the Milky Way}",
      journal = {\mnras},
         year = 2020,
        month = mar,
       volume = {493},
       number = {1},
        pages = {847-854},
          doi = {10.1093/mnras/staa245},
archivePrefix = {arXiv},
       eprint = {2002.01512},
 primaryClass = {astro-ph.GA},
       adsurl = {https://ui.adsabs.harvard.edu/abs/2020MNRAS.493..847F}
}

@article{GaiaMathias2023,
	author = {{Gaia Collaboration} and {Schultheis, M.} and {Zhao, H.} and {Zwitter, T.} and {Marshall, D. J.} and {Drimmel, R.} and {Frémat, Y.} and {Bailer-Jones, C. A. L.} and {Recio-Blanco, A.} and {Kordopatis, G.} and {de Laverny, P.} and {Andrae, R.} and {Dharmawardena, T. E.} and {Fouesneau, M.} and {Sordo, R.} and {Brown, A. G. A.} and {Vallenari, A.} and {Prusti, T.} and {de Bruijne, J. H. J.} and {Arenou, F.} and {Babusiaux, C.} and {Biermann, M.} and {Creevey, O. L.} and {Ducourant, C.} and {Evans, D. W.} and {Eyer, L.} and {Guerra, R.} and {Hutton, A.} and {Jordi, C.} and {Klioner, S. A.} and {Lammers, U. L.} and {Lindegren, L.} and {Luri, X.} and {Mignard, F.} and {Panem, C.} and {Pourbaix, D.} and {Randich, S.} and {Sartoretti, P.} and {Soubiran, C.} and {Tanga, P.} and {Walton, N. A.} and {Bastian, U.} and {Jansen, F.} and {Katz, D.} and {Lattanzi, M. G.} and {van Leeuwen, F.} and {Bakker, J.} and {Cacciari, C.} and {Castañeda, J.} and {De Angeli, F.} and {Fabricius, C.} and {Galluccio, L.} and {Guerrier, A.} and {Heiter, U.} and {Masana, E.} and {Messineo, R.} and {Mowlavi, N.} and {Nicolas, C.} and {Nienartowicz, K.} and {Pailler, F.} and {Panuzzo, P.} and {Riclet, F.} and {Roux, W.} and {Seabroke, G. M.} and {Thévenin, F.} and {Gracia-Abril, G.} and {Portell, J.} and {Teyssier, D.} and {Altmann, M.} and {Audard, M.} and {Bellas-Velidis, I.} and {Benson, K.} and {Berthier, J.} and {Blomme, R.} and {Burgess, P. W.} and {Busonero, D.} and {Busso, G.} and {Cánovas, H.} and {Carry, B.} and {Cellino, A.} and {Cheek, N.} and {Clementini, G.} and {Damerdji, Y.} and {Davidson, M.} and {de Teodoro, P.} and {Nuñez Campos, M.} and {Delchambre, L.} and {Dell’Oro, A.} and {Esquej, P.} and {Fernández-Hernández, J.} and {Fraile, E.} and {Garabato, D.} and {García-Lario, P.} and {Gosset, E.} and {Haigron, R.} and {Halbwachs, J.-L.} and {Hambly, N. C.} and {Harrison, D. L.} and {Hernández, J.} and {Hestroffer, D.} and {Hodgkin, S. T.} and {Holl, B.} and {Janßen, K.} and {Jevardat de Fombelle, G.} and {Jordan, S.} and {Krone-Martins, A.} and {Lanzafame, A. C.} and {Löffler, W.} and {Marchal, O.} and {Marrese, P. M.} and {Moitinho, A.} and {Muinonen, K.} and {Osborne, P.} and {Pancino, E.} and {Pauwels, T.} and {Reylé, C.} and {Riello, M.} and {Rimoldini, L.} and {Roegiers, T.} and {Rybizki, J.} and {Sarro, L. M.} and {Siopis, C.} and {Smith, M.} and {Sozzetti, A.} and {Utrilla, E.} and {van Leeuwen, M.} and {Abbas, U.} and {Ábrahám, P.} and {Abreu Aramburu, A.} and {Aerts, C.} and {Aguado, J. J.} and {Ajaj, M.} and {Aldea-Montero, F.} and {Altavilla, G.} and {Álvarez, M. A.} and {Alves, J.} and {Anders, F.} and {Anderson, R. I.} and {Anglada Varela, E.} and {Antoja, T.} and {Baines, D.} and {Baker, S. G.} and {Balaguer-Núñez, L.} and {Balbinot, E.} and {Balog, Z.} and {Barache, C.} and {Barbato, D.} and {Barros, M.} and {Barstow, M. A.} and {Bartolomé, S.} and {Bassilana, J.-L.} and {Bauchet, N.} and {Becciani, U.} and {Bellazzini, M.} and {Berihuete, A.} and {Bernet, M.} and {Bertone, S.} and {Bianchi, L.} and {Binnenfeld, A.} and {Blanco-Cuaresma, S.} and {Boch, T.} and {Bombrun, A.} and {Bossini, D.} and {Bouquillon, S.} and {Bragaglia, A.} and {Bramante, L.} and {Breedt, E.} and {Bressan, A.} and {Brouillet, N.} and {Brugaletta, E.} and {Bucciarelli, B.} and {Burlacu, A.} and {Butkevich, A. G.} and {Buzzi, R.} and {Caffau, E.} and {Cancelliere, R.} and {Cantat-Gaudin, T.} and {Carballo, R.} and {Carlucci, T.} and {Carnerero, M. I.} and {Carrasco, J. M.} and {Casamiquela, L.} and {Castellani, M.} and {Castro-Ginard, A.} and {Chaoul, L.} and {Charlot, P.} and {Chemin, L.} and {Chiaramida, V.} and {Chiavassa, A.} and {Chornay, N.} and {Comoretto, G.} and {Contursi, G.} and {Cooper, W. J.} and {Cornez, T.} and {Cowell, S.} and {Crifo, F.} and {Cropper, M.} and {Crosta, M.} and {Crowley, C.} and {Dafonte, C.} and {Dapergolas, A.} and {David, P.} and {De Luise, F.} and {De March, R.} and {De Ridder, J.} and {de Souza, R.} and {de Torres, A.} and {del Peloso, E. F.} and {del Pozo, E.} and {Delbo, M.} and {Delgado, A.} and {Delisle, J.-B.} and {Demouchy, C.} and {Diakite, S.} and {Diener, C.} and {Distefano, E.} and {Dolding, C.} and {Enke, H.} and {Fabre, C.} and {Fabrizio, M.} and {Faigler, S.} and {Fedorets, G.} and {Fernique, P.} and {Figueras, F.} and {Fournier, Y.} and {Fouron, C.} and {Fragkoudi, F.} and {Gai, M.} and {Garcia-Gutierrez, A.} and {Garcia-Reinaldos, M.} and {García-Torres, M.} and {Garofalo, A.} and {Gavel, A.} and {Gavras, P.} and {Gerlach, E.} and {Geyer, R.} and {Giacobbe, P.} and {Gilmore, G.} and {Girona, S.} and {Giuffrida, G.} and {Gomel, R.} and {Gomez, A.} and {González-Núñez, J.} and {González-Santamaría, I.} and {González-Vidal, J. J.} and {Granvik, M.} and {Guillout, P.} and {Guiraud, J.} and {Gutiérrez-Sánchez, R.} and {Guy, L. P.} and {Hatzidimitriou, D.} and {Hauser, M.} and {Haywood, M.} and {Helmer, A.} and {Helmi, A.} and {Sarmiento, M. H.} and {Hidalgo, S. L.} and {Hładczuk, N.} and {Hobbs, D.} and {Holland, G.} and {Huckle, H. E.} and {Jardine, K.} and {Jasniewicz, G.} and {Jean-Antoine Piccolo, A.} and {Jiménez-Arranz, Ó.} and {Juaristi Campillo, J.} and {Julbe, F.} and {Karbevska, L.} and {Kervella, P.} and {Khanna, S.} and {Korn, A. J.} and {Kóspál, Á.} and {Kostrzewa-Rutkowska, Z.} and {Kruszyńska, K.} and {Kun, M.} and {Laizeau, P.} and {Lambert, S.} and {Lanza, A. F.} and {Lasne, Y.} and {Le Campion, J.-F.} and {Lebreton, Y.} and {Lebzelter, T.} and {Leccia, S.} and {Leclerc, N.} and {Lecoeur-Taibi, I.} and {Liao, S.} and {Licata, E. L.} and {Lindstrøm, H. E. P.} and {Lister, T. A.} and {Livanou, E.} and {Lobel, A.} and {Lorca, A.} and {Loup, C.} and {Madrero Pardo, P.} and {Magdaleno Romeo, A.} and {Managau, S.} and {Mann, R. G.} and {Manteiga, M.} and {Marchant, J. M.} and {Marconi, M.} and {Marcos, J.} and {Marcos Santos, M. M. S.} and {Marín Pina, D.} and {Marinoni, S.} and {Marocco, F.} and {Martin Polo, L.} and {Martín-Fleitas, J. M.} and {Marton, G.} and {Mary, N.} and {Masip, A.} and {Massari, D.} and {Mastrobuono-Battisti, A.} and {Mazeh, T.} and {McMillan, P. J.} and {Messina, S.} and {Michalik, D.} and {Millar, N. R.} and {Mints, A.} and {Molina, D.} and {Molinaro, R.} and {Molnár, L.} and {Monari, G.} and {Monguió, M.} and {Montegriffo, P.} and {Montero, A.} and {Mor, R.} and {Mora, A.} and {Morbidelli, R.} and {Morel, T.} and {Morris, D.} and {Muraveva, T.} and {Murphy, C. P.} and {Musella, I.} and {Nagy, Z.} and {Noval, L.} and {Ocaña, F.} and {Ogden, A.} and {Ordenovic, C.} and {Osinde, J. O.} and {Pagani, C.} and {Pagano, I.} and {Palaversa, L.} and {Palicio, P. A.} and {Pallas-Quintela, L.} and {Panahi, A.} and {Payne-Wardenaar, S.} and {Peñalosa Esteller, X.} and {Penttilä, A.} and {Pichon, B.} and {Piersimoni, A. M.} and {Pineau, F.-X.} and {Plachy, E.} and {Plum, G.} and {Poggio, E.} and {Prša, A.} and {Pulone, L.} and {Racero, E.} and {Ragaini, S.} and {Rainer, M.} and {Raiteri, C. M.} and {Ramos, P.} and {Ramos-Lerate, M.} and {Re Fiorentin, P.} and {Regibo, S.} and {Richards, P. J.} and {Rios Diaz, C.} and {Ripepi, V.} and {Riva, A.} and {Rix, H.-W.} and {Rixon, G.} and {Robichon, N.} and {Robin, A. C.} and {Robin, C.} and {Roelens, M.} and {Rogues, H. R. O.} and {Rohrbasser, L.} and {Romero-Gómez, M.} and {Rowell, N.} and {Royer, F.} and {Ruz Mieres, D.} and {Rybicki, K. A.} and {Sadowski, G.} and {Sáez Núñez, A.} and {Sagristà Sellés, A.} and {Sahlmann, J.} and {Salguero, E.} and {Samaras, N.} and {Sanchez Gimenez, V.} and {Sanna, N.} and {Santoveña, R.} and {Sarasso, M.} and {Sciacca, E.} and {Segol, M.} and {Segovia, J. C.} and {Ségransan, D.} and {Semeux, D.} and {Shahaf, S.} and {Siddiqui, H. I.} and {Siebert, A.} and {Siltala, L.} and {Silvelo, A.} and {Slezak, E.} and {Slezak, I.} and {Smart, R. L.} and {Snaith, O. N.} and {Solano, E.} and {Solitro, F.} and {Souami, D.} and {Souchay, J.} and {Spagna, A.} and {Spina, L.} and {Spoto, F.} and {Steele, I. A.} and {Steidelmüller, H.} and {Stephenson, C. A.} and {Süveges, M.} and {Surdej, J.} and {Szabados, L.} and {Szegedi-Elek, E.} and {Taris, F.} and {Taylor, M. B.} and {Teixeira, R.} and {Tolomei, L.} and {Tonello, N.} and {Torra, F.} and {Torra, J.} and {Torralba Elipe, G.} and {Trabucchi, M.} and {Tsounis, A. T.} and {Turon, C.} and {Ulla, A.} and {Unger, N.} and {Vaillant, M. V.} and {van Dillen, E.} and {van Reeven, W.} and {Vanel, O.} and {Vecchiato, A.} and {Viala, Y.} and {Vicente, D.} and {Voutsinas, S.} and {Weiler, M.} and {Wevers, T.} and {Wyrzykowski, Ł.} and {Yoldas, A.} and {Yvard, P.} and {Zorec, J.} and {Zucker, S.}},
	title = {Gaia Data Release 3 - Exploring and mapping the diffuse interstellar band at 862 nm},
	DOI= "10.1051/0004-6361/202243283",
	url= "https://doi.org/10.1051/0004-6361/202243283",
	journal = {\aap},
	year = 2023,
	volume = 674,
	pages = "A40",
}

@ARTICLE{Bailer-Jones2021,
       author = {{Bailer-Jones}, C.~A.~L. and {Rybizki}, J. and {Fouesneau}, M. and {Demleitner}, M. and {Andrae}, R.},
        title = "{Estimating Distances from Parallaxes. V. Geometric and Photogeometric Distances to 1.47 Billion Stars in Gaia Early Data Release 3}",
      journal = {\aj},
         year = 2021,
        month = mar,
       volume = {161},
       number = {3},
          eid = {147},
        pages = {147},
          doi = {10.3847/1538-3881/abd806},
archivePrefix = {arXiv},
       eprint = {2012.05220},
 primaryClass = {astro-ph.SR},
       adsurl = {https://ui.adsabs.harvard.edu/abs/2021AJ....161..147B}
}

@ARTICLE{Gran2022,
       author = {{Gran}, F. and {Zoccali}, M. and {Saviane}, I. and {Valenti}, E. and {Rojas-Arriagada}, A. and {Contreras Ramos}, R. and {Hartke}, J. and {Carballo-Bello}, J.~A. and {Navarrete}, C. and {Rejkuba}, M. and {Olivares Carvajal}, J.},
        title = "{Hidden in the haystack: low-luminosity globular clusters towards the Milky Way bulge}",
      journal = {\mnras},
         year = 2022,
        month = feb,
       volume = {509},
       number = {4},
        pages = {4962-4981},
          doi = {10.1093/mnras/stab2463},
archivePrefix = {arXiv},
       eprint = {2108.11922},
 primaryClass = {astro-ph.GA},
       adsurl = {https://ui.adsabs.harvard.edu/abs/2022MNRAS.509.4962G}
}

@article{Bica2024,
	author = {{Bica}, E. and {Ortolani, S.} and {Barbuy, B.} and {Oliveira, R. A. P.}},
	title = {A census of new globular clusters in the Galactic bulge},
	DOI= "10.1051/0004-6361/202346377",
	url= "https://doi.org/10.1051/0004-6361/202346377",
	journal = {\aap},
	year = 2024,
	volume = 687,
	pages = "A201",
}

@ARTICLE{Belokurov2024,
       author = {{Belokurov}, Vasily and {Kravtsov}, Andrey},
        title = "{In-situ versus accreted Milky Way globular clusters: a new classification method and implications for cluster formation}",
      journal = {\mnras},
         year = 2024,
        month = feb,
       volume = {528},
       number = {2},
        pages = {3198-3216},
          doi = {10.1093/mnras/stad3920},
archivePrefix = {arXiv},
       eprint = {2309.15902},
 primaryClass = {astro-ph.GA},
       adsurl = {https://ui.adsabs.harvard.edu/abs/2024MNRAS.528.3198B}
}

@ARTICLE{Marconi2015,
       author = {{Marconi}, M. and {Coppola}, G. and {Bono}, G. and {Braga}, V. and {Pietrinferni}, A. and {Buonanno}, R. and {Castellani}, M. and {Musella}, I. and {Ripepi}, V. and {Stellingwerf}, R.~F.},
        title = "{On a New Theoretical Framework for RR Lyrae Stars. I. The Metallicity Dependence}",
      journal = {\apj},
         year = 2015,
        month = jul,
       volume = {808},
       number = {1},
          eid = {50},
        pages = {50},
          doi = {10.1088/0004-637X/808/1/50},
archivePrefix = {arXiv},
       eprint = {1505.02531},
 primaryClass = {astro-ph.SR},
       adsurl = {https://ui.adsabs.harvard.edu/abs/2015ApJ...808...50M}
}

@ARTICLE{Ibata2019,
       author = {{Ibata}, Rodrigo A. and {Bellazzini}, Michele and {Malhan}, Khyati and {Martin}, Nicolas and {Bianchini}, Paolo},
        title = "{Identification of the long stellar stream of the prototypical massive globular cluster {\ensuremath{\omega}} Centauri}",
      journal = {\nat},
         year = 2019,
        month = apr,
       volume = {3},
        pages = {667-672},
          doi = {10.1038/s41550-019-0751-x},
archivePrefix = {arXiv},
       eprint = {1902.09544},
 primaryClass = {astro-ph.GA},
       adsurl = {https://ui.adsabs.harvard.edu/abs/2019NatAs...3..667I}
}

@ARTICLE{Sollima2020,
       author = {{Sollima}, A.},
        title = "{The eye of Gaia on globular clusters structure: tidal tails}",
      journal = {\mnras},
         year = 2020,
        month = jun,
       volume = {495},
       number = {2},
        pages = {2222-2233},
          doi = {10.1093/mnras/staa1209},
archivePrefix = {arXiv},
       eprint = {2004.13754},
 primaryClass = {astro-ph.GA},
       adsurl = {https://ui.adsabs.harvard.edu/abs/2020MNRAS.495.2222S}
}

@ARTICLE{Deason2024,
       author = {{Deason}, Alis J. and {Belokurov}, Vasily},
        title = "{Galactic Archaeology with Gaia}",
      journal = {\nar},
         year = 2024,
        month = dec,
       volume = {99},
          eid = {101706},
        pages = {101706},
          doi = {10.1016/j.newar.2024.101706},
archivePrefix = {arXiv},
       eprint = {2402.12443},
 primaryClass = {astro-ph.GA},
       adsurl = {https://ui.adsabs.harvard.edu/abs/2024NewAR..9901706D}
}

@ARTICLE{Lambas2003,
       author = {{Lambas}, Diego G. and {Tissera}, Patricia B. and {Alonso}, M. Sol and {Coldwell}, Georgina},
        title = "{Galaxy pairs in the 2dF survey - I. Effects of interactions on star formation in the field}",
      journal = {\mnras},
         year = 2003,
        month = dec,
       volume = {346},
       number = {4},
        pages = {1189-1196},
          doi = {10.1111/j.1365-2966.2003.07179.x},
archivePrefix = {arXiv},
       eprint = {astro-ph/0212222},
 primaryClass = {astro-ph},
       adsurl = {https://ui.adsabs.harvard.edu/abs/2003MNRAS.346.1189L}
}

@ARTICLE{Hopkins2008,
       author = {{Hopkins}, Philip F. and {Hernquist}, Lars and {Cox}, Thomas J. and {Kere{\v{s}}}, Du{\v{s}}an},
        title = "{A Cosmological Framework for the Co-Evolution of Quasars, Supermassive Black Holes, and Elliptical Galaxies. I. Galaxy Mergers and Quasar Activity}",
      journal = {\apjs},
         year = 2008,
        month = apr,
       volume = {175},
       number = {2},
        pages = {356-389},
          doi = {10.1086/524362},
archivePrefix = {arXiv},
       eprint = {0706.1243},
 primaryClass = {astro-ph},
       adsurl = {https://ui.adsabs.harvard.edu/abs/2008ApJS..175..356H}
}

@article{Gran2024,
	author = {{Gran}, F. and {Kordopatis, G.} and {Zoccali, M.} and {Hill, V.} and {Saviane, I.} and {Navarrete, C.} and {Rojas-Arriagada, A.} and {Carballo-Bello, J.} and {Hartke, J.} and {Valenti, E.} and {Contreras Ramos, R.} and {De Leo, M.} and {Fabbro, S.}},
	title = {The treasure behind the haystack: MUSE analysis of five recently discovered globular clusters⋆},
	DOI= "10.1051/0004-6361/202347915",
	url= "https://doi.org/10.1051/0004-6361/202347915",
	journal = {\aap},
	year = 2024,
	volume = 683,
	pages = "A167",
}

@article{Saito2012,
	author = {{Saito}, R. K. and {Hempel, M.} and {Minniti, D.} and {Lucas, P. W.} and {Rejkuba, M.} and {Toledo, I.} and {Gonzalez, O. A.} and {Alonso-García, J.} and {Irwin, M. J.} and {Gonzalez-Solares, E.} and {Hodgkin, S. T.} and {Lewis, J. R.} and {Cross, N.} and {Ivanov, V. D.} and {Kerins, E.} and {Emerson, J. P.} and {Soto, M.} and {Amôres, E. B.} and {Gurovich, S.} and {Dékány, I.} and {Angeloni, R.} and {Beamin, J. C.} and {Catelan, M.} and {Padilla, N.} and {Zoccali, M.} and {Pietrukowicz, P.} and {Moni Bidin, C.} and {Mauro, F.} and {Geisler, D.} and {Folkes, S. L.} and {Sale, S. E.} and {Borissova, J.} and {Kurtev, R.} and {Ahumada, A. V.} and {Alonso, M. V.} and {Adamson, A.} and {Arias, J. I.} and {Bandyopadhyay, R. M.} and {Barbá, R. H.} and {Barbuy, B.} and {Baume, G. L.} and {Bedin, L. R.} and {Bellini, A.} and {Benjamin, R.} and {Bica, E.} and {Bonatto, C.} and {Bronfman, L.} and {Carraro, G.} and {Chenè, A. N.} and {Clariá, J. J.} and {Clarke, J. R. A.} and {Contreras, C.} and {Corvillón, A.} and {de Grijs, R.} and {Dias, B.} and {Drew, J. E.} and {Fariña, C.} and {Feinstein, C.} and {Fernández-Lajús, E.} and {Gamen, R. C.} and {Gieren, W.} and {Goldman, B.} and {González-Fernández, C.} and {Grand, R. J. J.} and {Gunthardt, G.} and {Hambly, N. C.} and {Hanson, M. M.} and {Hełminiak, K. G.} and {Hoare, M. G.} and {Huckvale, L.} and {Jordán, A.} and {Kinemuchi, K.} and {Longmore, A.} and {López-Corredoira, M.} and {Maccarone, T.} and {Majaess, D.} and {Martín, E. L.} and {Masetti, N.} and {Mennickent, R. E.} and {Mirabel, I. F.} and {Monaco, L.} and {Morelli, L.} and {Motta, V.} and {Palma, T.} and {Parisi, M. C.} and {Parker, Q.} and {Peñaloza, F.} and {Pietrzyński, G.} and {Pignata, G.} and {Popescu, B.} and {Read, M. A.} and {Rojas, A.} and {Roman-Lopes, A.} and {Ruiz, M. T.} and {Saviane, I.} and {Schreiber, M. R.} and {Schröder, A. C.} and {Sharma, S.} and {Smith, M. D.} and {Sodré, L.} and {Stead, J.} and {Stephens, A. W.} and {Tamura, M.} and {Tappert, C.} and {Thompson, M. A.} and {Valenti, E.} and {Vanzi, L.} and {Walton, N. A.} and {Weidmann, W.} and {Zijlstra, A.}},
	title = {VVV DR1: The first data release of the Milky Way bulge and
          southern plane from the near-infrared ESO public survey VISTA variables in the Vía
            Láctea⋆},
	DOI= "10.1051/0004-6361/201118407",
	url= "https://doi.org/10.1051/0004-6361/201118407",
	journal = {\aap},
	year = 2012,
	volume = 537,
	pages = "A107",
	month = "",
}

@ARTICLE{Bono2001,
       author = {{Bono}, G. and {Caputo}, F. and {Castellani}, V. and {Marconi}, M. and {Storm}, J.},
        title = "{Theoretical insights into the RR Lyrae K-band period-luminosity relation}",
      journal = {\mnras},
         year = 2001,
        month = sep,
       volume = {326},
       number = {3},
        pages = {1183-1190},
          doi = {10.1046/j.1365-8711.2001.04655.x},
archivePrefix = {arXiv},
       eprint = {astro-ph/0105481},
 primaryClass = {astro-ph},
       adsurl = {https://ui.adsabs.harvard.edu/abs/2001MNRAS.326.1183B}
}

@ARTICLE{Wang2022,
       author = {{Wang}, F. and {Zhang}, H. -W. and {Xue}, X. -X. and {Huang}, Y. and {Liu}, G. -C. and {Zhang}, L. and {Yang}, C. -Q.},
        title = "{Probing the Galactic halo with RR Lyrae stars - II. The substructures of the Milky Way}",
      journal = {\mnras},
         year = 2022,
        month = jun,
       volume = {513},
       number = {2},
        pages = {1958-1971},
          doi = {10.1093/mnras/stac874},
archivePrefix = {arXiv},
       eprint = {2203.17032},
 primaryClass = {astro-ph.GA},
       adsurl = {https://ui.adsabs.harvard.edu/abs/2022MNRAS.513.1958W}
}

@ARTICLE{Cabrera-Garcia2024,
       author = {{Cabrera Garcia}, Jonathan and {Beers}, Timothy C. and {Huang}, Yang and {Li}, Xin-Yi and {Liu}, Gaochao and {Zhang}, Huawei and {Hong}, Jihye and {Lee}, Young Sun and {Shank}, Derek and {Gudin}, Dmitrii and {Hirai}, Yutaka and {Komater}, Dante},
        title = "{Probing the Galactic halo with RR Lyrae stars - V. Chemistry, kinematics, and dynamically tagged groups}",
      journal = {\mnras},
         year = 2024,
        month = jan,
       volume = {527},
       number = {3},
        pages = {8973-8990},
          doi = {10.1093/mnras/stad3674},
archivePrefix = {arXiv},
       eprint = {2307.09572},
 primaryClass = {astro-ph.GA},
       adsurl = {https://ui.adsabs.harvard.edu/abs/2024MNRAS.527.8973C}
}

@ARTICLE{Sheng2024,
       author = {{Sheng}, Yanjun and {Ting}, Yuan-Sen and {Xue}, Xiang-Xiang and {Chang}, Jiang and {Tian}, Hao},
        title = "{Uncovering the first-infall history of the LMC through its dynamical impact in the Milky Way halo}",
      journal = {\mnras},
         year = 2024,
        month = nov,
       volume = {534},
       number = {3},
        pages = {2694-2714},
          doi = {10.1093/mnras/stae2259},
archivePrefix = {arXiv},
       eprint = {2404.08975},
 primaryClass = {astro-ph.GA},
       adsurl = {https://ui.adsabs.harvard.edu/abs/2024MNRAS.534.2694S}
}

@INPROCEEDINGS{DBSCAN,
  author={Khan, Kamran and Rehman, Saif Ur and Aziz, Kamran and Fong, Simon and Sarasvady, S.},
  booktitle={The Fifth International Conference on the Applications of Digital Information and Web Technologies (ICADIWT 2014)}, 
  title={DBSCAN: Past, present and future}, 
  year={2014},
  volume={},
  number={},
  pages={232-238},
  doi={10.1109/ICADIWT.2014.6814687}}

@ARTICLE{Starkenburg2009,
       author = {{Starkenburg}, Else and {Helmi}, Amina and {Morrison}, Heather L. and {Harding}, Paul and {van Woerden}, Hugo and {Mateo}, Mario and {Olszewski}, Edward W. and {Sivarani}, Thirupathi and {Norris}, John E. and {Freeman}, Kenneth C. and {Shectman}, Stephen A. and {Dohm-Palmer}, R.~C. and {Frey}, Lucy and {Oravetz}, Dan},
        title = "{Mapping the Galactic Halo. VIII. Quantifying Substructure}",
      journal = {\apj},
         year = 2009,
        month = jun,
       volume = {698},
       number = {1},
        pages = {567-579},
          doi = {10.1088/0004-637X/698/1/567},
archivePrefix = {arXiv},
       eprint = {0903.3043},
 primaryClass = {astro-ph.GA},
       adsurl = {https://ui.adsabs.harvard.edu/abs/2009ApJ...698..567S}
}

@ARTICLE{CruzReyes2024,
       author = {{Cruz Reyes}, Mauricio and {Anderson}, Richard I. and {Johansson}, Lucas and {Netzel}, Henryka and {Medaric}, Zo{\'e}},
        title = "{Variable stars in galactic globular clusters. I. The population of RR Lyrae stars}",
      journal = {\aap},
         year = 2024,
        month = apr,
       volume = {684},
          eid = {A173},
        pages = {A173},
          doi = {10.1051/0004-6361/202348961},
archivePrefix = {arXiv},
       eprint = {2402.08843},
 primaryClass = {astro-ph.SR},
       adsurl = {https://ui.adsabs.harvard.edu/abs/2024A&A...684A.173C}
}

@article{Saito2024,
	author = {{Saito}, R. K. and {Hempel, M.} and {Alonso-García, J.} and {Lucas, P. W.} and {Minniti, D.} and {Alonso, S.} and {Baravalle, L.} and {Borissova, J.} and {Caceres, C.} and {Chené, A. N.} and {Cross, N. J. G.} and {Duplancic, F.} and {Garro, E. R.} and {Gómez, M.} and {Ivanov, V. D.} and {Kurtev, R.} and {Luna, A.} and {Majaess, D.} and {Navarro, M. G.} and {Pullen, J. B.} and {Rejkuba, M.} and {Sanders, J. L.} and {Smith, L. C.} and {Albino, P. H. C.} and {Alonso, M. V.} and {Amôres, E. B.} and {Angeloni, R.} and {Arias, J. I.} and {Arnaboldi, M.} and {Barbuy, B.} and {Bayo, A.} and {Beamin, J. C.} and {Bedin, L. R.} and {Bellini, A.} and {Benjamin, R. A.} and {Bica, E.} and {Bonatto, C. J.} and {Botan, E.} and {Braga, V. F.} and {Brown, D. A.} and {Cabral, J. B.} and {Camargo, D.} and {Caratti o Garatti, A.} and {Carballo-Bello, J. A.} and {Catelan, M.} and {Chavero, C.} and {Chijani, M. A.} and {Clariá, J. J.} and {Coldwell, G. V.} and {Peña, C. Contreras} and {Ramos, R. Contreras} and {Corral-Santana, J. M.} and {Cortés, C. C.} and {Cortés-Contreras, M.} and {Cruz, P.} and {Daza-Perilla, I. V.} and {Debattista, V. P.} and {Dias, B.} and {Donoso, L.} and {D’Souza, R.} and {Emerson, J. P.} and {Federle, S.} and {Fermiano, V.} and {Fernandez, J.} and {Fernández-Trincado, J. G.} and {Ferreira, T.} and {Lopes, C. E. Ferreira} and {Firpo, V.} and {Flores-Quintana, C.} and {Fraga, L.} and {Froebrich, D.} and {Galdeano, D.} and {Gavignaud, I.} and {Geisler, D.} and {Gerhard, O. E.} and {Gieren, W.} and {Gonzalez, O. A.} and {Gramajo, L. V.} and {Gran, F.} and {Granitto, P. M.} and {Griggio, M.} and {Guo, Z.} and {Gurovich, S.} and {Hilker, M.} and {Jones, H. R. A.} and {Kammers, R.} and {Kuhn, M. A.} and {Kumar, M. S. N.} and {Kundu, R.} and {Lares, M.} and {Libralato, M.} and {Lima, E.} and {Maccarone, T. J.} and {Cortés, P. Marchant} and {Martin, E. L.} and {Masetti, N.} and {Matsunaga, N.} and {Mauro, F.} and {McDonald, I.} and {Mejías, A.} and {Mesa, V.} and {Milla-Castro, F. P.} and {Minniti, J. H.} and {Bidin, C. Moni} and {Montenegro, K.} and {Morris, C.} and {Motta, V.} and {Navarete, F.} and {Molina, C. Navarro} and {Nikzat, F.} and {Castellón, J. L. Nilo} and {Obasi, C.} and {Ortigoza-Urdaneta, M.} and {Palma, T.} and {Parisi, C.} and {Ramírez, K. Pena} and {Pereyra, L.} and {Perez, N.} and {Petralia, I.} and {Pichel, A.} and {Pignata, G.} and {Alegría, S. Ramírez} and {Rojas, A. F.} and {Rojas, D.} and {Roman-Lopes, A.} and {Rovero, A. C.} and {Saroon, S.} and {Schmidt, E. O.} and {Schröder, A. C.} and {Schultheis, M.} and {Sgró, M. A.} and {Solano, E.} and {Soto, M.} and {Stecklum, B.} and {Steeghs, D.} and {Tamura, M.} and {Tissera, P.} and {Valcarce, A. A. R.} and {Valotto, C. A.} and {Vasquez, S.} and {Villalon, C.} and {Villanova, S.} and {Cádiz, F. Vivanco} and {Bacigalupo, R. Zelada} and {Zijlstra, A.} and {Zoccali, M.}},
	title = {The VISTA Variables in the Vía Láctea extended (VVVX) ESO public survey: Completion of the observations and legacy★},
	DOI= "10.1051/0004-6361/202450584",
	url= "https://doi.org/10.1051/0004-6361/202450584",
	journal = {\aap},
	year = 2024,
	volume = 689,
	pages = "A148",
}

@ARTICLE{Minniti2021VVVCL160,
       author = {{Minniti}, Dante and {Fern{\'a}ndez-Trincado}, Jos{\'e} G. and {G{\'o}mez}, Mat{\'\i}as and {Smith}, Leigh C. and {Lucas}, Philip W. and {Contreras Ramos}, Rodrigo},
        title = "{Discovery of a new nearby globular cluster with extreme kinematics located in the extension of a halo stream}",
      journal = {\aap},
         year = 2021,
        month = jun,
       volume = {650},
          eid = {L11},
        pages = {L11},
          doi = {10.1051/0004-6361/202141129},
archivePrefix = {arXiv},
       eprint = {2106.01383},
 primaryClass = {astro-ph.GA},
       adsurl = {https://ui.adsabs.harvard.edu/abs/2021A&A...650L..11M}
}

@ARTICLE{Molnar2022,
       author = {{Molnar}, Thomas A. and {Sanders}, Jason L. and {Smith}, Leigh C. and {Belokurov}, Vasily and {Lucas}, Philip and {Minniti}, Dante},
        title = "{Variable star classification across the Galactic bulge and disc with the VISTA Variables in the V{\'\i}a L{\'a}ctea survey}",
      journal = {\mnras},
         year = 2022,
        month = jan,
       volume = {509},
       number = {2},
        pages = {2566-2592},
          doi = {10.1093/mnras/stab3116},
archivePrefix = {arXiv},
       eprint = {2110.15371},
 primaryClass = {astro-ph.SR},
       adsurl = {https://ui.adsabs.harvard.edu/abs/2022MNRAS.509.2566M}
}

@article{Hunt2021,
	author = {{Hunt}, Emily L. and {Reffert}, Sabine},
	title = {Improving the open cluster census - I. Comparison of clustering algorithms applied to Gaia DR2 data⋆},
	DOI= "10.1051/0004-6361/202039341",
	url= "https://doi.org/10.1051/0004-6361/202039341",
	journal = {\aap},
	year = 2021,
	volume = 646,
	pages = "A104",
}

@ARTICLE{Kovacs1995,
       author = {{Kovacs}, G. and {Zsoldos}, E.},
        title = "{A new method for the determination of [Fe/H] in RR Lyrae stars.}",
      journal = {\aap},
         year = 1995,
        month = jan,
       volume = {293},
        pages = {L57-L60},
       adsurl = {https://ui.adsabs.harvard.edu/abs/1995A&A...293L..57K}
}

@ARTICLE{Baker2015,
       author = {{Baker}, Mariah and {Willman}, Beth},
        title = "{Charting Unexplored Dwarf Galaxy Territory with RR Lyrae}",
      journal = {\aj},
         year = 2015,
        month = nov,
       volume = {150},
       number = {5},
          eid = {160},
        pages = {160},
          doi = {10.1088/0004-6256/150/5/160},
archivePrefix = {arXiv},
       eprint = {1507.00734},
 primaryClass = {astro-ph.GA},
       adsurl = {https://ui.adsabs.harvard.edu/abs/2015AJ....150..160B}
}

@ARTICLE{OGLE-IV,
       author = {{Udalski}, A. and {Szyma{\'n}ski}, M.~K. and {Szyma{\'n}ski}, G.},
        title = "{OGLE-IV: Fourth Phase of the Optical Gravitational Lensing Experiment}",
      journal = {\actaa},
         year = 2015,
        month = mar,
       volume = {65},
       number = {1},
        pages = {1-38},
          doi = {10.48550/arXiv.1504.05966},
archivePrefix = {arXiv},
       eprint = {1504.05966},
 primaryClass = {astro-ph.SR},
       adsurl = {https://ui.adsabs.harvard.edu/abs/2015AcA....65....1U}
}

@BOOK{Smith2004,
       author = {{Smith}, Horace A.},
        title = "{RR Lyrae Stars}",
         year = 2004,
       adsurl = {https://ui.adsabs.harvard.edu/abs/2004rrls.book.....S}
}

@article{Seaborn,
    doi = {10.21105/joss.03021},
    url = {https://doi.org/10.21105/joss.03021},
    year = {2021},
    publisher = {The Open Journal},
    volume = {6},
    number = {60},
    pages = {3021},
    author = {{Waskom}, Michael L.},
    title = {seaborn: statistical data visualization},
    journal = {Journal of Open Source Software}
 }

@ARTICLE{Vasiliev2021,
       author = {{Vasiliev}, Eugene and {Belokurov}, Vasily and {Erkal}, Denis},
        title = "{Tango for three: Sagittarius, LMC, and the Milky Way}",
      journal = {\mnras},
         year = 2021,
        month = feb,
       volume = {501},
       number = {2},
        pages = {2279-2304},
          doi = {10.1093/mnras/staa3673},
archivePrefix = {arXiv},
       eprint = {2009.10726},
 primaryClass = {astro-ph.GA},
       adsurl = {https://ui.adsabs.harvard.edu/abs/2021MNRAS.501.2279V}
}

@ARTICLE{Clarke2019,
       author = {{Clarke}, Jonathan P. and {Wegg}, Christopher and {Gerhard}, Ortwin and {Smith}, Leigh C. and {Lucas}, Phil W. and {Wylie}, Shola M.},
        title = "{The Milky Way bar/bulge in proper motions: a 3D view from VIRAC and Gaia}",
      journal = {\mnras},
         year = 2019,
        month = nov,
       volume = {489},
       number = {3},
        pages = {3519-3538},
          doi = {10.1093/mnras/stz2382},
archivePrefix = {arXiv},
       eprint = {1903.02003},
 primaryClass = {astro-ph.GA},
       adsurl = {https://ui.adsabs.harvard.edu/abs/2019MNRAS.489.3519C}
}

@ARTICLE{Harris2010,
       author = {{Harris}, William E.},
        title = "{A New Catalog of Globular Clusters in the Milky Way}",
      journal = {arXiv e-prints},
         year = 2010,
        month = dec,
          eid = {arXiv:1012.3224},
        pages = {arXiv:1012.3224},
          doi = {10.48550/arXiv.1012.3224},
archivePrefix = {arXiv},
       eprint = {1012.3224},
 primaryClass = {astro-ph.GA},
       adsurl = {https://ui.adsabs.harvard.edu/abs/2010arXiv1012.3224H}
}

@ARTICLE{Grillmair2025,
       author = {{Grillmair}, Carl J.},
        title = "{The Multiple Extended Tidal Tails of NGC 288}",
      journal = {\apj},
         year = 2025,
        month = jan,
       volume = {979},
       number = {1},
          eid = {75},
        pages = {75},
          doi = {10.3847/1538-4357/ada2ea},
archivePrefix = {arXiv},
       eprint = {2409.17361},
 primaryClass = {astro-ph.GA},
       adsurl = {https://ui.adsabs.harvard.edu/abs/2025ApJ...979...75G}
}

@ARTICLE{Starkman2020,
       author = {{Starkman}, Nathaniel and {Bovy}, Jo and {Webb}, Jeremy J.},
        title = "{An extended Pal 5 stream in Gaia DR2}",
      journal = {\mnras},
         year = 2020,
        month = apr,
       volume = {493},
       number = {4},
        pages = {4978-4986},
          doi = {10.1093/mnras/staa534},
archivePrefix = {arXiv},
       eprint = {1909.03048},
 primaryClass = {astro-ph.GA},
       adsurl = {https://ui.adsabs.harvard.edu/abs/2020MNRAS.493.4978S}
}

@ARTICLE{Valenzuela2024,
       author = {{Valenzuela}, Lucas M. and {Remus}, Rhea-Silvia and {McKenzie}, Madeleine and {Forbes}, Duncan A.},
        title = "{Galaxy archaeology for wet mergers: Globular cluster age distributions in the Milky Way and nearby galaxies}",
      journal = {\aap},
         year = 2024,
        month = jul,
       volume = {687},
          eid = {A104},
        pages = {A104},
          doi = {10.1051/0004-6361/202348010},
archivePrefix = {arXiv},
       eprint = {2309.11545},
 primaryClass = {astro-ph.GA},
       adsurl = {https://ui.adsabs.harvard.edu/abs/2024A&A...687A.104V}
}

@ARTICLE{Mateu2018,
       author = {{Mateu}, Cecilia and {Read}, Justin I. and {Kawata}, Daisuke},
        title = "{Fourteen candidate RR Lyrae star streams in the inner Galaxy}",
      journal = {\mnras},
         year = 2018,
        month = mar,
       volume = {474},
       number = {3},
        pages = {4112-4129},
          doi = {10.1093/mnras/stx2937},
archivePrefix = {arXiv},
       eprint = {1711.03967},
 primaryClass = {astro-ph.GA},
       adsurl = {https://ui.adsabs.harvard.edu/abs/2018MNRAS.474.4112M}
}

@ARTICLE{Medina2024,
       author = {{Medina}, Gustavo E. and {Mu{\~n}oz}, Ricardo R. and {Carlin}, Jeffrey L. and {Vivas}, A. Katherina and {Grebel}, Eva K. and {Mart{\'\i}nez-V{\'a}zquez}, Clara E. and {Hansen}, Camilla J.},
        title = "{Taking the pulse of the outer Milky Way with the Halo Outskirts With Variable Stars (HOWVAST) survey: an RR Lyrae density profile out to >200 kpc}",
      journal = {\mnras},
         year = 2024,
        month = jul,
       volume = {531},
       number = {4},
        pages = {4762-4780},
          doi = {10.1093/mnras/stae1137},
archivePrefix = {arXiv},
       eprint = {2402.14055},
 primaryClass = {astro-ph.GA},
       adsurl = {https://ui.adsabs.harvard.edu/abs/2024MNRAS.531.4762M}
}

@ARTICLE{Muraveva2025,
       author = {{Muraveva}, Tatiana and {Giannetti}, Andrea and {Clementini}, Gisella and {Garofalo}, Alessia and {Monti}, Lorenzo},
        title = "{Metallicity of RR Lyrae stars from the Gaia Data Release 3 catalogue computed with Machine Learning algorithms}",
      journal = {\mnras},
         year = 2025,
        month = jan,
       volume = {536},
       number = {3},
        pages = {2749-2769},
          doi = {10.1093/mnras/stae2679},
archivePrefix = {arXiv},
       eprint = {2407.05815},
 primaryClass = {astro-ph.SR},
       adsurl = {https://ui.adsabs.harvard.edu/abs/2025MNRAS.536.2749M}
}

@ARTICLE{VIRAC2,
       author = {{Smith}, Leigh C. and {Lucas}, Philip W. and {Koposov}, Sergey E. and {Gonzalez-Fernandez}, Carlos and {Alonso-Garc{\'\i}a}, Javier and {Minniti}, Dante and {Sanders}, Jason L. and {Bedin}, Luigi R. and {Belokurov}, Vasily and {Evans}, N. Wyn and {Hempel}, Maren and {Ivanov}, Valentin D. and {Kurtev}, Radostin G. and {Saito}, Roberto K.},
        title = "{VIRAC2: NIR astrometry and time series photometry for 500M+ stars from the VVV and VVVX surveys}",
      journal = {\mnras},
         year = 2025,
        month = feb,
       volume = {536},
       number = {4},
        pages = {3707-3738},
          doi = {10.1093/mnras/stae2797},
archivePrefix = {arXiv},
       eprint = {2501.06295},
 primaryClass = {astro-ph.GA},
       adsurl = {https://ui.adsabs.harvard.edu/abs/2025MNRAS.536.3707S}
}

@ARTICLE{AlonsoGarcia2025,
       author = {{Alonso-Garc{\'\i}a}, Javier and {Smith}, Leigh C. and {Sanders}, Jason L. and {Minniti}, Dante and {Catelan}, M{\'a}rcio and {Aravena Rojas}, Gonzalo and {Carballo-Bello}, Julio A. and {Fern{\'a}ndez-Trincado}, Jos{\'e} G. and {Ferreira Lopes}, Carlos E. and {Garro}, Elisa R. and {Guo}, Zhen and {Hempel}, Maren and {Lucas}, Philip W. and {Majaess}, Daniel and {Saito}, Roberto K. and {Vivas}, A. Katherina},
        title = "{Variable stars in the VVV globular clusters: III. RR Lyrae stars in the inner Galactic globular clusters}",
      journal = {\aap},
         year = 2025,
        month = mar,
       volume = {695},
          eid = {A47},
        pages = {A47},
          doi = {10.1051/0004-6361/202453558},
archivePrefix = {arXiv},
       eprint = {2502.06504},
 primaryClass = {astro-ph.GA},
       adsurl = {https://ui.adsabs.harvard.edu/abs/2025A&A...695A..47A}
}

@ARTICLE{Matplotlib,
       author = {{Hunter}, John D.},
        title = "{Matplotlib: A 2D Graphics Environment}",
      journal = {Computing in Science and Engineering},
         year = 2007,
        month = may,
       volume = {9},
       number = {3},
        pages = {90-95},
          doi = {10.1109/MCSE.2007.55},
       adsurl = {https://ui.adsabs.harvard.edu/abs/2007CSE.....9...90H}
}

\begin{appendix}

\section{Globular clusters used for parameter calibration}
\label{app:GC_members}

In order to retrieve confirmed RRab members of known GCs within our \textit{Gaia} sample, we use the catalog of RR Lyrae stars in Galactic GCs presented in \citet{CruzReyes2024}. Specifically, we use their ``Final'' sample, which was obtained considering angular distance, parallaxes, PMs, and RVs (when available), and only includes variables located in the HB of their parent clusters.
On the other hand, for our VIVACE and Z24 samples, we use the catalog of RR Lyrae stars in inner Galactic GCs presented in \citet{AlonsoGarcia2025}, which makes use of VVV and VVVx data. We only use stars considered as bona fide or dubious cluster members by their analysis (i.e., those with Flag 0 or Flag 1 in their work).
A crossmatch is done between these catalogs and our respective final samples to identify the GC members that satisfied our quality cuts (see Sect.~\ref{sec:quality_cuts}). 

Analysis of the cluster members revealed that some GCs include RRab stars with small angular distance from the cluster center, but with remarkably different PMs and/or heliocentric distances from the mean values of the GC. Even if these stars are bona fide cluster members, their inclusion in the calibration of the clustering parameters could lead to an overestimation of the Galactic coordinates' relevance and a corresponding underestimation of the contribution of PMs and distances.
Therefore, we adopt a criterion based on PMs and distances to remove these stars from our calibrating samples. First, we keep stars for which the total PM difference from the cluster's PM was consistent within $3 \sigma$, with a threshold of $0.5$ mas yr$^{-1}$. Specifically, a star was retained if it satisfied the condition

\begin{equation}
    \sqrt{(\mu_{\ell}^{\mathrm{star}} - \mu_{\ell}^{\mathrm{GC}})^2 + (\mu_{b}^{\mathrm{star}} - \mu_{b}^{\mathrm{GC}})^2} < 0.5 + 3 \, \sigma_{\mu},
    \label{eq:PM_cut}
\end{equation}

\noindent where $\sigma_{\mu}$ is the uncertainty in the stars’ total PM, propagated as

\begin{equation}
    \sigma_{\mu} = \sqrt{\sigma_{\mu_{\ell}}^2 + \sigma_{\mu_{b}}^2}.
    \label{eq:PM_error}
\end{equation}

For the distance-based criterion, we keep stars whose distances are within $3 \sigma$ of a relative threshold of $7.5\%$ around the cluster’s distance. That is, stars were retained if their distances satisfied 

\begin{equation}
    |d_{\mathrm{star}} - d_{\mathrm{GC}}| < 0.075 \, d_{\mathrm{GC}} + 3 \, \sigma_{d_{star}}.
    \label{eq:dist_cut}
\end{equation}

The PMs of the clusters were retrieved from the GGCD, except for 2MASS-GC02, for which we use the PMs provided by \citet[][see their Sect.~6 for details on the possibly inaccurate PMs derived based on \textit{Gaia} data]{AlonsoGarcia2021}.
The distances of the GCs were also retrieved from the GGCD.

This cleaning procedure removed $18\%$, $26\%$, and $13\%$ of the GC members included in our final \textit{Gaia}, Z24, and VIVACE samples, respectively. The GCs found with at least $2$ RRab members in our final samples, after the PM and distance cleaning, are used in the calibration of our clustering parameters. These clusters are included in Table~\ref{tab:calibrating_GCs}, in addition to their adopted distances and metallicities (both of which are also used in Sect.~\ref{sec:GC_distanceFeH}) from the literature, and their references.

\begin{table}[h]
    \caption{GCs used for the calibration of the clustering parameters in each sample.}
    \resizebox{\columnwidth}{!}{%
    \centering
    \begin{tabular}{ccccc}
        \hline
        \hline
        Sample & Cluster & $N_{\rm RRab}$\tablefootmark{a} & [Fe/H]$_{\rm lit}$\tablefootmark{b} & $d_{\rm lit}$\tablefootmark{c} \\
         &  &  & (dex) & (kpc) \\
        \hline
        \textit{Gaia} & NGC 3201 & $62$ & $-1.59$ & $4.74$ \\
        \textit{Gaia} & NGC 6266 (M62) & $47$ & $-1.18$ & $6.03$ \\
        \textit{Gaia} & NGC 6715 (M54) & $32$ & $-1.49$ & $26.28$ \\
        \textit{Gaia} & NGC 6402 (M14) & $18$ & $-1.28$ & $9.14$ \\
        \textit{Gaia} & NGC 6401 & $13$ & $-1.02$ & $7.44$ \\
        \textit{Gaia} & Rup 106 & $10$ & $-1.68$ & $20.71$ \\
        \textit{Gaia} & NGC 4833 & $4$ & $-1.85$ & $6.48$ \\
        \textit{Gaia} & NGC 6642 & $4$ & $-1.26$ & $8.05$ \\
        \textit{Gaia} & NGC 6333 (M9) & $3$ & $-1.77$ & $8.30$ \\
        \textit{Gaia} & NGC 6441 & $3$ & $-0.46$ & $12.73$ \\
        \textit{Gaia} & BH 140 & $3$ & $-1.72^{1}$ & $4.81$ \\
        \textit{Gaia} & NGC 6284 & $3$ & $-1.26$ & $14.21$ \\
        \textit{Gaia} & NGC 6712 & $3$ & $-1.02$ & $7.38$ \\
        \textit{Gaia} & NGC 5986 & $2$ & $-1.59$ & $10.54$ \\
        \textit{Gaia} & NGC 6626 (M28) & $2$ & $-1.32$ & $5.37$ \\
        \textit{Gaia} & NGC 6235 & $2$ & $-1.28$ & $11.94$ \\
        \textit{Gaia} & FSR 1758 & $2$ & $-1.50^{2}$ & $10.19$ \\
        VIVACE & NGC 6401 & $19$ & $-1.02$ & $7.44$ \\
        VIVACE & NGC 6441 & $9$ & $-0.46$ & $12.73$ \\
        VIVACE & NGC 6626 (M28) & $8$ & $-1.32$ & $5.37$ \\
        VIVACE & NGC 6642 & $8$ & $-1.26$ & $8.05$ \\
        VIVACE & NGC 6638 & $7$ & $-0.95$ & $9.78$ \\
        VIVACE & NGC 6656 (M22) & $6$ & $-1.70$ & $3.30$ \\
        VIVACE & NGC 6380 & $5$ & $-0.75$ & $9.61$ \\
        VIVACE & NGC 6558 & $5$ & $-1.32$ & $7.79$ \\
        VIVACE & FSR 1716 & $5$ & $-1.38^{3}$ & $7.43$ \\
        VIVACE & NGC 6522 & $3$ & $-1.34$ & $7.29$ \\
        VIVACE & NGC 6569 & $3$ & $-0.76$ & $10.53$ \\
        VIVACE & FSR 1735 & $2$ & $-0.90^{4}$ & $9.08$ \\
        VIVACE & NGC 6453 & $2$ & $-1.50$ & $10.07$ \\
        VIVACE & NGC 6540 & $2$ & $-1.35$ & $5.91$ \\
        Z24 & Terzan 10 & $6$ & $-1.00$ & $10.21$ \\
        Z24 & HP 1 & $4$ & $-1.00$ & $7.00$ \\
        Z24 & 2MASS-GC02 & $4$ & $-1.08$ & $5.50$ \\
        Z24 & Djorg 2 & $3$ & $-0.65$ & $8.76$ \\
        Z24 & NGC 6544 & $3$ & $-1.40$ & $2.58$ \\
        Z24 & Terzan 1 & $3$ & $-1.03$ & $5.67$ \\
        Z24 & VVV-CL160 & $2$ & $-1.40^{5}$ & $6.80$ \\
        \hline
    \end{tabular}}
    \tablefoot{
    \tablefoottext{a}{Number of RRab members the GCs have in the respective final sample, after the PM and distance cleaning detailed in the text.}
    \tablefoottext{b}{Literature [Fe/H] values adopted for our study. Retrieved from the \citet{Harris2010} catalog. For GCs not included in this catalog, the specific references for the [Fe/H] values are provided.}
    \tablefoottext{c}{Literature distances adopted for our study. Retrieved from the GGCD for all GCs.}
    }
    \tablebib{(1) \citet{Prudil2024b}; (2) \citet{Barbá2019}; (3) \citet{Koch2017}; (4) \citet{Carballo-Bello2016}; (5) \citet{Minniti2021VVVCL160}.}
    \label{tab:calibrating_GCs}
\end{table}

\section{New members of known GCs}
\label{app:new_GC_members}

To check if we recover new RRab members of known GCs, for each of them we combine all of the groups including at least one confirmed member, and look for stars not included in the literature \citep[i.e.,][]{Clement2001, CruzReyes2024, AlonsoGarcia2025} as members that satisfy our PM and distance criteria described in Appendix~\ref{app:GC_members}.
We do not include a criterion based on the angular distance from the cluster center so that possible extra-tidal stars recovered by our clustering algorithm are not discarded.

\begin{figure}
    \begin{center}
    \includegraphics[width=0.9\columnwidth]{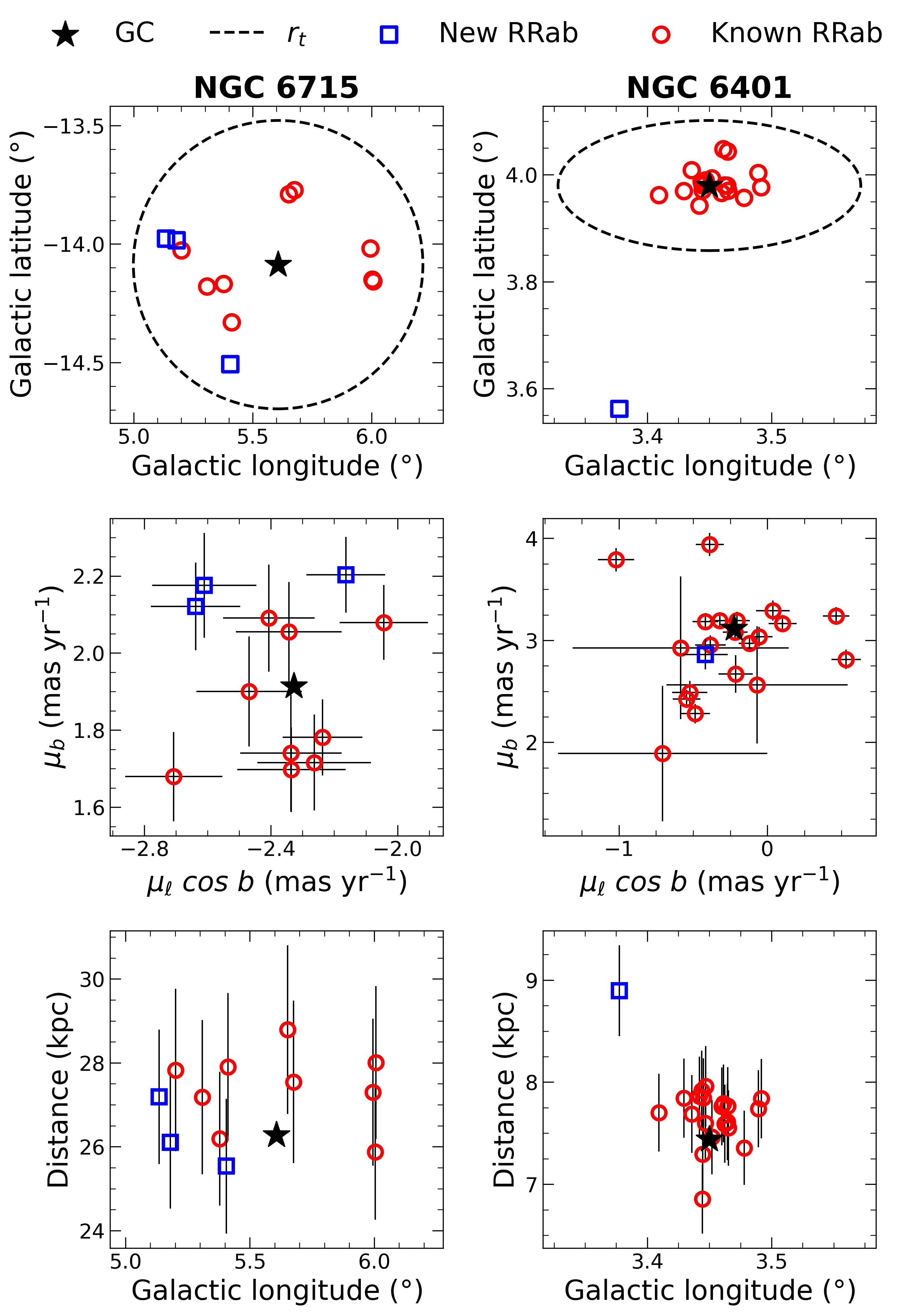}
    \end{center}
    \caption{Overview of the proposed new GC members found with our clustering procedure. The parameters of the GCs are shown as black stars, while their known RRab members in our quality samples are depicted with red circles and their proposed new RRab members as blue squares. Additionally, the tidal radii of the GCs ($r_t$, from the GGCD) are represented by black-dashed circles in the top panels.}
    \label{fig:new_members_properties}
\end{figure}

\begin{figure}
    \begin{center}
    \includegraphics[width=0.9\columnwidth]{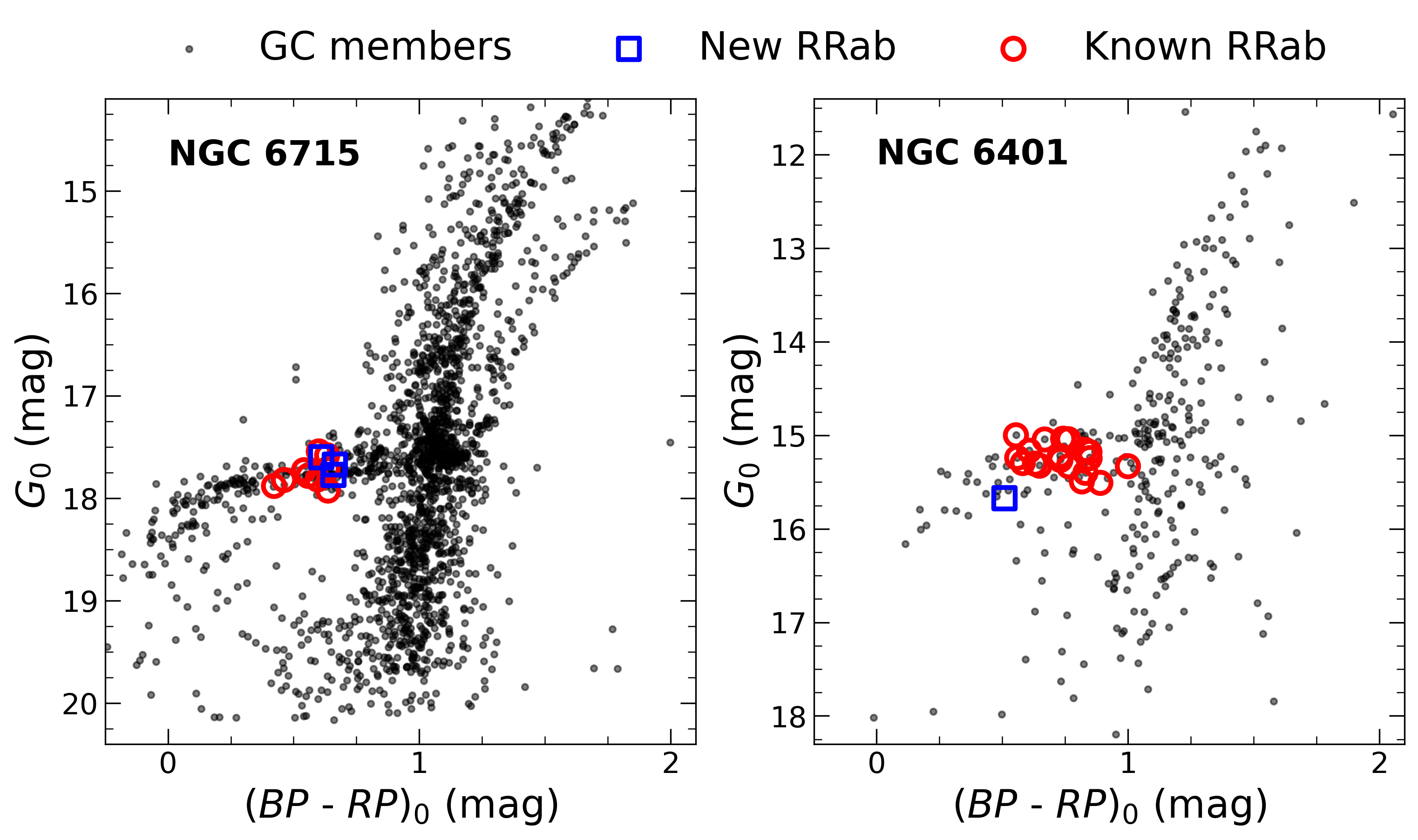}
    \end{center}
    \caption{Extinction-corrected \textit{Gaia} CMD of NGC~6715 (left panel) and NGC~6401 (right panel), with the loci of their known (red circles) and proposed new (blue squares) RRab members shown.}
    \label{fig:new_members_CMDs}
\end{figure}

With these criteria, we find three possible new members of NGC~6715~= M54 (in the \textit{Gaia} sample) and one possible new member in both NGC~6401 and FSR~1716 (both in the VIVACE sample). However, the possible new member in FSR~1716 showed significantly different PMs from the mean values for the GC (by $\sim3.5$ mas yr$^{-1}$; it only satisfied our criteria due to large PM uncertainties), so we discard it.
The positions, PMs, and distances of these proposed new members, compared to the previously known members \citep[extracted from the Galactic Globular Cluster Database\footnote{\url{https://people.smp.uq.edu.au/HolgerBaumgardt/globular/}}, hereafter GGCD;][March 2023 update]{Baumgardt2021}, are shown in Fig.~\ref{fig:new_members_properties}. 
Additionally, to check if these stars are located in the HB of their respective GC, in Fig.~\ref{fig:new_members_CMDs} we compare their positions in the extinction-corrected color-magnitude diagram (CMD).
GC member stars in \textit{Gaia} DR3 are retrieved by performing a circular query centered in the GC coordinates, with a radius of $5$ arcminutes, a PM radius of $0.5$ mas yr$^{-1}$, and a maximum $G$-band magnitude of $20.5$ mag. We do note that our query for NGC~6715 is slighly contaminated by Sagittarius (Sgr) dSph stars, which form an RGB sequence redder than that of the GC in the CMD shown in Fig.~\ref{fig:new_members_CMDs}.
The extinction and reddening values are adopted from the SFD dust map \citep[recalibrated as recommended in][]{Schlafly2011} and transformed to $E(BP - RP)$ and $\mathcal{A}_G$ values using the relative extinction values presented in Table~3 of \citet{Wang2019}.

As shown in Figs.~\ref{fig:new_members_properties}~and~\ref{fig:new_members_CMDs}, the three stars found as possible new members of NGC~6715\footnote{Their \textit{Gaia} DR3 identifiers are $6760319996108611456$, $6760320515827606912$, and $6760355837612159104$.} are inside its tidal radius (retrieved from the GGCD), they show very similar kinematics and distances with the GC, and they are located on top of the HB in the CMD. All this evidence suggests that these stars are indeed members of NGC~6715.

On the other hand, the possible new member of NGC~6401\footnote{Its \textit{Gaia} DR3 identifier is $4068294216580911488$.} shows similar PMs to the GC, but it lies outside of its tidal radius and is $\sim 1.5$ kpc further from the Sun than the GC. Additionally, this star is located slightly below the HB of NGC~6401; however, this difference in apparent $G$-band magnitude can be explained by the distance difference between this star and the GC (which translates to a $\Delta G \simeq 0.33$ mag). Thus, it is possible that this star is a former member of the GC which became unbound in a previous stage of its orbit. To assess this possibility, we compared the properties of this star with the mock extra-tidal stars of NGC~6401 presented in \citet{Grondin2024}. Although we found that some mock stars share similar positions and distances to our RRab star, these stars have significantly different PMs (by $\sim 6$ mas yr$^{-1}$). Therefore, without an estimate of the RV for this star, there is not enough evidence to confirm its association to the cluster.

\section{Globular cluster distances in the Z24 and VIVACE samples}
\label{app:GC_dists_VVV}

As shown in Fig.~\ref{fig:ClusterFeHsDistances}, we systematically overestimate the distances of GCs present in our Z24 and VIVACE samples by $\sim$$10\%$, when compared to estimates provided by \citet{Baumgardt2021}. This overestimation may be caused by the systematic offset present in the photometric metallicities of the Z24 and VIVACE samples \citepalias[obtained with the relation from][see Sect.~\ref{sec:GC_distanceFeH}]{Dekany2022}. To explore this, we recompute the distances of these GCs' members following the steps listed in Sect.~\ref{sec:dist}, but adopting the literature metallicity of their host cluster (see Sect.~\ref{sec:GC_distanceFeH} for the origin of these metallicities), instead of their individual RRab photometric metallicities, when using the PLZ relations from \citet{Prudil2024}.

\begin{figure}
    \begin{center}
    \includegraphics[width=\columnwidth]{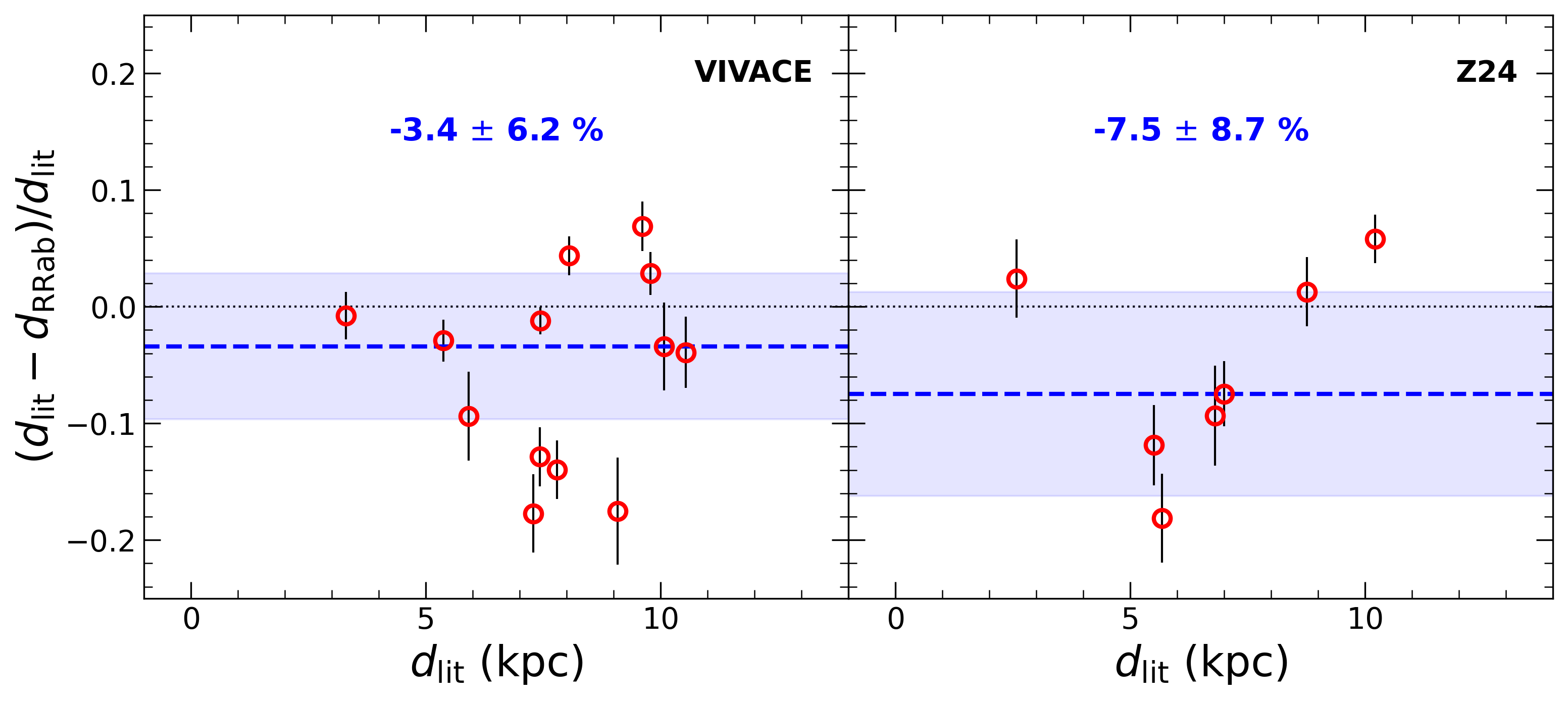}
    \end{center}
    \caption{Same as the lower-left and lower-middle panels of Fig.~\ref{fig:ClusterFeHsDistances}, but adopting the literature metallicity of the stars' host cluster (instead of their photometric metallicities) when using the PLZ relations from \citet{Prudil2024} to derive distances.}
    \label{fig:GC_dists_using_literature_FeHs}
\end{figure}

In Fig.~\ref{fig:GC_dists_using_literature_FeHs} we compare the mean values of the GCs' recomputed distances against the estimates provided by \citet{Baumgardt2021}. The median relative difference in the distances decreased by $\sim$ $3.5\%$ in both the Z24 and VIVACE samples, which shows that part of the offset observed in the distances of these samples originates from using the photometric [Fe/H] relation provided by \citetalias{Dekany2022}.
Notably, when adopting the [Fe/H] values for the GCs from the literature, several of them still exhibit relative distance differences larger than $10 \%$. This may be attributed to the fact that, for some of these GCs, distance estimates in the literature can vary significantly across different studies. Notable cases include (but are not limited to) FSR~1716 and NGC~6522, which have a $2-3$ kpc difference between the mean distance value provided in the GGCD and the distances obtained by \citet{Baumgardt2021} based on \textit{Gaia} early DR3 data.
With this in mind, we believe our estimates can be used to constrain the distances of the RRab groups found within our Z24 and VIVACE samples with at least $10\%$ precision.

\section{Using a larger \texttt{min\_cluster\_size}}
\label{app:min_3}

As argued in Sect.~\ref{sec:clustering}, even a pair of RR Lyrae stars close to each other in the sky, with similar movements and distances, can be a significant indicator of an underlying substructure. However, choosing a value of $2$ for the \texttt{min\_cluster\_size} may in part be causing the identification of an excess number of groups with our clustering algorithm.
To test this hypothesis, we ran our clustering algorithm on the \textit{Gaia} sample using a value of $3$ for this parameter (with the other parameters, shown in Table~\ref{tab:HDBSCAN_parameters}, kept unchanged), and inspected the consequences this change has on the quantity and properties of the groups obtained.

\begin{figure}
    \begin{center}
    \includegraphics[width=1.\columnwidth]{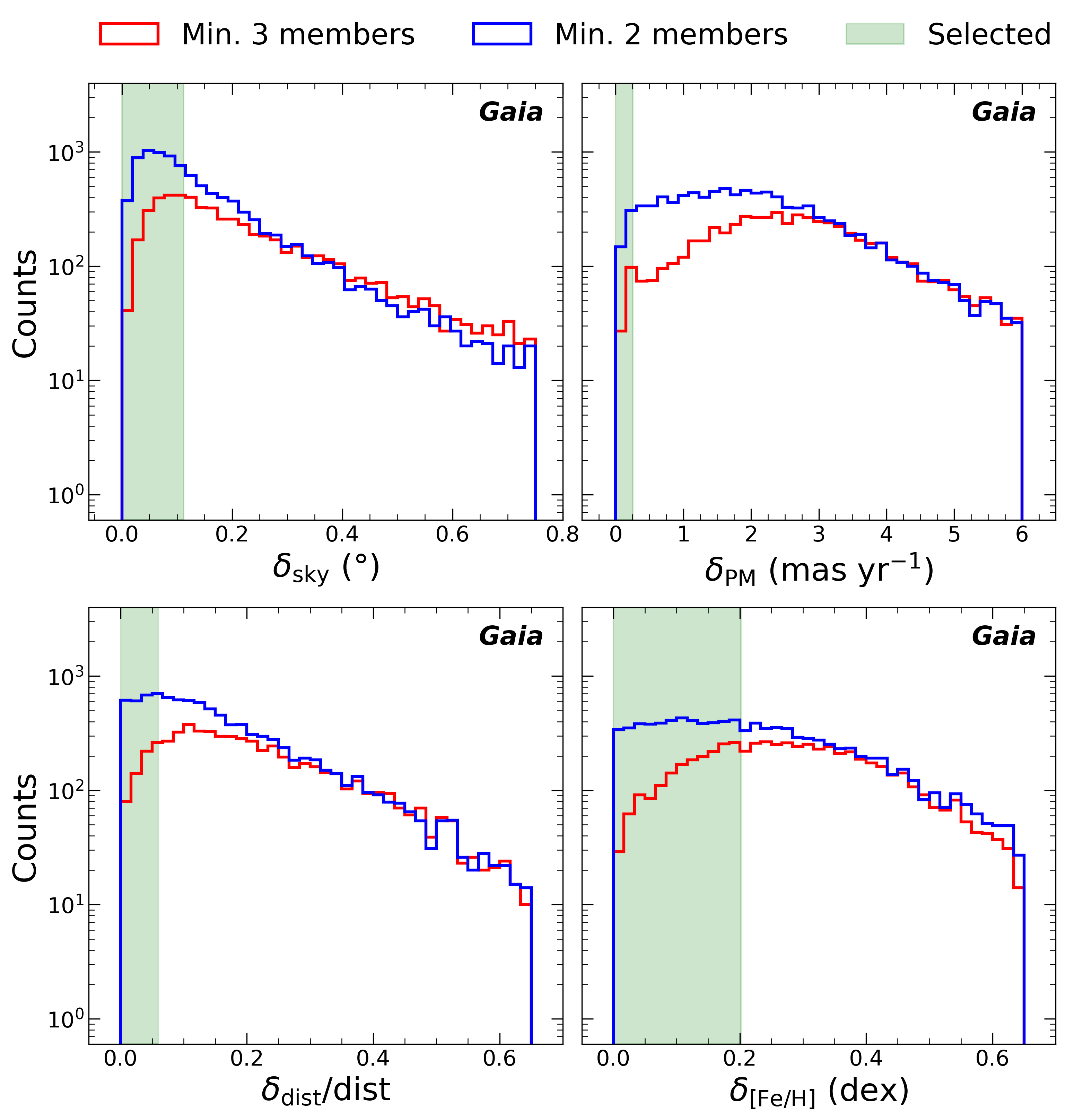}
    \end{center}
    \caption{Same as Fig.~\ref{fig:metric_distributions}, but comparing the metric distributions obtained for the \textit{Gaia} sample when using \texttt{min\_cluster\_size} values of $2$ (blue histograms) and $3$ (red histograms).}
    \label{fig:metrics_min_two_and_three}
\end{figure}

This clustering run returns a total of $5951$ groups ($\sim 60 \%$ as many groups as found with a \texttt{min\_cluster\_size} of $2$), $339$ of which are pairs, as these can still be formed by the post processing done after running HDBSCAN in the $1000$ Monte Carlo resamplings of the data (see Sect.\ref{sec:clustering}). To inspect the properties of these groups, in Fig.~\ref{fig:metrics_min_two_and_three} we compare their distributions in the compactness metrics to those of the groups found when using a \texttt{min\_cluster\_size} of $2$.

As shown here, when using a \texttt{min\_cluster\_size} of $3$, we recover practically the same amount of groups showing large scatter within the 6D space studied, while significantly reducing the number of compact groups obtained.
This shows that, although increasing the value of \texttt{min\_cluster\_size} reduces the number of groups obtained, it leads to the loss of the most compact groups with few members, while not improving the properties (i.e., how compact they are in the 6D space) of the larger groups that are still recovered.

\section{Properties of the HDBSCAN groups}
\label{app:compact_groups_pairs_tables}


The number of RRab stars ($N_{\rm RRab}$) and the mean Galactic coordinates, PMs, (photometric) [Fe/H], and distance of the compact groups and pairs we found are presented in Tables~\ref{tab:compact_groups}~and~\ref{tab:compact_pairs}, respectively.

\begin{table}
    \caption{Properties of the most compact groups found in our RRab samples with at least three stars.}
    \resizebox{1\columnwidth}{!}{%
    \centering
    \begin{tabular}{cccccccc}
        \hline
        \hline
        ID\tablefootmark{a} & N$_{\rm RRab}$ & $\langle \ell \rangle$ & $\langle b \rangle$ & $\langle \mu_{\ell}$ cos b$\rangle$ & $\langle \mu_b \rangle$ & $\langle$[Fe/H]$\rangle$ & $\langle d \rangle$ \\
         &  & $(\degree)$ & $(\degree)$ & (mas yr$^{-1}$) & (mas yr$^{-1}$) & (dex) & (kpc) \\
        \hline
        G1 & $3$ & $5.00$ & $-8.50$ & $-2.37$ & $2.12$ & $-1.43$ & $25.5$ \\
        G2 & $3$ & $7.72$ & $-10.94$ & $-2.43$ & $1.66$ & $-1.44$ & $27.0$ \\
        G3 & $3$ & $5.82$ & $-13.05$ & $-2.45$ & $2.07$ & $-1.47$ & $26.8$ \\
        G4 & $3$ & $5.45$ & $-13.56$ & $-2.42$ & $1.64$ & $-1.33$ & $26.2$ \\
        G5 & $5$ & $5.90$ & $-12.60$ & $-2.37$ & $1.84$ & $-1.58$ & $26.9$ \\
        G6 & $3$ & $3.57$ & $-11.92$ & $-2.45$ & $1.92$ & $-1.53$ & $26.5$ \\
        G7 & $4$ & $5.43$ & $-11.67$ & $-2.44$ & $1.85$ & $-1.37$ & $25.8$ \\
        G8 & $3$ & $6.31$ & $-11.76$ & $-2.42$ & $1.97$ & $-1.67$ & $25.5$ \\
        G9 & $4$ & $5.01$ & $-12.08$ & $-2.20$ & $1.94$ & $-1.27$ & $27.7$ \\
        G10 & $3$ & $7.51$ & $-14.57$ & $-2.35$ & $1.80$ & $-1.49$ & $26.9$ \\
        G11 & $3$ & $5.30$ & $-10.88$ & $-2.4$ & $2.05$ & $-1.25$ & $24.8$ \\
        G12 & $3$ & $5.28$ & $-11.58$ & $-2.32$ & $1.93$ & $-1.46$ & $28.5$ \\
        Z1 & $3$ & $3.47$ & $2.29$ & $-7.52$ & $0.88$ & $-1.38$ & $8.2$ \\
        Z2 & $3$ & $2.98$ & $-1.76$ & $-8.90$ & $-1.59$ & $-1.28$ & $8.1$ \\
        Z3 & $3$ & $0.23$ & $-1.93$ & $-9.71$ & $2.48$ & $-1.35$ & $8.4$ \\
        Z4 & $3$ & $-2.58$ & $-2.33$ & $-7.77$ & $2.03$ & $-1.42$ & $5.9$ \\
        Z5 & $3$ & $1.75$ & $-1.30$ & $-6.93$ & $3.38$ & $-1.29$ & $8.4$ \\
        Z6 & $3$ & $3.48$ & $-1.28$ & $-6.87$ & $-1.76$ & $-1.28$ & $8.2$ \\
        Z7 & $3$ & $2.26$ & $-1.84$ & $-8.73$ & $-0.51$ & $-1.40$ & $8.6$ \\
        Z8 & $3$ & $1.05$ & $-1.64$ & $-4.59$ & $-0.69$ & $-1.32$ & $8.3$ \\
        Z9 & $3$ & $3.82$ & $-0.75$ & $-4.61$ & $0.03$ & $-1.41$ & $8.0$ \\
        Z10 & $3$ & $0.55$ & $2.65$ & $-6.45$ & $-1.92$ & $-1.52$ & $8.1$ \\
        V1 & $3$ & $-2.74$ & $-7.20$ & $-8.48$ & $-1.12$ & $-1.19$ & $9.2$ \\
        V2 & $3$ & $4.79$ & $3.76$ & $-7.33$ & $-0.87$ & $-1.44$ & $7.5$ \\
        V3 & $3$ & $-0.53$ & $3.56$ & $-6.28$ & $-1.04$ & $-1.34$ & $8.7$ \\
        V4 & $3$ & $1.95$ & $-5.25$ & $-4.97$ & $-0.87$ & $-1.45$ & $8.3$ \\
        V5 & $3$ & $-0.56$ & $-3.57$ & $-8.24$ & $2.71$ & $-1.28$ & $9.9$ \\
        \hline
    \end{tabular}}
    \tablefoot{
    \tablefoottext{a}{Groups in the \textit{Gaia}, Z24, and VIVACE samples have G, Z, and V as the first character in their ID, respectively, and are numbered according to their stability (e.g., the most stable group in the \textit{Gaia} sample is G1).}
    }
    \label{tab:compact_groups}
\end{table}

\begin{table}
    \caption{Same as Table~\ref{tab:compact_groups}, but for the compact pairs that were not identified as potentially associated to the Galactic bulge or the Sgr dSph.}
    \resizebox{1\columnwidth}{!}{%
    \centering
    \begin{tabular}{cccccccc}
        \hline
        \hline
        ID & N$_{\rm RRab}$ & $\langle \ell \rangle$ & $\langle b \rangle$ & $\langle \mu_{\ell}$ cos b$\rangle$ & $\langle \mu_b \rangle$ & $\langle$[Fe/H]$\rangle$ & $\langle d \rangle$ \\
         &  & $(\degree)$ & $(\degree)$ & (mas yr$^{-1}$) & (mas yr$^{-1}$) & (dex) & (kpc) \\
        \hline
        G1p & $2$ & $37.76$ & $-5.54$ & $-1.59$ & $-0.22$ & $-1.09$ & $15.8$ \\
        G2p & $2$ & $43.89$ & $13.17$ & $-1.73$ & $0.27$ & $-1.26$ & $23.6$ \\
        G3p & $2$ & $57.98$ & $10.22$ & $-0.28$ & $0.43$ & $-1.71$ & $13.2$ \\
        G4p & $2$ & $-114.38$ & $5.44$ & $0.82$ & $-0.15$ & $-1.60$ & $21.2$ \\
        G5p & $2$ & $46.97$ & $-4.72$ & $-1.01$ & $-0.44$ & $-1.20$ & $16.8$ \\
        G6p & $2$ & $-18.25$ & $-14.34$ & $-1.65$ & $0.13$ & $-1.30$ & $28.7$ \\
        G7p & $2$ & $-116.88$ & $12.42$ & $0.64$ & $-0.58$ & $-1.38$ & $26.1$ \\
        G8p & $2$ & $-7.98$ & $9.50$ & $-4.01$ & $-1.76$ & $-1.09$ & $14.0$ \\
        G9p & $2$ & $-41.66$ & $-6.43$ & $-2.25$ & $-0.04$ & $-1.40$ & $21.1$ \\
        G10p & $2$ & $40.33$ & $-13.07$ & $-1.26$ & $-0.12$ & $-1.19$ & $23.7$ \\
        G11p & $2$ & $-62.55$ & $8.80$ & $-0.74$ & $-0.20$ & $-1.39$ & $22.6$ \\
        G12p & $2$ & $63.54$ & $-8.58$ & $-0.82$ & $-0.39$ & $-1.32$ & $16.9$ \\
        G13p & $2$ & $-31.96$ & $-10.74$ & $-3.06$ & $-0.55$ & $-1.33$ & $16.3$ \\
        G14p & $2$ & $1.85$ & $9.21$ & $-7.12$ & $2.42$ & $-1.25$ & $11.9$ \\
        G15p & $2$ & $-26.77$ & $9.98$ & $-2.14$ & $-0.27$ & $-1.28$ & $27.6$ \\
        G16p & $2$ & $-8.07$ & $7.10$ & $-2.30$ & $0.26$ & $-1.52$ & $16.0$ \\
        G17p & $2$ & $-13.63$ & $-6.81$ & $-3.16$ & $0.29$ & $-1.05$ & $18.8$ \\
        Z1p & $2$ & $0.61$ & $1.60$ & $-13.94$ & $0.28$ & $-1.32$ & $8.6$ \\
        Z2p & $2$ & $0.42$ & $2.49$ & $-7.28$ & $6.42$ & $-1.90$ & $8.6$ \\
        Z3p & $2$ & $5.85$ & $2.25$ & $-6.90$ & $-0.55$ & $-1.28$ & $11.7$ \\
        Z4p & $2$ & $5.12$ & $-1.70$ & $0.25$ & $-1.69$ & $-1.36$ & $8.9$ \\
        Z5p & $2$ & $-0.88$ & $-1.77$ & $-7.17$ & $6.53$ & $-1.36$ & $8.5$ \\
        Z6p & $2$ & $4.99$ & $-2.38$ & $-12.04$ & $7.64$ & $-1.77$ & $8.2$ \\
        Z7p & $2$ & $1.49$ & $2.11$ & $-12.57$ & $-0.61$ & $-1.27$ & $9.5$ \\
        Z8p & $2$ & $0.99$ & $-1.46$ & $-6.72$ & $6.11$ & $-1.20$ & $7.9$ \\
        Z9p & $2$ & $1.75$ & $1.08$ & $-0.41$ & $2.67$ & $-1.32$ & $8.5$ \\
        Z10p & $2$ & $1.66$ & $-1.84$ & $-1.48$ & $-3.56$ & $-1.41$ & $7.6$ \\
        Z11p & $2$ & $0.99$ & $-1.93$ & $-6.75$ & $-6.00$ & $-1.27$ & $7.8$ \\
        V1p & $2$ & $-6.57$ & $-4.44$ & $-5.96$ & $-1.95$ & $-1.33$ & $13.6$ \\
        V2p & $2$ & $3.62$ & $-3.33$ & $-6.87$ & $-6.36$ & $-1.18$ & $7.4$ \\
        V3p & $2$ & $-4.55$ & $3.56$ & $-8.23$ & $6.52$ & $-1.50$ & $8.9$ \\
        V4p & $2$ & $-1.14$ & $4.32$ & $-6.87$ & $0.70$ & $-1.62$ & $12.1$ \\
        V5p & $2$ & $3.23$ & $4.25$ & $0.82$ & $0.21$ & $-1.21$ & $7.6$ \\
        V6p & $2$ & $0.89$ & $3.90$ & $-0.37$ & $1.82$ & $-1.51$ & $8.2$ \\
        V7p & $2$ & $-7.12$ & $-4.50$ & $-3.48$ & $-0.30$ & $-1.68$ & $14.1$ \\
        V8p & $2$ & $-3.67$ & $3.55$ & $-4.98$ & $1.70$ & $-1.39$ & $11.3$ \\
        V9p & $2$ & $-4.95$ & $-3.29$ & $-12.33$ & $3.21$ & $-1.61$ & $7.2$ \\
        \hline
    \end{tabular}}
    \label{tab:compact_pairs}
\end{table}

\section{Retrieving Galactic bulge and Sgr dSph stars}
\label{app:contours}

To compute the sky and PM contours of Sgr dSph stars (shown in the left panels of Fig.~\ref{fig:compact_groups}), we adopt the catalog produced by \citet{Vasiliev2021}, and only use stars which have $-5 < \ell(\degree) < 15 $ and $-20 < b(\degree) < 0 $ (to focus on the core of Sgr, where our compact groups reside).

For the Galactic bulge stars (whose contours are shown in the middle and right panels of Figs.~\ref{fig:compact_groups}~and~\ref{fig:compact_pairs_Z24_VIVACE}), on the other hand, we perform a circular query on the \textit{Gaia} archive, centered at ($\ell$, $b$) = ($0 \degree$, $0 \degree$), with a radius of $3 \degree$, including a $G$-band magnitude upper limit of $18$ mag, and a minimum distance \citep[using geometric distances from][]{Bailer-Jones2021} of $3$ kpc. This query yields $\sim 2 \times 10^6$ stars, from which we randomly selected $10^5$ sources to compute the contours.

\end{appendix}

\end{document}